\documentclass[aps,prc,twocolumn,superscriptaddress,nofootinbib,preprintnumbers]{revtex4-2}

\usepackage{xcolor}
\usepackage{amsmath}
\usepackage{amssymb}
\usepackage{graphicx}
\usepackage{multirow}
\allowdisplaybreaks

\begin{document}
\preprint{INT-PUB-24-033}

\title{Bulk viscosity of nuclear matter with pions in the neutrino-trapped regime}



\author{Steven P.~Harris}
\affiliation{Center for the Exploration of Energy and Matter and Department of Physics,
Indiana University, Bloomington, IN 47405, USA}
\affiliation{Institute for Nuclear Theory, University of Washington, Seattle, WA 98195, USA}
\author{Bryce Fore}
\affiliation{Physics Division, Argonne National Laboratory, Argonne, Illinois 60439, USA}
\author{Sanjay Reddy}
\affiliation{Institute for Nuclear Theory, University of Washington, Seattle, WA 98195, USA}

\date{December 29, 2024}
\begin{abstract}
Recent work [B.~Fore and S.~Reddy, Phys.~Rev.~C 101 035809 (2020)] has shown that a population of thermal pions could modify the equation of state and transport properties of hot and dense neutron-rich matter and introduce new reaction pathways to change the proton fraction. In this article we study their impact on the bulk viscosity of dense matter, focusing on the neutrino-trapped regime that would be realized in neutron star mergers and supernovae.  We find that the presence of a thermal population of pions alters the bulk viscosity by modifying the EoS (via the susceptibilities) and by providing new reaction pathways to achieve beta equilibrium.  In neutron star merger conditions, the bulk viscosity in neutrino-trapped $npe\mu$ matter (without pions) has its peak at temperatures of at most a couple MeV and is quite small at temperatures of tens of MeV.  We find that thermal pions enhance the low-temperature peak of the bulk viscosity by a factor of a few and shift it to slightly lower temperatures.  At higher temperatures, where the pion abundance is large but the bulk viscosity is traditionally small, pions can increase the bulk viscosity by an order of magnitude or more, although it is still orders of magnitude smaller than its peak value. 
\end{abstract}
\maketitle 
\section{Introduction}
Transport properties of hot and dense matter are expected to play a role in neutron star mergers and core-collapse supernovae. At high densities and temperatures encountered in these extreme astrophysical phenomena, neutrinos dominate energy and momentum transport \cite{Alford:2017rxf}. Long ago, bulk viscosity due to out-of-equilibrium weak reactions involving nucleons was shown to damp density oscillations on dynamical timescales relevant to neutron star oscillations \cite{Sawyer:1980wp,Sawyer:1989dp}.  Following initial estimates of the bulk viscosity by Alford \textit{et al.~}\cite{Alford:2017rxf}, there has been renewed interest in detailed calculations of the bulk viscosity in hot and dense nuclear matter with \cite{Alford:2021lpp,Alford:2019kdw} and without \cite{Alford:2023gxq,Yang:2023ogo,Alford:2023uih,Alford:2019qtm} trapped neutrinos.  Two of these studies included muons in addition to the usual neutrons, protons, and electrons.  Other recent studies have looked at exotic phases of matter including dense matter with hyperons \cite{Alford:2020pld} and quark matter \cite{Alford:2024tyj,CruzRojas:2024etx,Hernandez:2024rxi}.  For a review of bulk viscosity in neutron star environments, see the book chapter \cite{Harris:2024evy}.  Favorable results in these calculations, including the prediction of millisecond timescale bulk viscous damping in neutrino-transparent matter \cite{Alford:2019qtm}, have lead to the inclusion of weak-interaction driven bulk viscosity in neutron star merger simulations \cite{Camelio:2022fds,Most:2022yhe,Chabanov:2023blf,Espino:2023dei,Chabanov:2023abq,Radice:2021jtw,Zappa:2022rpd,Most:2021zvc}, which has been facilitated by improvements in neutrino transport schemes \cite{Zappa:2022rpd} and the relativistic hydrodynamics of multicomponent fluids \cite{Celora:2022nbp,Camelio:2022ljs,Gavassino:2023eoz,Yang:2023ogo,Gavassino:2020kwo}.

In addition to neutrons, protons, electrons, muons, and trapped neutrinos, a thermal population of negatively charged pions could be relevant at the high temperatures and densities encountered in neutron star mergers \cite{Fore:2019wib}. Thermal pions have been shown to soften the equation of state (EoS), enhance the proton content, and modify the neutrino opacity in dense matter \cite{Fore:2019wib}. The study also mentioned that pions provide a new pathway to equilibrate the proton fraction, thereby altering the out-of-equilibrium weak reactions that give rise to the bulk viscosity. 

In this work, we study the effect of thermal pions on bulk viscosity, considering their effects on out-of-equilibrium chemical reactions and on the EoS. We will fully develop the formalism for bulk viscosity in neutrino-trapped nuclear matter containing neutrons, protons, electrons, muons, and a thermal population of pions.  In $npe$ matter, the proton fraction is adequate to specify the composition since the neutron and electron densities can be determined via baryon number conservation and charge neutrality.  There is one chemical equilibration channel and therefore the bulk viscosity exhibits a single resonant peak. 
 For a given density, the resonance occurs at a temperature when the equilibration rate becomes comparable to the oscillation frequency \cite{Alford:2010gw,Alford:2019qtm}.  Systems with several independent particle species can have a bulk viscosity with a complicated temperature dependence with several distinct or overlapping resonances.  The rule of thumb is one resonance for each independent equilibration channel.  This complexity has been studied in earlier work for the case of $npe\mu$ matter \cite{Alford:2022ufz,Alford:2021lpp,Alford:2023uih,Harris:2024evy}, and in quark matter \cite{Alford:2006gy,Dong:2007mb,Sad:2007afd,Wang:2010ydb,Shovkovy:2010xk,Hernandez:2024rxi}. Here we perform a similar analysis for $npe\mu$ matter containing a population of thermal pions, taking into account the effect of pions on the out-of-equilibrium reactions and the thermodynamic properties such as the susceptibilities that influence the bulk viscosity. 

In Sec.~2, we discuss the model of nuclear matter and the treatment of the pion-nucleon interactions and pion dispersion relation.  In Sec.~3, we develop the formalism to calculate the bulk viscosity with pions and trapped neutrinos.  In Sec.~4, we discuss the results.  

We work in natural units, where $\hbar=c=k_B=1$. 
\section{Nuclear matter with thermal pions.}\label{sec:EoS}
The population of negatively charged thermal pions in neutron-rich matter at high density is enhanced due to a rapid increase in their associated chemical potential. The negative charge chemical potential, denoted by $\hat{\mu}=\mu_n-\mu_p$ where $\mu_n$ and $\mu_p$ are the neutron and proton chemical potentials, respectively, is needed to ensure electric charge neutrality and its magnitude (which depends on the nuclear symmetry energy) becomes comparable to the vacuum pion mass $m_\pi$. Consequently, in the absence of interactions, the pion number density increases exponentially as $ \exp{\left[(\hat{\mu}-m_\pi)/T\right]}$. Calculations in \cite{Fore:2019wib}, which includes pion-nucleon interactions that are based on the virial expansion, found that interactions increase the pion fraction at finite temperatures.  This study found that although s-wave interactions are repulsive, a strong and attractive p-wave interaction can greatly enhance the pion number at a modest temperature, motivating the study of the presence of pions on properties of hot, dense matter.

We consider matter composed of neutrons $n$, protons $p$, electrons $e^-$, muons $\mu^-$, and pions $\pi^-$, as well as neutrinos and anti-neutrinos of electron and muon flavors.   The system at a particular baryon density $n_B$, temperature $T$, conserved electron-type lepton fraction $Y_{Le}\equiv (n_e+n_{\nu_e})/n_B$ (number densities $n$ refer to the net densities, particle minus antiparticle), and conserved muon-type lepton fraction $Y_{L\mu}\equiv (n_{\mu}+n_{\nu_{\mu}})/n_B$ has three independent particle fractions, chosen to be $x_p$, $x_{\mu}$, and $x_{\pi}$.  In beta equilibrium, these particle fractions are functions of $n_B$, $T$, $Y_{Le}$, and $Y_{L\mu}$.  We treat the electron and muon lepton numbers to be separately conserved because we neglect neutrino oscillations, which are suppressed in dense matter \cite{Wolfenstein:1979ni}.

The neutron and proton sector of the equation of state is described here with a nonrelativistic Skyrme model called NRAPR, described in \cite{Fore:2019wib,Steiner:2004fi}.  To model interactions between pions and nucleons we use a construction of the pion self-energy, $\Sigma_{\pi^-}(p)$ which is structurally similar to the one introduced in \cite{Fore:2019wib}. The pion dispersion relation is given by

\begin{equation} 
    E_{\pi^-}(p) = \sqrt{p^2+m_\pi^2} + \Sigma_{\pi^-}(p)\,, 
    \label{eq:pion_dispersion}
\end{equation}

with the pion self-energy given by

\begin{align}
    &\Sigma_{\pi^-}(p) = \int \frac{d^3k}{(2\pi)^3}~\sum_{N=n,p} f_N(E_N(k)) ~V^{ps}_{N\pi^-}(p_{cm})\\
    &V^{(ps)}_{N\pi^-} (p_{\rm cm}) = - \sum_{I,l,\nu} \alpha_l (2l+1) \frac{ 2\pi\delta_{l,\nu}^{I}}{ {\bar m} ~p_{\rm cm}}, \nonumber
\end{align}
where the sum is over the isospin $I$, the angular momentum $l$, and the nucleon spin $\nu$.  The symbols $\bar{m}$ and $p_{\text{cm}}$ represent the reduced mass of the pion-nucleon system and the center of mass momentum, respectively.  The term $\delta_{l,\nu}^{I}$ is the phase shift in the given channel.  The modified pseudopotential now has two parameters $\alpha_l$ rather than the single value used in \cite{Fore:2019wib}. This change is motivated by the fact that the model under-represents the s-wave repulsion present in pion-nucleon interactions by suppressing the s-wave and p-wave interactions equally. To prevent this we set the parameter $\alpha_0$=1 and fit the value of the remaining parameter $\alpha_1$ such that the model produces the same pion fraction as the virial calculation presented in \cite{Fore:2019wib} at a baryon density equal to nuclear saturation density, a temperature of 30 MeV, and with both electron-type and muon-type conserved lepton fractions equal to 0.1. Unlike in \cite{Fore:2019wib}, where the value of $\alpha$ was explored at various temperatures and densities, our fitting parameter, $\alpha_1$, has a fixed value of 0.23596 for all conditions considered.

\begin{figure}\centering
\includegraphics[width=0.4\textwidth]{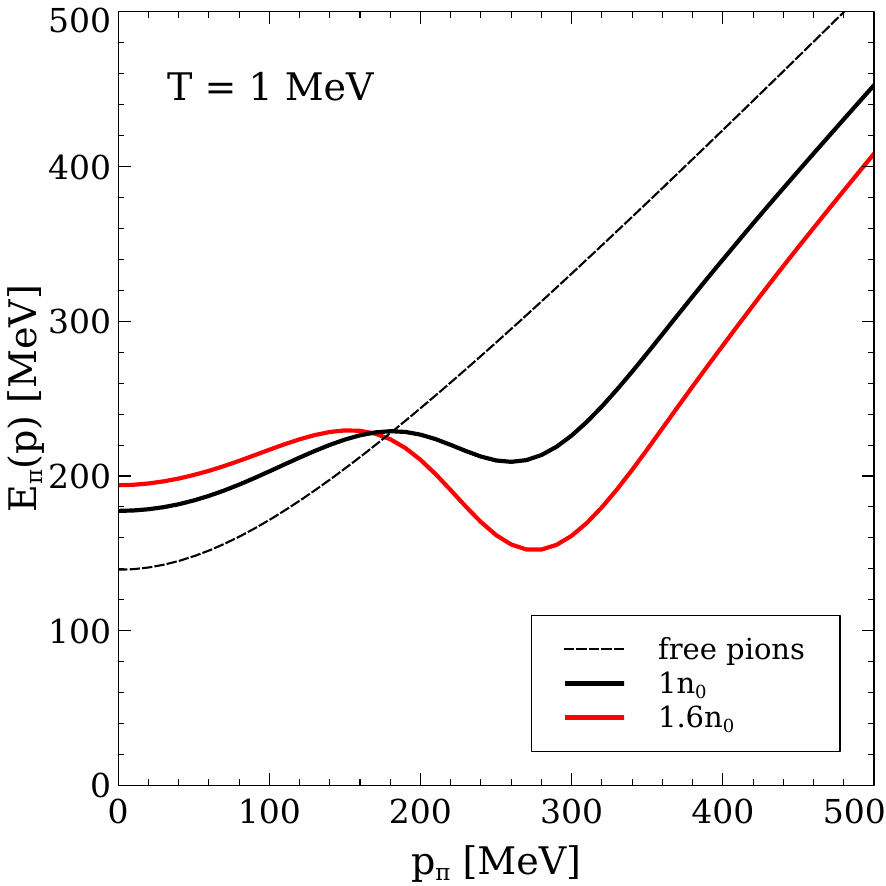}
\caption{The pion dispersion relation in vacuum (dashed line) and in dense matter, as predicted by our EoS (solid lines).  The pion ``mass'' [$E_{\pi^-}(p=0)$] is shifted from its vacuum value (139.6 MeV) to 177.4 MeV at $1n_0$ and to 194.0 MeV at $1.6n_0$.  The temperature is taken to be 1 MeV.}
\label{fig:pion_disp}
\end{figure}

The pion dispersion relation predicted by Eq.~\ref{eq:pion_dispersion} is shown in Fig.~\ref{fig:pion_disp}. As noted earlier, the shape of the dispersion curve can be understood as arising due to the competition between s-wave repulsion and strong p-wave attraction. The self-energy at zero momentum leads to a positive shift of the pion ``mass'' [$E_{\pi^-}(p=0)$] of about $40$ MeV at $n_B=n_0$ and by about $54$ MeV at $n_B=1.6~n_0$. These large shifts are consistent with recent calculations of the pion mass in dense neutron-rich matter reported in Ref.~\cite{Fore:2023gwv}. With increasing momentum, the self-energy becomes negative due to attractive p-wave interactions and produces the distinct non-monotonic behavior characterized by the maximum and minimum seen in the figure. Since the typical pion momentum $p_\pi \simeq 3 T$, with increasing temperature the low lying states at higher momentum will be preferentially occupied and the non-monotonic dispersion relation leads to a stronger temperature dependence of the thermal pion population as will be discussed below.        

Since we aim to cover the neutrino-trapped conditions that exist in supernovae and neutron star mergers, we choose a variety of $Y_{L,i}$ to mimic these conditions.  No neutron star merger simulation has, to this point, dynamically included muons, and therefore we refer to a post-processing analysis \cite{Loffredo:2022prq} which predicts that matter above saturation density could have $Y_{Le}$ in the range of 0.04-0.10 and $Y_{L\mu}$ in the range 0.01 - 0.07.  Supernovae simulations reliably predict $Y_{Le}$ to be between 0.3 and 0.4 \cite{Burrows:1990ts}.  Muons have been included in recent supernovae simulations \cite{Fischer:2020vie,Bollig:2017lki}, and since the net muon number $Y_{L\mu}$ in the supernova progenitor is close to zero, it remains that way throughout the duration of the supernova.  To study these different physical situations, we will focus on the two configurations $\{Y_{Le}, Y_{L\mu}\} = \{0.05,0\}$ (a good description of neutron star merger conditions) and $\{Y_{Le}, Y_{L\mu}\} = \{0.3,0\}$ (a good approximation of supernova conditions).  In the merger case, the two merging neutron stars are cold and do not contain a large muon population.  Upon rapid heating at the contact of the two stars, the muon number is frozen due to the trapping of neutrinos.  For supernovae, the matter is initially at low density and therefore does not contain a large muon population, and once the matter becomes dense enough to trap neutrinos, the muon number becomes frozen in.  While it is almost certainly not \textit{zero} in either physical situation, it is reasonable to choose zero as a representative case.  In the main text, we discuss the particle content, susceptibilities, equilibration rates, and bulk viscosity for matter with the above two choices of lepton fractions, and put the corresponding results for equal values of the conserved electron and muon fraction $\{Y_{Le}, Y_{L\mu}\} = \{0.05,0.05\} \text{ and } \{0.3,0.3\}$ in Appendix \ref{appendix:bv_finite_YLmu}. 

\begin{figure*}\centering
\includegraphics[width=0.4\textwidth]{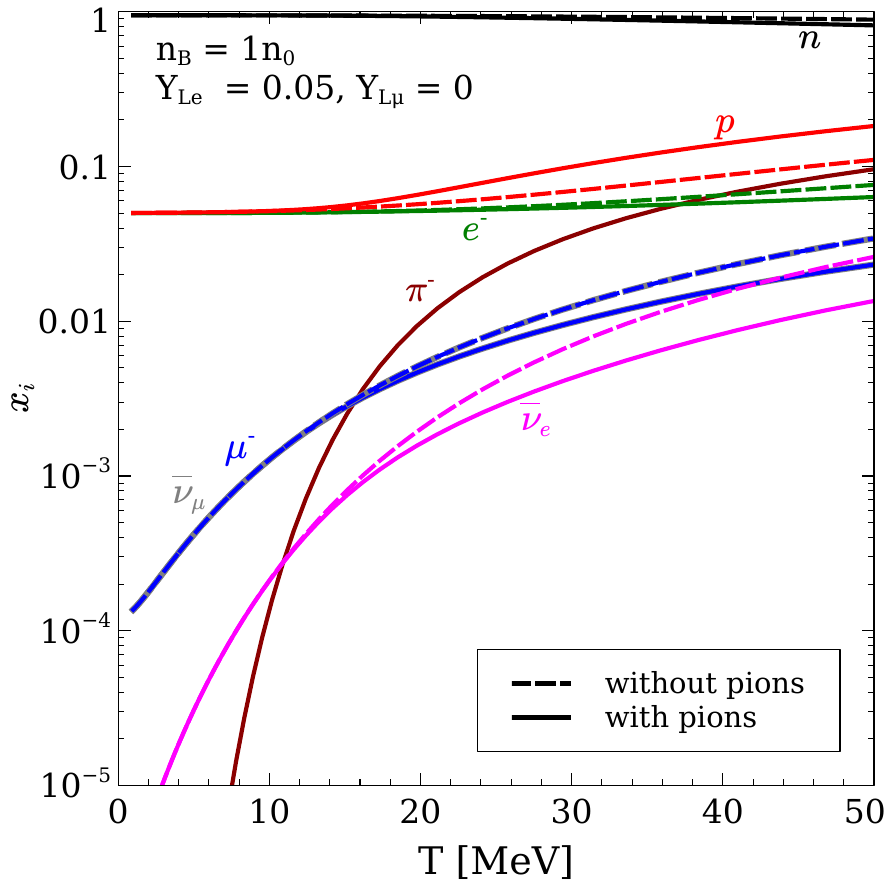}
\includegraphics[width=0.4\textwidth]{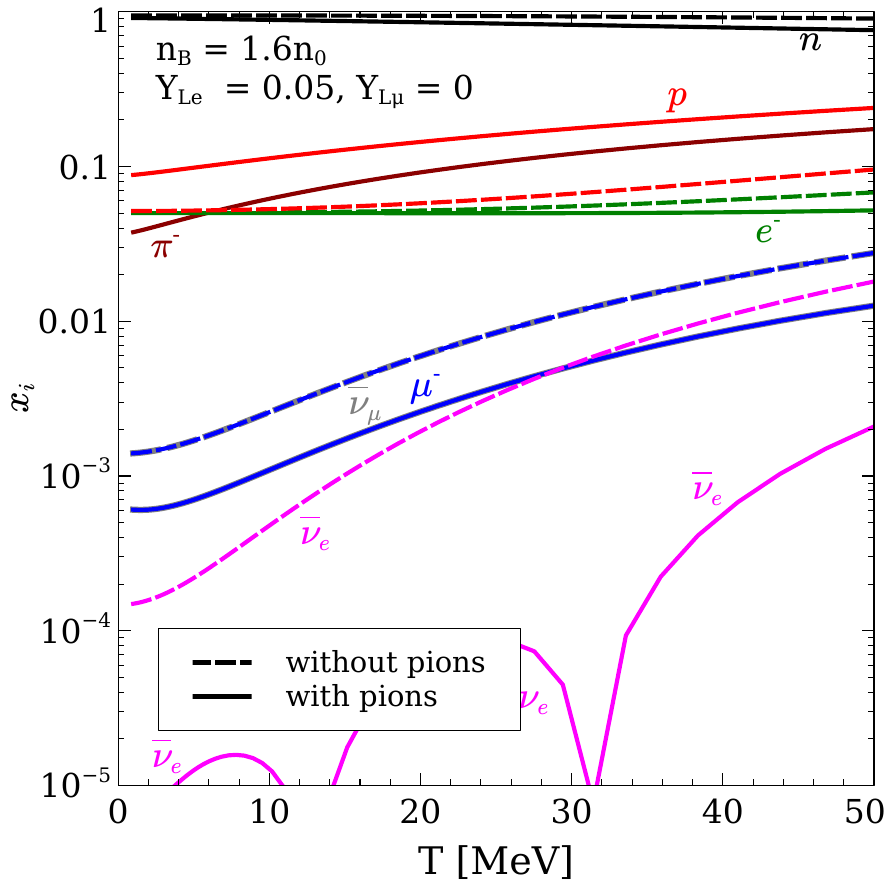}
\includegraphics[width=0.4\textwidth]{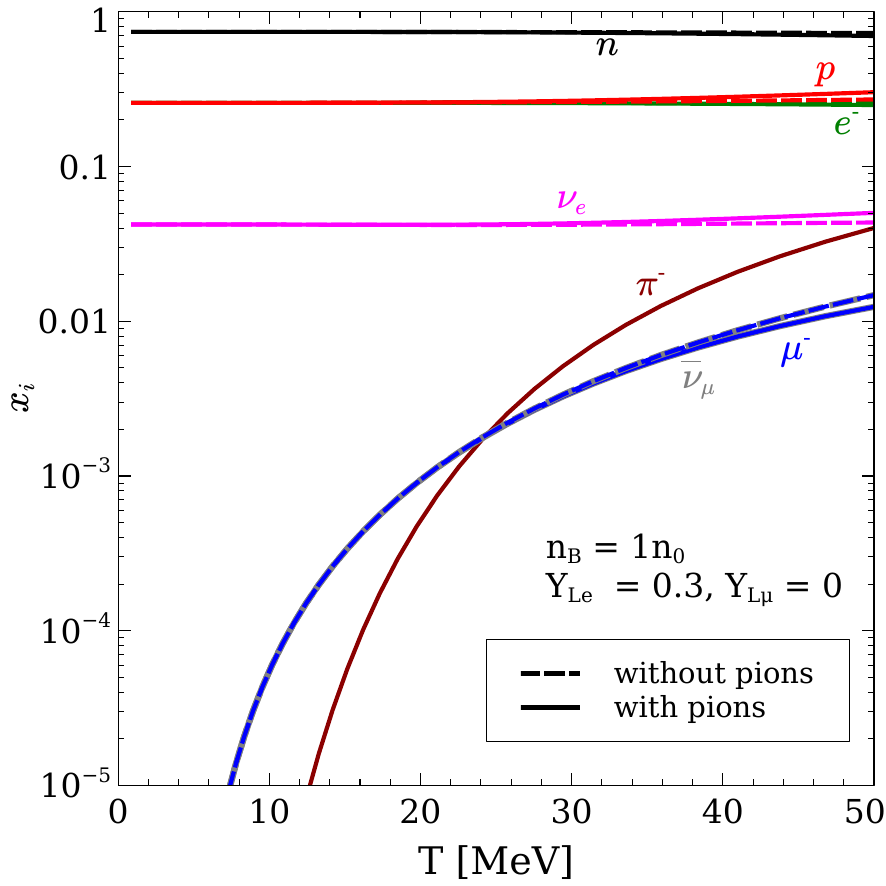}
\includegraphics[width=0.4\textwidth]{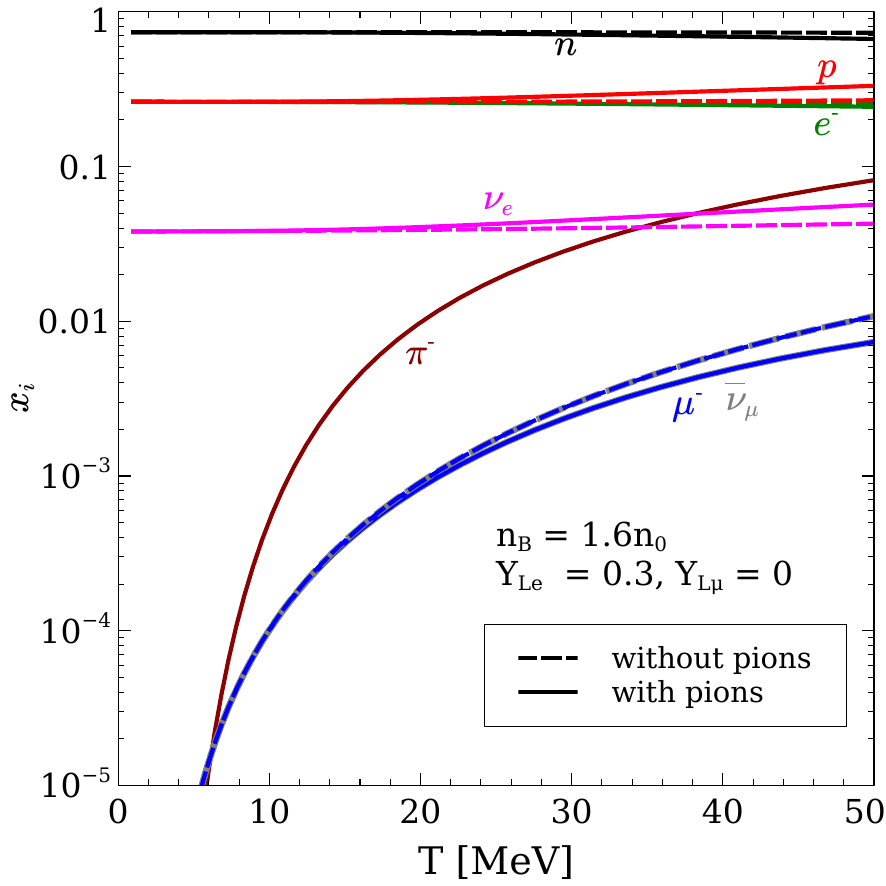}
\caption{Particle content in beta equilibrium in various thermodynamic conditions expected to exist in neutron star mergers (top two panels) and supernovae (bottom two panels).  The left column depicts matter at $1n_0$, while the right column shows matter at $1.6n_0$.  Particle densities are defined as particles minus antiparticles.  In the case of neutrinos, the absolute value of the particle fraction is plotted (thus a sharp dip in the neutrino fraction indicates that the net density passes through zero, indicating a switchover in neutrino or antineutrino dominance).  Particle fractions in the pionless EoS are shown with dashed lines, while the fractions in the EoS with pions are depicted with solid lines.}
\label{fig:xi}
\end{figure*}

The particle fractions predicted by our EoS are plotted in Fig.~\ref{fig:xi}.  The top panel depicts matter in typical neutron star merger conditions ($\{Y_{Le}, Y_{L\mu}\} = \{0.05,0\}$), and the bottom panel typical supernovae conditions ($\{Y_{Le}, Y_{L\mu}\} = \{0.3,0\}$).  The dashed curves represent the EoS without pions, while the solid curves include thermal pions.    

In the merger conditions plotted above, the charge-neutral, beta-equilibrated pionless matter is neutron-dominated, having a proton fraction of about 0.05, which rises slightly as the temperature increases.  The electron population is greater than the muon population due to the large mass of the muon, but together they balance out the positive charge of the proton.  Focusing on $1n_0$ (top left panel) for now, when thermal pions are included in the EoS, their population at low temperatures is small, but increases exponentially with temperature.  As their population becomes appreciable at $T\gtrsim 10\text{ MeV}$, the proton fraction is pushed upward (compared to the pionless case) while the lepton populations decrease (compared to the pionless case).  This effect of the pions was seen in the original version of this EoS \cite{Fore:2019wib}.  At a higher density, $1.6n_0$ (top right panel), the sourcing chemical potential $\hat{\mu}$ is larger, and thus the pion population is much larger, even at low temperatures.  Again, the pion population grows with temperature, surpassing the lepton population at temperatures of just a few MeV.  Thus, in merger conditions, the pions play a significant role in the EoS at temperatures above 10 MeV, or even lower temperatures as the density rises above saturation density.  

In supernovae (bottom two panels of Fig.~\ref{fig:xi}), the lepton number is much higher when the neutrinos are trapped, since trapping occurs well before deleptonization \cite{Burrows:1981zz,2017hsn..book.1575J}, and thus supernova matter has a much larger conserved lepton fraction $Y_{Le}$.  When pion are included in the EoS, their population is suppressed because under these conditions $\hat{\mu}=\mu_n-\mu_p$ is small compared to $m_\pi$ and the isospin asymmetry is smaller. The pion population, and thus the alteration to the EoS is small unless the temperature exceeds several tens of MeV.  
\section{Bulk viscosity}\label{sec:bulkviscosity}
We derive an expression for the bulk viscosity of nuclear matter containing a thermal population of pions.  We will consider a fluid element undergoing small-amplitude baryon density oscillations of the form
\begin{align}
    n_B(t) &= n_B + \Re{(\delta n_B e^{i\omega t})}\\
    &=n_B + \delta n_B \cos{(\omega t)},\nonumber
\end{align}
where $\delta n_B \ll n_B$ and we have chosen $\delta n_B$ to be real.  Here, the real and imaginary parts of a quantity $X$ are denoted $\Re{(X)}$ and $\Im{(X)}$, respectively.  A change in density can push the nuclear matter fluid element out of beta equilibrium by an amount $\delta\mu$ (to be precisely defined later in Eqs.~\ref{eq:dmu1definition}-\ref{eq:dmu3definition}).  In this study, we assume the density oscillation amplitude is small enough to only push the system slightly out of chemical equilibrium, such that $\delta\mu\ll T$.  The resulting bulk viscosity calculated under these conditions is called ``subthermal'' \cite{Haensel:2002qw}.  Below, we extend the subthermal bulk viscosity derivation in $npe$ matter with one equilibrating chemical potential $\delta\mu$ \cite{Alford:2010gw,Harris:2020rus,Harris:2024evy} to neutrino-trapped $npe\mu\pi^-$ matter, which will turn out to have three equilibrating chemical potentials $\left\{\delta\mu_1,\delta\mu_2,\delta\mu_3\right\}$.

As the fluid element in question undergoes a density oscillation, we will assume that the thermal conductivity (due to trapped neutrinos \cite{Alford:2017rxf,Goodwin:1982hy,vandenHorn:1984zz}) is high enough to keep the matter at constant temperature throughout an oscillation.  Such an oscillation in baryon density pushes the three particle fractions $\{x_p,x_{\mu},x_{\pi}\}$ away from their values in chemical equilibrium.  Flavor-changing interactions (these can be weak or strong interactions, as we will see) push the particle fractions back into chemical equilibrium, which takes a certain amount of time.  The pressure of nuclear matter, which depends on both the baryon density and the three independent particle fractions $\{x_p,x_{\mu},x_{\pi}\}$, changes due to chemical equilibration and thus will oscillate out of phase with the baryon density oscillation due to the finite rate of the weak interactions.  The pressure can be written as
\begin{align}\label{eq:P_of_t}
    P(t) &= P_0 + \Re{(\delta P e^{i\omega t})}\\
    &= P_0+ \Re{(\delta P)}\cos{(\omega t)} - \Im{(\delta P)}\sin{(\omega t)}.\nonumber
\end{align}
The $\Im{(\delta P)}$ term that represents the phase lag between the pressure $P(t)$ and the baryon density $n_B(t)$ which will give rise to bulk viscous energy dissipation via the mechanism of $PdV$ work.

The bulk viscosity is defined through its contribution to the energy dissipation of a density oscillation 
\begin{equation}
    \dfrac{\mathop{d\varepsilon}_{\text{osc}}}{\mathop{dt}} = - \zeta (\nabla \cdot \mathbf{v})^2.
\end{equation}
We use the continuity equation for a Lagrangian fluid element
\begin{equation}
    \dfrac{\mathop{dn_B}(t)}{\mathop{dt}} + n_B(t) \nabla \cdot \mathbf{v}=0,
\end{equation}
and then average over one oscillation period
\begin{align}
    \left\langle\dfrac{\mathop{d\varepsilon}_{\text{osc}}}{\mathop{dt}}\right\rangle &= -\dfrac{1}{2}\left(\dfrac{\delta n_B}{n_B(t)}\right)^2\omega^2\zeta\nonumber\\
    &\approx   -\dfrac{1}{2}\left(\dfrac{\delta n_B}{n_B}\right)^2\omega^2\zeta. \label{eq:dedt1}
\end{align}
The energy lost due to $PdV$ work is
\begin{align}
    \mathop{dE_{\text{diss}}}=\mathop{d\left(\varepsilon V\right)} = -P\mathop{dV}.
\end{align}
We can express this as the energy \textit{density} lost due to $PdV$ work
\begin{equation}
    \mathop{d\varepsilon_{\text{diss}}} = \left(\dfrac{\varepsilon+P}{n_B}\right)\mathop{dn_B},
\end{equation}
where we used the fact that baryon number is conserved $\mathop{dN_B}=\mathop{d\left(n_BV\right)}=n_B\mathop{dV}+V\mathop{dn_B}=0$.  Averaging over a cycle, 
\begin{align}
    \left\langle \dfrac{\mathop{d\varepsilon}_{\text{diss}}}{\mathop{dt}}\right\rangle &= \dfrac{\omega}{2\pi}\int_0^{2\pi/\omega}\mathop{dt}\left[\dfrac{P(t)+\varepsilon(t)}{n_B(t)}\right]\dfrac{\mathop{dn_B}(t)}{\mathop{dt}}\label{eq:dedt2}\\
    &\approx\dfrac{\omega^2}{2\pi}\left(\dfrac{\delta n_B}{n_B}\right)\Im{(\delta P)}\int_0^{2\pi/\omega}\mathop{dt}\sin^2{\left(\omega t\right)}\nonumber\\
    &\approx\dfrac{\omega}{2}\left(\dfrac{\delta n_B}{n_B}\right)\Im{(\delta P)},\nonumber
\end{align}
where we used Eq.~\ref{eq:P_of_t} for the time-dependence of the pressure.  We drop the energy density term, because the energy density oscillates in-phase with the baryon density and therefore drops out of the time-average\footnote{This fact can be seen from the Taylor expansion of the pressure (Eq.~\ref{eq:P_taylor_Exp}).  The partial derivatives of the pressure with respect to particle fractions have finite values (related to the susceptibilities, discussed later), which gives the pressure $P(t)$ terms that are proportional to sine which are thus of phase with the density oscillation.  An analogous expansion of the energy density around beta equilibrium would yield partial derivatives of the energy density with respect to particle fractions, which are zero in beta equilibrium.  Thus the energy density is in phase with the baryon density for small amplitude oscillations around beta equilibrium and it drops out of Eq.~\ref{eq:dedt2}.}.

The energy lost by the oscillation is gained by the fluid element, and thus Eqs.~\ref{eq:dedt1} and \ref{eq:dedt2} sum to zero and we find
\begin{equation}\label{eq:zeta_imP}
    \zeta = \left(\dfrac{n_B}{\delta n_B}\right)\dfrac{\Im{(\delta P)}}{\omega}.
\end{equation}
This expression is independent of the nature of the chemical equilibration (assuming deviations from equilibrium are small), as thus matches the expression for $npe$ matter given in \cite{Harris:2020rus,Harris:2024evy} and in other phases of matter \cite{Alford:2007rw,Alford:2010jf}.  The physics of the chemical equilibration (such as the number of equilibrating chemical potentials $\delta\mu$) will affect the form of $\Im{\left(\delta P\right)}$, which we will now determine.  

Driven by a harmonically oscillating density, the particle fractions $x_p,x_{\mu},x_{\pi}$ are also harmonically oscillating, 
\begin{align}
    x_i(t) &= x_i^0 + \Re{(\delta x_i e^{i\omega t})}\label{eq:xi_exp}\\
    &= x_i^0 + \Re{(\delta x_i)}\cos{(\omega t)} - \Im{(\delta x_i)}\sin{(\omega t)},\nonumber
\end{align}
potentially out of phase with $n_B(t)$.  Now, the pressure $P\left(n_B,T,x_p,x_{\mu},x_{\pi}\right)$ can be expanded around its beta equilibrium value (keeping $T$ fixed throughout the oscillation because of the high thermal conductivity)
\begin{widetext}
\begin{equation}
    P = P_0 + \dfrac{\partial P}{\partial n_B}\bigg\vert_{T,x_p,x_{\mu},x_{\pi}}\Re{(\delta n_B e^{i\omega t})}+\dfrac{\partial P}{\partial x_p}\bigg\vert_{T,n_B,x_{\mu},x_{\pi}}\Re{(\delta x_p e^{i\omega t})}+\dfrac{\partial P}{\partial x_{\mu}}\bigg\vert_{T,n_B,x_p,x_{\pi}}\Re{(\delta x_{\mu} e^{i\omega t})}+\dfrac{\partial P}{\partial x_{\pi}}\bigg\vert_{T,n_B,x_{p},x_{\mu}}\Re{(\delta x_{\pi} e^{i\omega t})}.\label{eq:P_taylor_Exp}
\end{equation}
Evidently (from Eqs.~\ref{eq:P_of_t}, \ref{eq:xi_exp}, and \ref{eq:P_taylor_Exp})
\begin{equation}
    \Im{(\delta P)} = \dfrac{\partial P}{\partial x_p}\bigg\vert_{T,n_B,x_{\mu},x_{\pi}}\Im{(\delta x_p)}+\dfrac{\partial P}{\partial x_{\mu}}\bigg\vert_{T,n_B,x_p,x_{\pi}}\Im{(\delta x_{\mu})}+\dfrac{\partial P}{\partial x_{\pi}}\bigg\vert_{T,n_B,x_{p},x_{\mu}}\Im{(\delta x_{\pi})},\label{eq:ImP}
\end{equation}    
\end{widetext}
indicating that to calculate the bulk viscosity (Eq.~\ref{eq:zeta_imP}), we need the imaginary part of the particle fraction oscillation terms.  
\subsection{Classification of chemical reactions in neutrino-trapped $npe\mu\pi^-$ matter}
At the high temperatures ($T \gtrsim 5\text{ MeV}$) considered here, all constituent particle species are in thermal equilibrium, even the neutrinos (as their mean free path is well below 1 km \cite{Alford:2018lhf}).  This is in contrast to the often-studied case where the matter is neutrino-transparent and therefore any reaction with neutrinos or antineutrinos in the initial state is forbidden, and thus the neutrinos are not in statistical equilibrium \cite{Alford:2021ogv,Alford:2018lhf}.  In the neutrino-trapped case, studied here, all weak processes proceed in both directions.  The processes proceed in a manner that would balance the chemical potentials on each side of the reaction,
\begin{equation}
    \sum_{i \in \text{LHS}}\mu_i = \sum_{i \in \text{RHS}}\mu_i, \label{eq:beta_eq_mu}
\end{equation}
indicating chemical equilibrium \cite{Shapiro:1983du,greiner1995thermodynamics}.  

There are six classes\footnote{Reactions within each class are related by ``crossing'' (e.g.~direct Urca neutron decay and electron capture) or by the addition of a spectator particle to both sides of the reaction (e.g.~direct Urca neutron decay and modified Urca neutron decay).} of weak reactions that can occur in neutrino-trapped $npe\mu\pi$ matter, and therefore six chemical potential differences $\delta\mu_i$ that are zero in chemical equilibrium.  However, three of them are redundant and can be written in terms of three independent equilibrating chemical potentials
\begin{subequations}
\begin{align}
\delta\mu_1 &\equiv \mu_n+\mu_{\nu_e}-\mu_p-\mu_e\label{eq:dmu1definition}\\
\delta\mu_2 &\equiv \mu_n+\mu_{\nu_{\mu}}-\mu_p-\mu_{\mu}\label{eq:dmu2definition}\\
\delta\mu_3 &\equiv \mu_n-\mu_p-\mu_{\pi}.\label{eq:dmu3definition}
\end{align}
\end{subequations}
We assume in this analysis that particles are chemically equilibrated with their respective antiparticles, $\mu_X = -\mu_{\bar{X}}$.  This is a good assumption for the hadrons and charged leptons, though for neutrinos it is merely a simplifying assumption\footnote{If neutrinos and antineutrinos were not equilibrated with each other, as is likely the case except at quite large temperatures, we would have to assign distinct chemical potentials $\mu_{\nu_{e}},\mu_{\bar{\nu}_{e}},\mu_{\nu_{\mu}},\mu_{\bar{\nu}_{\mu}}$ and consider reactions such as $n+n\leftrightarrow n+n+\nu_e+\bar{\nu}_e$ or $\nu_{e}+\bar{\nu}_{e}\leftrightarrow \nu_{\mu}+\bar{\nu}_{\mu}$.}.  Given this, the six classes of reactions are listed in Table \ref{table:allreactions}.
\begin{widetext}
\begin{center}
\begin{table}[]
\begin{tabular}{l|l|l}
\multicolumn{1}{c|}{Equilibrating $\delta\mu$} & \multicolumn{1}{c|}{Reactions} & \multicolumn{1}{c}{$\overrightarrow{\Gamma} - \overleftarrow{\Gamma}$ (subthermal)} \\ \hline
\multirow{2}{*}{1) $\delta\mu_1 \equiv \mu_n+\mu_{\nu_e}-\mu_p-\mu_e$} & $n \leftrightarrow p + e^- + \bar{\nu}_e$ & $\lambda_a \delta\mu_1$ \\
 & $n + \nu_e \leftrightarrow e^- + p$ & $\lambda_b \delta\mu_1$ \\ \hline
\multirow{2}{*}{2) $\delta\mu_2\equiv \mu_n + \mu_{\nu_{\mu}} - \mu_p -\mu_{\mu}$} & $n\leftrightarrow p + \mu^- + \bar{\nu}_{\mu}$ & $\lambda_c \delta\mu_2$ \\
 & $n+\nu_{\mu}\leftrightarrow \mu^- + p$ & $\lambda_d \delta\mu_2$ \\ \hline
3) $\delta\mu_3 \equiv \mu_n-\mu_p-\mu_{\pi}$ & $n\leftrightarrow p + \pi^-$ & $\lambda_3 \delta\mu_3$ \\ \hline
\multirow{2}{*}{4) $\delta\mu_4 \equiv \mu_{\pi}+\mu_{\nu_e}-\mu_e = \delta\mu_1-\delta\mu_3$} & $\pi^- \leftrightarrow e^-+\bar{\nu}_e$ & $\lambda_e (\delta\mu_1-\delta\mu_3)$ \\
 & $\pi^-+\nu_e\leftrightarrow e^-$ & $\lambda_f (\delta\mu_1-\delta\mu_3)$ \\ \hline
\multirow{2}{*}{5) $\delta\mu_5 \equiv \mu_{\pi}+\mu_{\nu_\mu}-\mu_{\mu}=\delta\mu_2-\delta\mu_3$} & $\pi^-\leftrightarrow \mu^-+\bar{\nu}_{\mu}$ & $\lambda_g (\delta\mu_2-\delta\mu_3)$ \\
 & $\pi^-+\nu_{\mu}\leftrightarrow \mu^-$ & $\lambda_h (\delta\mu_2-\delta\mu_3)$ \\ \hline
\multirow{4}{*}{6) $\delta\mu_6 \equiv \mu_{\mu}+\mu_{\nu_e}-\mu_e-\mu_{\nu_{\mu}} = \delta\mu_1 - \delta\mu_2$} & $\mu^- \leftrightarrow e^-+\bar{\nu}_e+\nu_{\mu}$ & $\lambda_i (\delta\mu_1-\delta\mu_2)$ \\
 & $\mu^-+\bar{\nu}_{\mu}\leftrightarrow e^-+\bar{\nu}_e$ & $\lambda_j (\delta\mu_1-\delta\mu_2)$ \\
 & $\mu^- + \nu_e \leftrightarrow e^- + \nu_{\mu}$ & $\lambda_k (\delta\mu_1-\delta\mu_2)$ \\
 & $\mu^-+\nu_e+\bar{\nu}_{\mu} \leftrightarrow e^-$ & $\lambda_l (\delta\mu_1-\delta\mu_2)$ \\ \hline
\end{tabular}
\caption{The chemical reactions that occur in neutrino-trapped $npe\mu\pi$ matter, split into six categories based on the $\delta\mu$ that they equilibrate.  Not included in the list are the ``modified'' versions of these reactions, where a spectator particle is in the initial and final state.  This table also serves to define the $\lambda_i$ ($i$ running through the alphabet from $a$ to $l$) corresponding to each process.}
\label{table:allreactions}
\end{table}
\end{center}
\end{widetext}

We list next to each reaction the relaxation rate towards beta equilibrium (the forward rate $\overrightarrow{\Gamma}$ minus the backward rate $\overleftarrow{\Gamma}$) and its approximation to linear order in $\delta\mu$, which we will use since we only consider small deviations from chemical equilibrium.  The coefficients 
\begin{equation}
    \lambda \equiv \dfrac{\partial (\overrightarrow{\Gamma}-\overleftarrow{\Gamma})}{\partial \left(\delta\mu\right)}\bigg\rvert_{\delta\mu=0}
\end{equation}
describe the rate of beta relaxation of a particular reaction.  The relaxation rate $\lambda$ of each class of processes is given by the sum of $\lambda$ for each reaction in the class
\begin{align}
    \lambda_1 &= \lambda_a+\lambda_b\nonumber\\
    \lambda_2 &= \lambda_c + \lambda_d \nonumber\\
    \lambda_4 &= \lambda_e+\lambda_f \\
    \lambda_5 &= \lambda_g+\lambda_h\nonumber\\
    \lambda_6 &= \lambda_i+\lambda_j+\lambda_k+\lambda_l. \nonumber
\end{align}
The processes shown above can also occur in the presence of spectator particles.  For example, the direct Urca neutron decay process $n\rightarrow p + e^- + \bar{\nu}_e$ is supplemented by the modified Urca process $n+N\rightarrow N+p+e^-+\bar{\nu}_e$ \cite{Yakovlev:2000jp}, the electron-muon conversion process $\mu^-+\nu_e+\bar{\nu}_{\mu}\leftrightarrow e^-$ can only proceed with a lepton spectator \cite{Alford:2010jf}, and $n\leftrightarrow p + \pi^-$ needs a nucleon spectator to proceed \cite{Fore:2019wib}.  

In this work, where we study nuclear matter at temperatures high enough ($T\gtrsim 5\text{ MeV}$) such that neutrinos are trapped, we will not need to consider spectator particles directly.  The reaction that proceeds with a spectator particle is usually slower than the reaction without the spectator particle, because of the additional phase space restriction from the spectator particle (in degenerate matter, by $(T/\mu)^2$) and in the case of weakly coupled theories like electromagnetism, the reaction is also suppressed by additional powers of the small coupling $\alpha$ \cite{Alford:2010jf}.  However, the direct (or, spectatorless) reaction is sometimes kinematically forbidden, and the only process that operates is that with the spectator. The classic example (in neutrino-transparent matter) is the direct Urca process, which is forbidden at low densities where the proton fraction is too small \cite{Lattimer:1991ib,Boguta:1981mw}, in which case the modified Urca process dominates.  At finite but low temperatures, the direct Urca process is not forbidden but is strongly Boltzmann suppressed.  As temperature rises, the Boltzmann suppression decreases, and at temperatures above 1 MeV, the direct Urca process, even below the threshold, dominates over modified Urca \cite{Alford:2021ogv,Alford:2018lhf}.  

Here, we consider neutrino-trapped nuclear matter, and thus the neutrino (or antineutrino) population has a finite Fermi momentum, which alters the kinematics of the Urca process from the neutrino-transparent case.  In neutrino-trapped matter, the direct Urca process $e^-+p\leftrightarrow n + \nu_e$ and its muon counterpart ($\lambda_1$ and $\lambda_2$) do not have a density threshold because the neutrinos have a finite Fermi momentum (see Figs.~13 and 14 in \cite{Alford:2019kdw}), and as such will dominate over the modified Urca processes.  

The process $n\leftrightarrow p + \pi^-$ needs a spectator to proceed, but even the spectator reaction $n+n\leftrightarrow n+p+\pi^-$ is very fast compared to all other timescales in the problem as long as $T\gtrsim 1\text{ MeV}$, so in our numerical calculations we will take the $\lambda_3\rightarrow\infty$ limit and never have to calculate the rate of $n+n\leftrightarrow n+p+\pi^-$.  The rate of this process, however, can be calculated using models such as the one developed in \cite{Engel:1996ic}.  

The rate of the pion decay processes in the presence of a nucleon spectator (for example, $n + \pi^- \rightarrow n + \mu^- + \bar{\nu}_{\mu}$) has not been rigorously calculated, as far as we know.  A rough estimate was given for neutrino transparent matter \cite{Bahcall:1965zzb}.  We cannot rule out the possibility that our calculation underestimates the rates $\lambda_4$ and $\lambda_5$ because we neglect the possibility of spectators particles in pion decays.  

Finally, the direct leptonic processes $\lambda_6$ are kinematically allowed since there is a trapped neutrino population and they will dominate over their counterparts with an electrically charged spectator particle.  Note that at our level of approximation, where we neglect spectator processes in the $\lambda_6$ subprocesses, $\lambda_l = 0$ because the process $\mu^-+\nu_e+\bar{\nu}_{\mu}\rightarrow e^-$ is kinematically forbidden.   
\subsection{Weak interaction rates}
We calculate the weak interaction rates $\Gamma$ (number of reactions per time per volume) with Fermi's golden rule, where the rate is determined by a multidimensional phase space integral of the product of a matrix element and the Fermi-Dirac or Bose-Einstein distributions of the involved particles \cite{Yakovlev:2000jp,Alford:2021ogv,Schmitt:2017efp}.  We perform the full phase space integration (as in \cite{Alford:2021ogv,Alford:2018lhf,Alford:2021lpp,Alford:2019kdw,Alford:2020pld,Alford:2019qtm,Alford:2020lla,Alford:2022ufz}) without making any approximation for the degeneracy of any particle species.  The rate calculations are briefly described in Appendix \ref{sec:rates}.

In beta equilibrium, the net rate $\overrightarrow{\Gamma} - \overleftarrow{\Gamma}$ of any particular flavor-changing interaction is zero.  When the system is pushed out of beta equilibrium by amount $\delta\mu$ ($\delta\mu$ measures the degree of violation of Eq.~\ref{eq:beta_eq_mu}), the net rate $\overrightarrow{\Gamma} - \overleftarrow{\Gamma}$ becomes nonzero, forcing the system back to chemical equilibrium (per Le Chatelier's principle).  The net rate can be calculated for $\delta\mu$ of arbitrary size (see analytic calculations for the direct and modified Urca processes in strongly degenerate $npe$ matter in \cite{Reisenegger:1994be}), though we consider only subthermal bulk viscosity where $\delta\mu\ll T$.  In this regime, the net rate is proportional to the size of the departure from beta equilibrium $\overrightarrow{\Gamma} - \overleftarrow{\Gamma} \approx \lambda \delta\mu$.  For a given reaction, if all involved particles are in thermal equilibrium, there is a simple relationship\footnote{This equation does not hold for reactions involving neutrinos in neutrino-transparent matter (not studied here) as the rate calculation does not contain neutrino distributions in the rate integral, since neutrinos are assumed to not build up a population in the neutron star due to their long mean free path.  In such a case, the calculation of $\lambda$ is more complicated \cite{Alford:2018lhf,Alford:2021ogv}.} between the rate of the process and $\lambda$ \cite{Alford:2021lpp,Alford:2019kdw}
\begin{equation}
    \overrightarrow{\Gamma}(\delta\mu=0) = \overleftarrow{\Gamma}(\delta\mu=0) = \lambda T.
\end{equation}
So, to obtain $\lambda$ we merely need to calculate $\Gamma$ in beta equilibrium and divide by the temperature.

At fixed density, the particle fractions evolve according to 
\begin{subequations}
\begin{align}
    n_B\dfrac{\mathop{dx_p}}{\mathop{dt}} &= \lambda_1\delta\mu_1+\lambda_2\delta\mu_2 + \lambda_3\delta\mu_3,\label{eq:nbdxp}\\
    n_B\dfrac{\mathop{dx_{\mu}}}{\mathop{dt}} &= -\lambda_6\delta\mu_1+(\lambda_2+\lambda_5+\lambda_6)\delta\mu_2 - \lambda_5\delta\mu_3,\label{eq:nbdxmu}\\
    n_B\dfrac{\mathop{dx_{\pi}}}{\mathop{dt}} &= -\lambda_4\delta\mu_1-\lambda_5\delta\mu_2 + (\lambda_3+\lambda_4+\lambda_5)\delta\mu_3.\label{eq:nbdxpi}
\end{align}
\end{subequations}
We will refrain from plotting $\lambda_i$, but instead, later (Fig.~\ref{fig:gamma}), plot the beta equilibration rates $\gamma_i$ which have a clearer physical interpretation.  But before we do that, we must discuss the susceptibilities of the EoS.  
\subsection{Susceptibilities}
In our study of dense matter slightly perturbed from beta equilibrium, it is convenient to introduce the susceptibilities 
\begin{subequations}
\begin{align}
    A_i &\equiv n_B \dfrac{\partial \delta\mu_i}{\partial n_B}\bigg\vert_{T,x_p,x_{\mu},x_{\pi}},\label{eq:suscA}\\
    B_i &\equiv \dfrac{1}{n_B}\dfrac{\partial\delta\mu_i}{\partial x_p}\bigg\vert_{T,n_B,x_{\mu},x_{\pi}},\label{eq:suscB}\\
    C_i &\equiv \dfrac{1}{n_B}\dfrac{\partial\delta\mu_i}{\partial x_{\mu}}\bigg\vert_{T,n_B,x_p,x_{\pi}},\label{eq:suscC}\\
    D_i &\equiv \dfrac{1}{n_B}\dfrac{\partial\delta\mu_i}{\partial x_{\pi}}\bigg\vert_{T,n_B,x_p,x_{\mu}},\label{eq:suscD}
\end{align}
\end{subequations}
where $i$ ranges from 1 to 3.  These susceptibilities are properties of the nuclear matter EoS and do not themselves depend on the reaction rates in the medium.  There appear to be 12 of these susceptibilities, but they are not all independent, and Maxwell relations
\begin{subequations}
\begin{align}
C_1 &= B_2 - B_1\label{eq:maxwell4}\\
D_1 &= B_3 - B_1\label{eq:maxwell5}\\
D_1 - D_2 &= C_1 - C_3\label{eq:maxwell6}.
\end{align}
\end{subequations}
can be used to eliminate 3 of them.  See Appendix \ref{sec:maxwell} for details on the Maxwell relations and thermodynamics of multicomponent systems.  Additionally, the susceptibility $C_3=0$ because the leptons and hadrons are independent.  With our EoS, $D_2=0$, because of the specifics of our treatment of the pion-nucleon interaction.  A different prescription for the interaction could lead to a nonzero value of $D_2$.  At this point, we can reduce the 12 susceptibilities down to 7 $\{A_1,A_2,A_3,B_1,B_2,C_2,D_3\}$.  We will keep $D_2$ and $C_3$ in our exact expression for the bulk viscosity (Eq.~\ref{eq:full_bulk_viscosity} and Appendix \ref{sec:full_expression_for_bulk_viscosity}), but set them to zero elsewhere.  

Finally, we note that susceptibilities $A_i$ (Eq.~\ref{eq:suscA}) can be written, using Maxwell relations (see Appendix \ref{sec:maxwell}), in terms of derivatives of the pressure.  We find 
\begin{subequations}
\begin{align}
    \dfrac{\partial P}{\partial x_p}\bigg\vert_{T,n_B,x_{\mu},x_{\pi}} &= -n_B A_1\label{eq:maxwell1}\\
    \dfrac{\partial P}{\partial x_{\mu}}\bigg\vert_{T,n_B,x_{p},x_{\pi}} &= n_B (A_1-A_2)\label{eq:maxwell2}\\
    \dfrac{\partial P}{\partial x_{\pi}}\bigg\vert_{T,n_B,x_{p},x_{\mu}} &= n_B(A_1-A_3).\label{eq:maxwell3}
\end{align}
\end{subequations}
Before we describe the nature of the susceptibilities and plot them for the EoS developed in the previous section, we write the Taylor expansions of $P$ and $\delta\mu$ in terms of the susceptibilities.  First, 
\begin{widetext}
\begin{align}
    \delta\mu_i &= A_i\dfrac{\delta n_B}{n_B}\cos{(\omega t)} + n_B \left[B_i \Re{(\delta x_p e^{i\omega t})}+C_i\Re{(\delta x_{\mu} e^{i\omega t})}+D_i\Re{(\delta x_{\pi}e^{i\omega t})} \right]\label{eq:dmu_exp}\\
    &=A_i\dfrac{\delta n_B}{n_B}\cos{(\omega t)}+ n_B\left[B_i\Re{(\delta x_p)}+C_i\Re{(\delta x_{\mu})}+D_i\Re{(\delta x_{\pi})}\right]\cos{\left(\omega t\right)}-n_B\left[B_i\Im{(\delta x_p)}+C_i\Im{(\delta x_{\mu})}+D_i\Im{(\delta x_{\pi})}\right]\sin{\left(\omega t\right)},\nonumber
\end{align}
with $i$ again taking values from 1 to 3.  The pressure (Eq.~\ref{eq:P_taylor_Exp}) can be rewritten as
\begin{align}
P &= P_0 + \dfrac{\delta n_B}{n_B}\kappa_T^{-1}\cos{\left(\omega t\right)}+n_B\cos{\left(\omega t\right)}\left[-A_1\Re{\left(\delta x_p\right)}+(A_1-A_2)\Re{\left(\delta x_{\mu}\right)}+(A_1-A_3)\Re{\left(\delta x_{\pi}\right)}\right]\nonumber\\
&+n_B\sin{\left(\omega t\right)}\left[A_1\Im{\left(\delta x_p\right)}-(A_1-A_2)\Im{\left(\delta x_{\mu}\right)}-(A_1-A_3)\Im{\left(\delta x_{\pi}\right)}\right],
\end{align}
\end{widetext}
where we have introduced the isothermal compressibility \cite{CompOSECoreTeam:2022ddl}
\begin{equation}\label{eq:compressibility}
    \kappa_T = \left( \left. n_B\dfrac{\partial P}{\partial n_B}\right\vert_{T,x_p,x_{\mu},x_{\pi}}  \right)^{-1}.
\end{equation}

The susceptibilities $A_i$ describe how far the system is pushed out of beta equilibrium when the system is compressed if the chemical fractions are frozen.  If $A_i = 0$, then the system does not depart from beta equilibrium when the density is increased, and so with respect to $\delta\mu_i$, the system is conformal.  However, the system may have other degrees of freedom that do change when the density is increased, and so we describe the situation where, say, $A_1=0$ as \textit{partially conformal}.  The Maxwell relations (Eqs.~\ref{eq:maxwell1}-\ref{eq:maxwell3}) indicate that the $A_i$ are related to the degree to which the pressure depends on a particular particle fraction.  As will be seen later, generally bulk viscosity grows with the increased magnitude of the $A_i$.  In particular, if all $A_i=0$, then bulk viscosity (of the type described here, related to chemical equilibration) should be zero.

\begin{figure*}\centering
\includegraphics[width=0.4\textwidth]{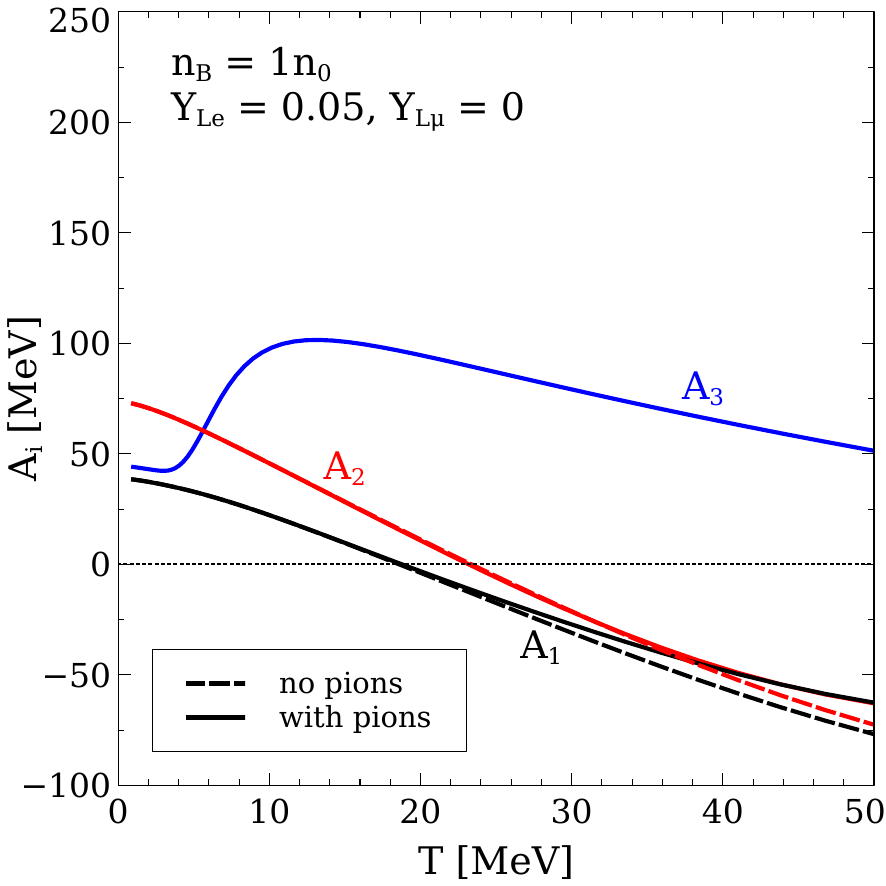}
\includegraphics[width=0.4\textwidth]{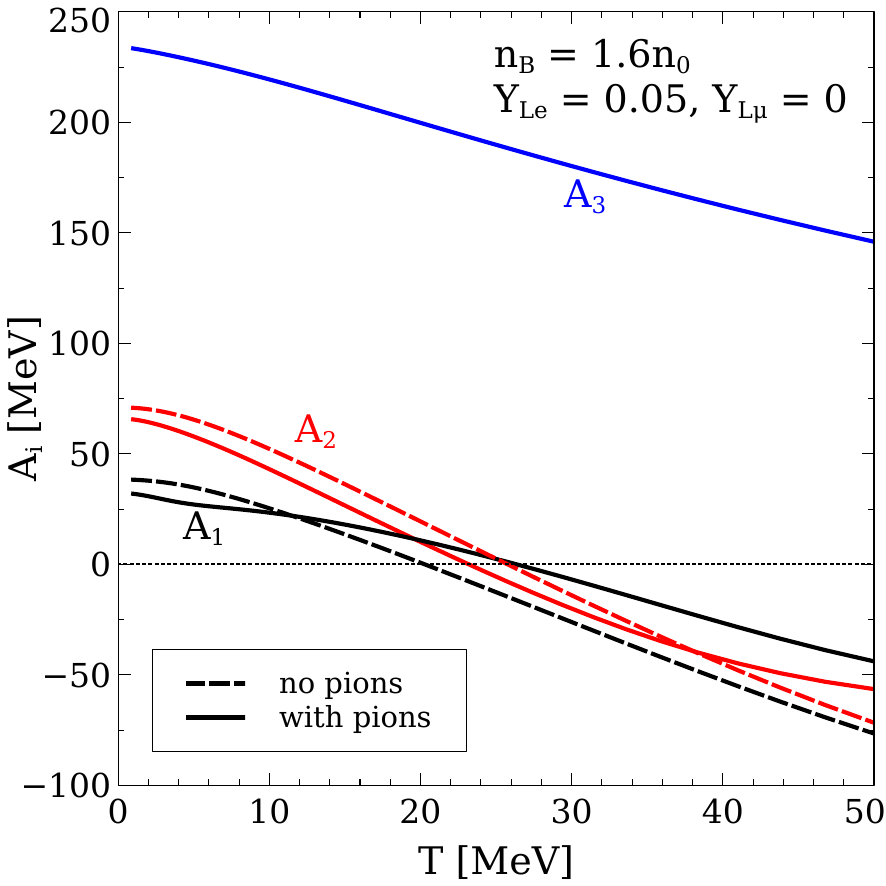}\\
\includegraphics[width=0.4\textwidth]{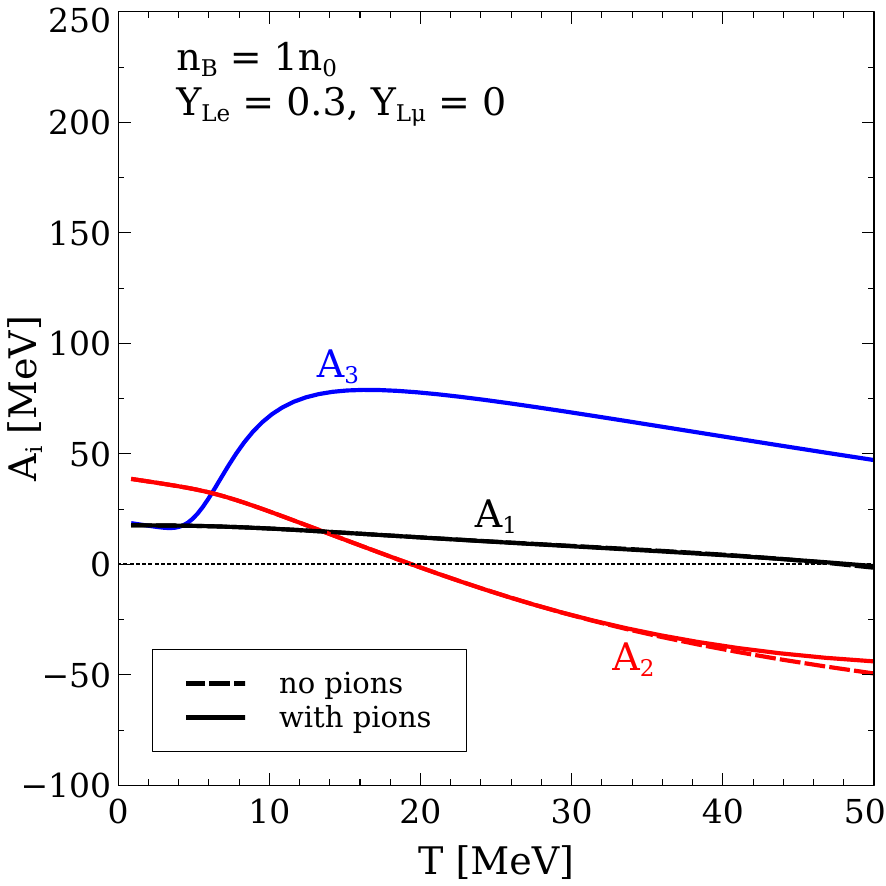}
\includegraphics[width=0.4\textwidth]{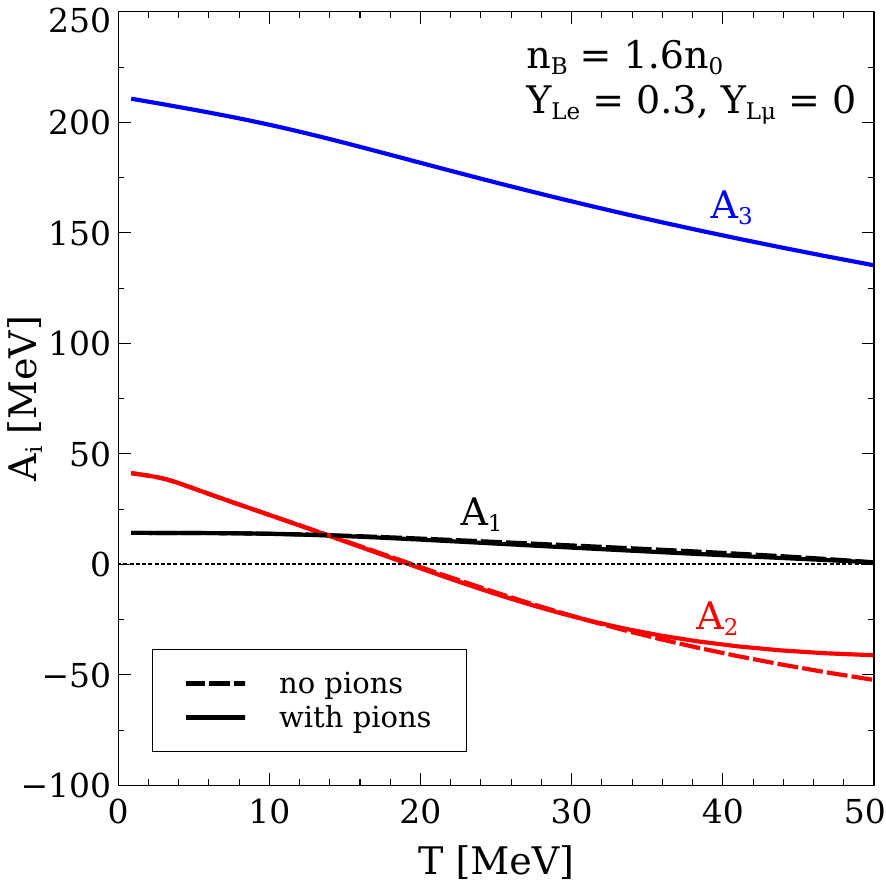}
\caption{Susceptibilities of the $A$ type from the EoS described in Sec.~\ref{sec:EoS} plotted in merger conditions (top panels) and in supernovae conditions (bottom panels).  The left column plots are at a density of $1n_0$, while the right column panels are at $1.6n_0$.  The dashed lines correspond to the EoS without pions, while the solid lines indicate the EoS with pions.  $A_3$ has no counterpart in a system without pions.}
\label{fig:A_susc}
\end{figure*}

We plot the susceptibilities $A_i$ in merger conditions (top two panels) and in supernovae conditions (bottom two panels) in Fig.~\ref{fig:A_susc}.  These susceptibilities are not tremendously different between merger and supernovae conditions.  $A_1$ and $A_2$ have magnitudes around a few tens of MeV, and can switch signs, indicating a partial conformal point.  In merger conditions, both $A_1$ and $A_2$ cross through zero as temperature rises beyond a couple tens of MeV.  In supernovae conditions, only $A_2$ crosses through zero in the displayed temperature range, though likely $A_1$ does too at a higher temperature.  Including pions in the EoS slightly modifies $A_1$ and $A_2$, mostly at high temperature where the pion population is large.  The main effect of including pions is to bring about a new susceptibility $A_3$, which will have important consequences for the bulk viscosity, discussed later.  The susceptibility $A_3$ is significantly larger than $A_1$ or $A_2$, especially at high temperatures or densities where the pion population is large.  

\begin{figure*}\centering
\includegraphics[width=0.4\textwidth]{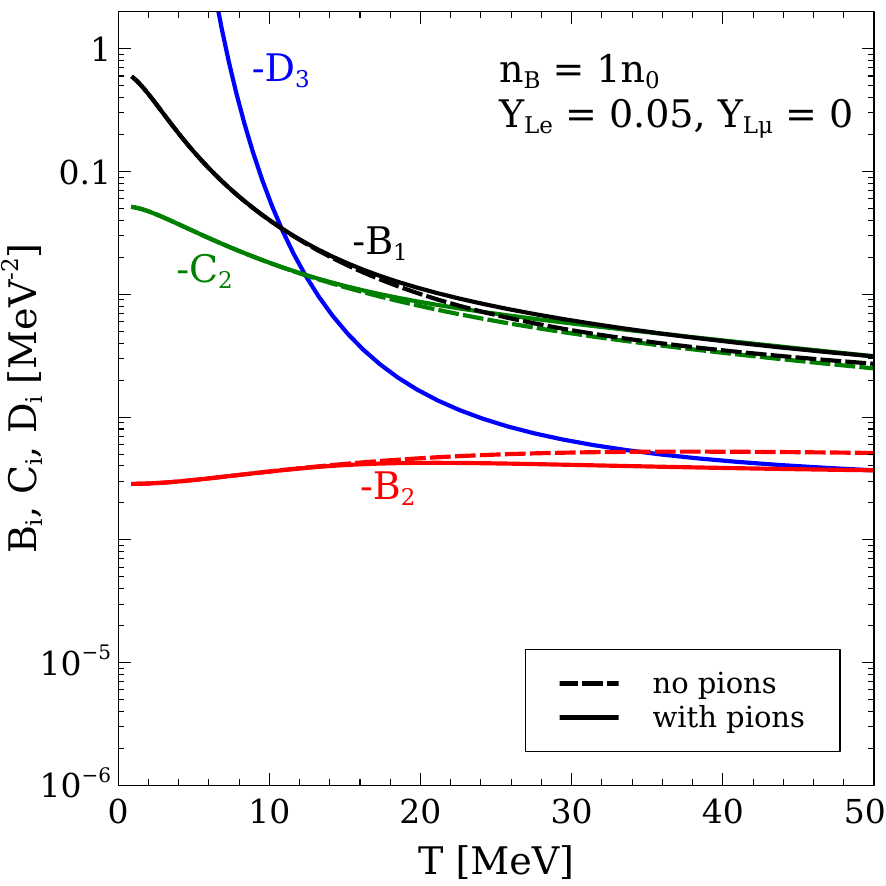}
\includegraphics[width=0.4\textwidth]{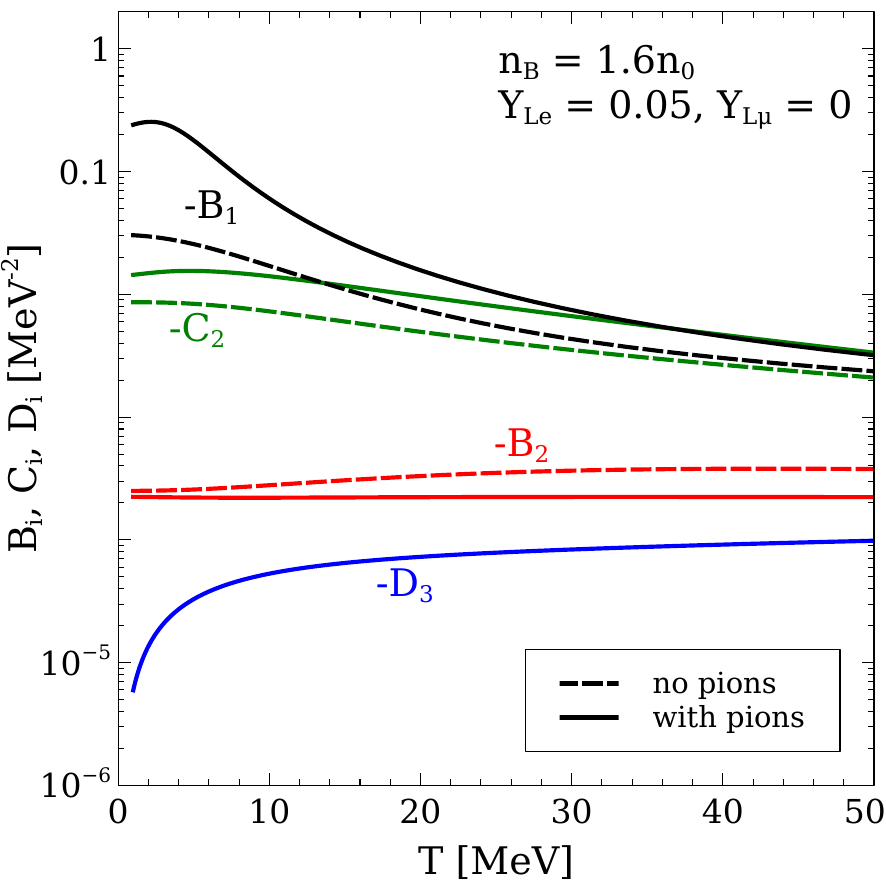}\\
\includegraphics[width=0.4\textwidth]{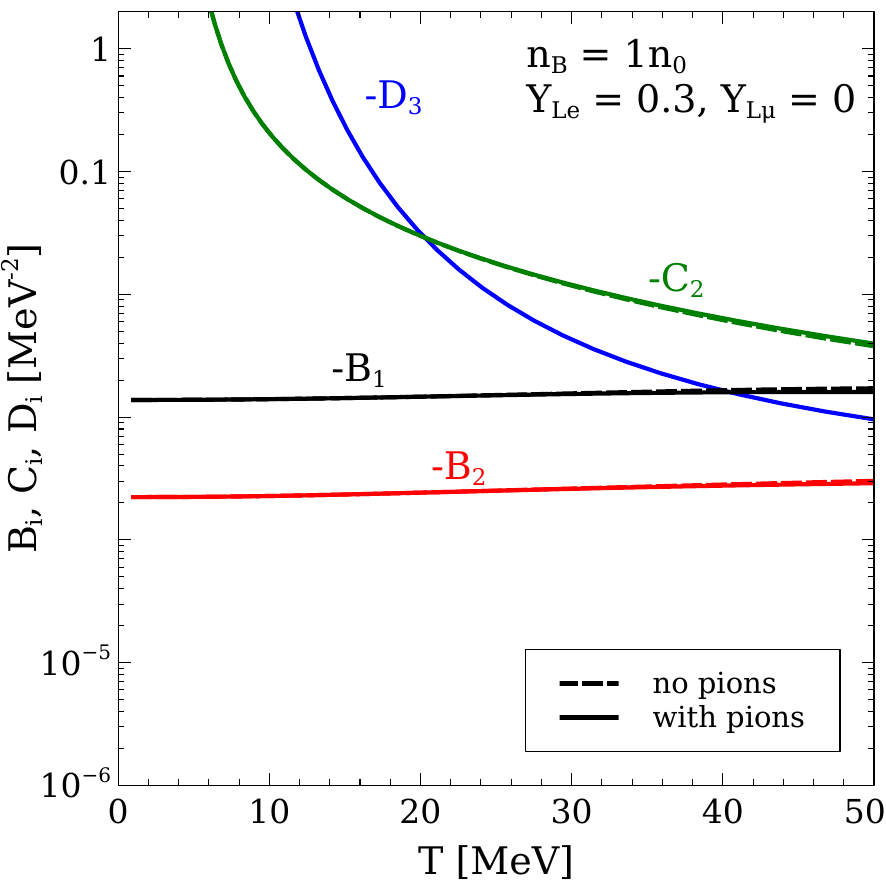}
\includegraphics[width=0.4\textwidth]{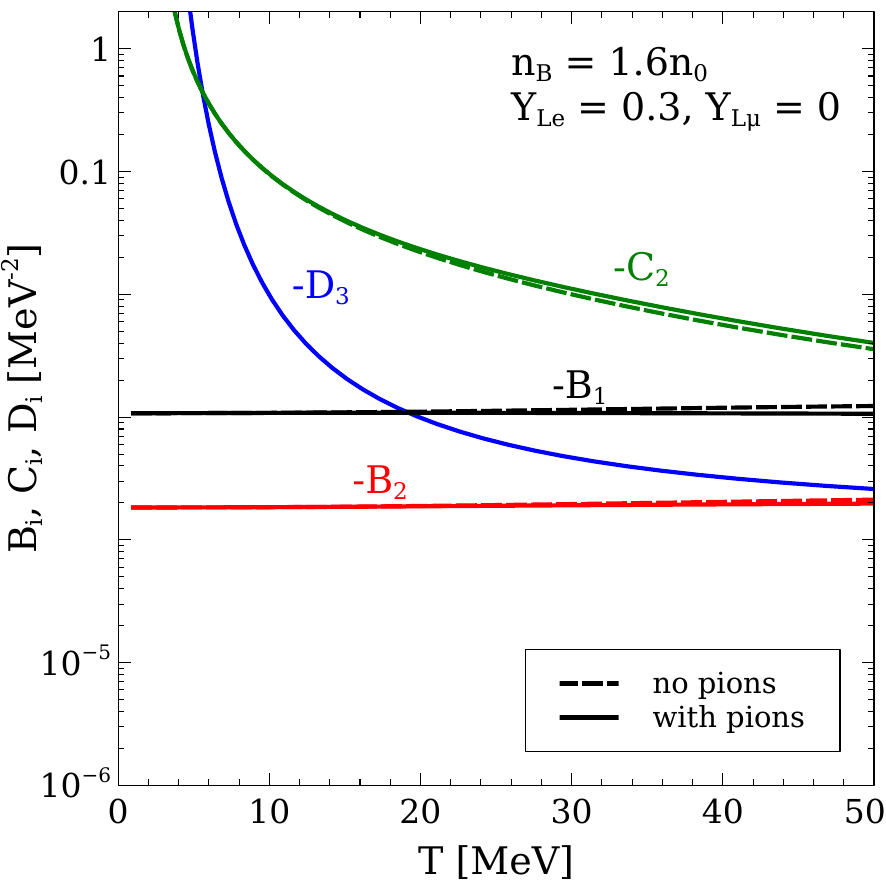}
\caption{Susceptibilities of the $B$, $C$, and $D$ types from the EoS described in Sec.~\ref{sec:EoS} plotted in merger conditions (top panels) and in supernovae conditions (bottom panels).  The dashed lines correspond to an EoS without pions, while the solid lines indicate an EoS with pions.  $D_3$ has no counterpart in a system without pions.}
\label{fig:BCD_susc}
\end{figure*}

The susceptibilities $B$, $C$, and $D$ represent, at a fixed density and temperature, how far the system is pushed from chemical equilibrium when just one of the particle fractions is adjusted, keeping the others fixed.  For example, $B_1$, $C_1$, and $D_1$ (roughly) represent the degree of dependence of $\delta\mu_1$ on $x_p$, $x_{\mu}$, and $x_{\pi}$ respectively.  The $B$, $C$, and $D$ susceptibilities convert $\lambda_i$ into $\gamma_i$, as we will see in a later section.  We plot the susceptibilities $B_i$, $C_i$, and $D_i$ in Fig.~\ref{fig:BCD_susc}.  These susceptibilities, at least in the conditions encountered in this work, never cross zero.  As we will see, while they do affect the strength of the maximum bulk viscosity, their most important role is their impact on the equilibration rates $\gamma_i$ which determine the location of the resonant maxima of the bulk viscosity.  
\subsection{Bulk viscosity at finite $\lambda_3$}
Combining Eqs.~\ref{eq:zeta_imP}, \ref{eq:ImP}, and \ref{eq:maxwell1}-\ref{eq:maxwell3}, we find the bulk viscosity is given by
\begin{align}\label{eq:zeta_intermediate_expression}
    \zeta &= \dfrac{1}{\omega}\dfrac{n_B^2}{\delta n_B}\big[-A_1\Im{(\delta x_p)}\\
    &+(A_1-A_2)\Im{(\delta x_{\mu})}+(A_1-A_3)\Im{(\delta x_{\pi})}    \big].  \nonumber
\end{align}
To obtain $\Im{(\delta x_i)}$, we plug Eq.~\ref{eq:xi_exp} and \ref{eq:dmu_exp} into Eqs.~\ref{eq:nbdxp}, \ref{eq:nbdxmu}, and \ref{eq:nbdxpi} and match the sine terms and the cosine terms.  This yields six equations with six variables ($\Re{(\delta x_i)}$ and $\Im{(\delta x_i)}$ for $i=p, \mu, \pi$).  The three real-part variables can be eliminated, yielding a set of 3 equations to be solved to obtain $\Im{(\delta x_p)}, \Im{(\delta x_{\mu})}, \Im{(\delta x_{\pi})}$.  Solving the system results in the expression for the bulk viscosity
\begin{equation}\label{eq:full_bulk_viscosity}
    \zeta = \dfrac{F + G\omega^2+H\omega^4}{J+K\omega^2+L\omega^4+\omega^6}.
\end{equation}
The expressions for $F,G,H,J,K$, and $L$ are very complicated and are given in Appendix \ref{sec:full_expression_for_bulk_viscosity}.  This formula is consistent with the trend that the bulk viscosity of a system with $n$ independent equilibration channels, when expressed as a combined fraction, has in the numerator a polynomial in $\omega$ of order $2n-2$ and in the denominator a polynomial in $\omega$ of order $2n$ (see, e.g.,~\cite{Alford:2019qtm} for $n=1$ and \cite{Alford:2006gy} for $n=2$).  This trend holds for $n$ independent equilibration channels $\zeta = \sum_{i=1}^n\gamma_i/(\gamma_i^2+\omega^2)$, but also seems to hold for systems where the bulk viscosity is not a simple sum of $n=1$ bulk viscosities. 

Before we present numerical results from Eq.~\ref{eq:full_bulk_viscosity}, let us consider some limiting cases first.  
\subsection{Bulk viscosity without pions}
We can obtain the formula for the bulk viscosity if pions are not present in the EoS by setting $\lambda_3=\lambda_4=\lambda_5=0$ in Eq.~\ref{eq:full_bulk_viscosity}.  The EoS would also have to be recomputed with the pions removed.  The susceptibilities $A_3$, $B_3$, $C_3$, $D_1$, $D_2$, and $D_3$ are all ill-defined in a system without pions.  The previous definitions for the other susceptibilities carry over to the pionless case, except for the derivatives are, of course, no longer at constant pion fraction.  The expression for bulk viscosity in neutrino-trapped $npe\mu$ matter that undergoes the reactions corresponding to $\lambda_1, \lambda_2$, and $\lambda_6$ is given by
\begin{equation}\label{eq:bv_without_pions}
    \zeta = \dfrac{G'+H'\omega^2}{K'+L'\omega^2+\omega^4}.
\end{equation}
where the expressions for $G'$, $H'$, $K'$, and $L'$ are given in Appendix \ref{sec:bv_without_pions_appendix}.  As a check, we can see that the susceptibilities $A_3, B_3, C_3,$ $D_1$, $D_2$, and $D_3$ do not appear in the expression.  Furthermore, we also note that the Maxwell relation (\ref{eq:maxwell4}) applies in pionless matter as well (see also \cite{Harris:2024evy}).  In pionless matter, the independent susceptibilities are $A_1$, $A_2$, $B_1$, $B_2$, and $C_2$.  In Appendix \ref{sec:bv_without_pions_appendix}, we present further simplifications of this expression.
\subsection{Bulk viscosity in the limit $\lambda_3\rightarrow\infty$}
We return now to the EoS with pions included.  The reaction $n \leftrightarrow p + \pi^-$ (mediated by a spectator nucleon) is a strong interaction and therefore\footnote{At temperatures below $T\approx 1\text{ MeV}$, the pion density becomes so low that this rate becomes comparable to weak interaction timescales.  But for the temperatures studied in this paper, the rate can be considered infinitely fast.} occurs on a much faster timescale than the weak interactions \{$\lambda_1,\lambda_2,\lambda_4,\lambda_5,\lambda_6$\} and the timescale of density oscillations in a merger.  Thus, we take $\lambda_3 \rightarrow \infty$ in Eq.~\ref{eq:full_bulk_viscosity}.  

The limit where $n\leftrightarrow p + \pi^-$ occurs infinitely quickly is subtle.  One might assume that $\delta\mu_3$ could simply be set to zero in Eq.~\ref{eq:nbdxp}, \ref{eq:nbdxmu}, and \ref{eq:nbdxpi}.  However, it is the product $\lambda_3\delta\mu_3$ that appears on the RHS of those equations, and even for very tiny deviations of $\delta\mu_3$ from zero, the product $\lambda_3\delta\mu_3$ could still be sizeable since $\lambda_3$ is very large.  One must keep $\lambda_3$ finite throughout the calculation (up until Eq.~\ref{eq:full_bulk_viscosity}) and then take the limit $\lambda_3\rightarrow \infty$.  This was pointed out by \cite{Gusakov:2008hv,Jones:2001ya}, and was applied to the case of hyperon bulk viscosity in \cite{Alford:2020pld}. 

The $\lambda_3\rightarrow\infty$ limit of Eq.~\ref{eq:full_bulk_viscosity} yields
\begin{equation}\label{eq:bulk_viscosity_l3infinity}
    \zeta = \dfrac{Q+R\omega^2}{U+W\omega^2+\omega^4}.
\end{equation}
The coefficients $Q,R,U$ and $W$ are given in Appendix \ref{appendix:bv_l3inf}.
\subsection{Partial bulk viscosities}
Before we discuss the bulk viscosity results, it is first useful to introduce the concept of partial bulk viscosities.  The bulk viscosity with only one equilibration rate $\lambda_i$ active, and all other reaction rates $\lambda_{j\neq i}$ set to zero is defined to be the partial bulk viscosity associated with the associated process.  In neutrino-trapped $npe\mu$ matter, there are three partial bulk viscosities, while when thermal pions are added into the mix, there are six (c.f.~Table \ref{table:allreactions}).  We emphasize that the total bulk viscosity is \textit{not} the sum of the partial bulk viscosities.  But, often the total bulk viscosity will ``track'' one of the partial bulk viscosity curves over a certain range of thermodynamic conditions, which is why it is useful to consider the partial bulk viscosities.  This ``tracking'' will be apparent in the subsequent section.  Each partial bulk viscosity will look like the bulk viscosity in matter with one equilibrating particle fraction, for example, $npe$ matter where the proton fraction equilibrates with rate $\gamma$.  

Throughout this paper, we have called both $\lambda$ and $\gamma$ ''equilibration rates''.  Each is useful in its own way.  The difference in the rates, schematically, can be written $\overrightarrow{\Gamma}-\overleftarrow{\Gamma} \sim \lambda\left(\delta\mu-\delta\mu^0\right)=\lambda\delta\mu$ or $\overrightarrow{\Gamma}-\overleftarrow{\Gamma} \sim \gamma\left(n-n^0\right)$, where the superscript 0 denotes the beta equilibrium value.  So, $\lambda$ describes the rate at which the chemical potential difference is forced to its beta equilibrium value (zero), while $\gamma$ describes how quickly the particle fraction is forced to its beta equilibrium value by Le Chatelier's principle.  The susceptibilities of type $B$, $C$, and $D$ facilitate the conversion between $\gamma$ and $\lambda$.  Note that both $\lambda_i$ and $\gamma_i$ are positive.  $\gamma$ has a clearer physical meaning, as it has dimension one, and thus can be written as an inverse timescale and directly compared with the density oscillation frequency $\omega$.

In neutrino-trapped $npe\mu$ matter, without pions, there are three partial bulk viscosities because there are three equilibration rates\footnote{The equilibration rate $\gamma_i$ is obtained by taking the full expression for the bulk viscosity and setting $\lambda_{j\neq i}=0$.  Then the resultant expression is written in the form $\zeta \propto \gamma_i/(\gamma_i^2+\omega^2)$, from which $\gamma_i$ is extracted.}
\begin{subequations}
    \begin{align}
        \gamma_1^{\text{no } \pi} &\equiv -B_1\lambda_1\\
        \gamma_2^{\text{no } \pi} &\equiv -(B_2+C_2)\lambda_2\\
        \gamma_6^{\text{no } \pi} &\equiv (B_2-B_1-C_2)\lambda_6.
    \end{align}
\end{subequations}

The three partial bulk viscosities are then
\begin{subequations}
    \begin{align}
        \zeta_1^{\text{no } \pi} &= -\dfrac{A_1^2}{B_1}\dfrac{\gamma_1^{\text{no } \pi}}{(\gamma_1^{\text{no } \pi})^2+\omega^2}\label{eq:bvnopions1}\\
        \zeta_2^{\text{no } \pi} &= -\dfrac{A_2^2}{B_2+C_2}\dfrac{\gamma_2^{\text{no } \pi}}{(\gamma_2^{\text{no } \pi})^2+\omega^2}\label{eq:bvnopions2}\\
        \zeta_6^{\text{no } \pi} &= \dfrac{(A_1-A_2)^2}{B_2-B_1-C_2}\dfrac{\gamma_6^{\text{no } \pi}}{(\gamma_6^{\text{no } \pi})^2+\omega^2}.\label{eq:bvnopions6}
    \end{align}
\end{subequations}
The partial bulk viscosities are positive.  At a fixed density, the bulk viscosity $\zeta_i$ as a function of temperature is maximum when $\gamma_i=\omega$.  Therefore, the value of the bulk viscosity at its resonant maximum is given by the susceptibility prefactor divided by $2\omega$.  Thus, for a given frequency density oscillation $\omega$, the weak interaction rates set the temperature of the resonant maximum bulk viscosity and properties of the EoS (the susceptibilities) set the value of the bulk viscosity at its maximum.  The susceptibility prefactors can be related to the compressibility of the dense matter.  We find that
\begin{subequations}
    \begin{align}
        \zeta_{1,\text{max}}^{\text{no } \pi} &= \dfrac{n_B}{2\omega}\left( \dfrac{\partial P}{\partial n_B}\bigg\vert_{x_p,x_{\mu},T}  - \dfrac{\partial P}{\partial n_B}\bigg\vert_{x_{\mu},\delta\mu_1,T} \right)\label{eq:zeta_nopion_partial_1}\\
        \zeta_{2,\text{max}}^{\text{no } \pi} &= \dfrac{n_B}{2\omega}\left( \dfrac{\partial P}{\partial n_B}\bigg\vert_{x_p,x_{\mu},T}  - \dfrac{\partial P}{\partial n_B}\bigg\vert_{x_p-x_{\mu},\delta\mu_2,T} \right)\label{eq:zeta_nopion_partial_2}\\
        \zeta_{6,\text{max}}^{\text{no } \pi} &= \dfrac{n_B}{2\omega}\left( \dfrac{\partial P}{\partial n_B}\bigg\vert_{x_p,x_{\mu},T}  - \dfrac{\partial P}{\partial n_B}\bigg\vert_{x_{p},\delta\mu_1-\delta\mu_2,T} \right).\label{eq:zeta_nopion_partial_3}
    \end{align}
\end{subequations}
See Appendix \ref{appendix:jacobians} for the thermodynamic calculations.  The maximum value of bulk viscosity, to the extent that it tracks a particular partial bulk viscosity near its resonant maxima, is related to the compressibility predicted by the EoS (c.f.~Eq.~\ref{eq:compressibility}) and its multicomponent generalizations.  

In neutrino-trapped $npe\mu\pi$ matter, we can define analogously six partial bulk viscosities, each of which with one nonzero $\lambda_i$ ($i$ ranging from 1 to 6).  The expressions for them are given in Appendix \ref{appendix:partialbv}.  We will predominantly consider instead versions of the partial bulk viscosities where the $\lambda_3\rightarrow\infty$ limit is taken (that is, the bulk viscosity $\zeta_i^{\lambda_3\rightarrow\infty}$ where $\lambda_i$ is kept finite, $\lambda_3\rightarrow\infty$, and $\lambda_{j\neq\{i,3\}}=0$).  We will see that the partial bulk viscosity expressions become, in some sense, ``renormalized'' by the infinitely fast $\lambda_3$ process.  

In the system containing pions, with $\lambda_3\rightarrow\infty$, the $\gamma_i$ are given by
\begin{subequations}
    \begin{align}
        \gamma_1^{\lambda_3\rightarrow\infty} &\equiv -\left(B_1-\dfrac{B_2^2}{B_2+D_3}\right)\lambda_1\\
        \gamma_2^{\lambda_3\rightarrow\infty} &\equiv -\left(C_2+\dfrac{B_2D_3}{B_2+D_3}\right)\lambda_2\\
        \gamma_4^{\lambda_3\rightarrow\infty} &\equiv -\left(B_1-\dfrac{B_2^2}{B_2+D_3}\right)\lambda_4\\
        \gamma_5^{\lambda_3\rightarrow\infty} &\equiv -\left(C_2+\dfrac{B_2D_3}{B_2+D_3}\right)\lambda_5\\
        \gamma_6^{\lambda_3\rightarrow\infty} &\equiv \left(B_2-B_1-C_2\right)\lambda_6.
    \end{align}
\end{subequations}
We leave out $\gamma_3$ because we have already taken $\lambda_3\rightarrow\infty$.  It is interesting that the susceptibility prefactors in $\gamma_1$ and $\gamma_4$ match, as do those of $\gamma_2$ and $\gamma_5$.  The partial bulk viscosities (with $\lambda_3\rightarrow\infty$) are
\begin{subequations}
\begin{align}
\zeta_1^{\lambda_3\rightarrow\infty}&=-\dfrac{\left(A_1-\dfrac{A_3B_2}{B_2+D_3}\right)^2}{B_1-\dfrac{B_2^2}{B_2+D_3}}\dfrac{\gamma_1^{\lambda_3\rightarrow\infty}}{(\gamma_1^{\lambda_3\rightarrow\infty})^2+\omega^2}\\
\zeta_2^{\lambda_3\rightarrow\infty}&=-\dfrac{\left(A_2-\dfrac{A_3B_2}{B_2+D_3}\right)^2}{C_2+\dfrac{B_2D_3}{B_2+D_3}}\dfrac{\gamma_2^{\lambda_3\rightarrow\infty}}{(\gamma_2^{\lambda_3\rightarrow\infty})^2+\omega^2}\\
\zeta_4^{\lambda_3\rightarrow\infty}&=-\dfrac{\left(A_1-\dfrac{A_3B_2}{B_2+D_3}\right)^2}{B_1-\dfrac{B_2^2}{B_2+D_3}}\dfrac{\gamma_4^{\lambda_3\rightarrow\infty}}{(\gamma_4^{\lambda_3\rightarrow\infty})^2+\omega^2}\\
\zeta_5^{\lambda_3\rightarrow\infty}&=-\dfrac{\left(A_2-\dfrac{A_3B_2}{B_2+D_3}\right)^2}{C_2+\dfrac{B_2D_3}{B_2+D_3}}\dfrac{\gamma_5^{\lambda_3\rightarrow\infty}}{(\gamma_5^{\lambda_3\rightarrow\infty})^2+\omega^2}\\
\zeta_6^{\lambda_3\rightarrow\infty}&=\dfrac{(A_1-A_2)^2}{B_2-B_1-C_2}\dfrac{\gamma_6^{\lambda_3\rightarrow\infty}}{(\gamma_6^{\lambda_3\rightarrow\infty})^2+\omega^2}.
\end{align}
\end{subequations}

It is not clear to us that the prefactors in the $\lambda_3\rightarrow\infty$ limit can be written in terms of differences of compressibilities.  But, it is useful to compare these expressions to their counterparts without pions (Eqs.~\ref{eq:bvnopions1}-\ref{eq:bvnopions6}) and with pions but without $\lambda_3\rightarrow\infty$ (Eqs.~\ref{eq:bv1_l3zero}-\ref{eq:bv6_l3zero}).  

\begin{figure*}\centering
\includegraphics[width=0.4\textwidth]{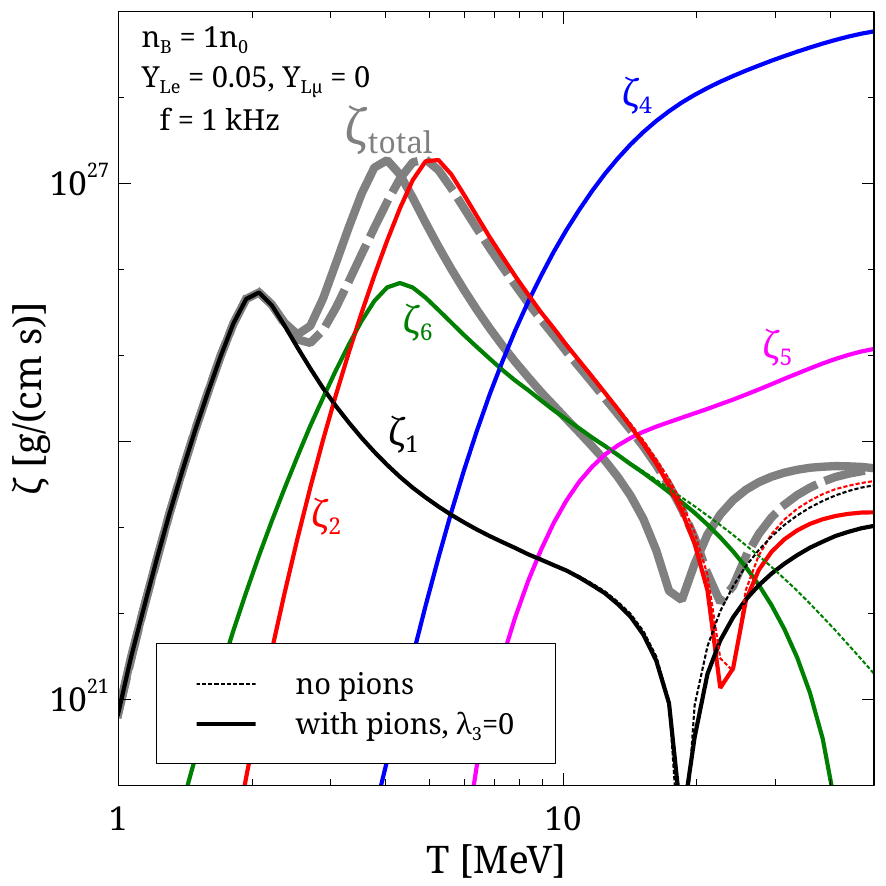}
\includegraphics[width=0.4\textwidth]{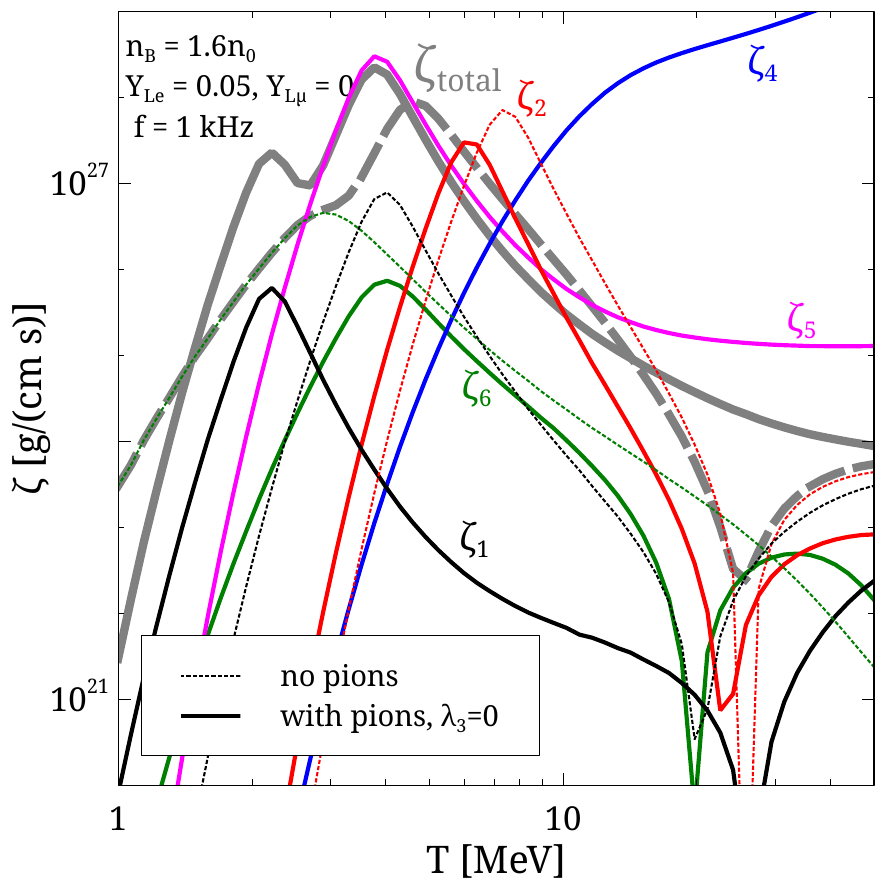}
\includegraphics[width=0.4\textwidth]{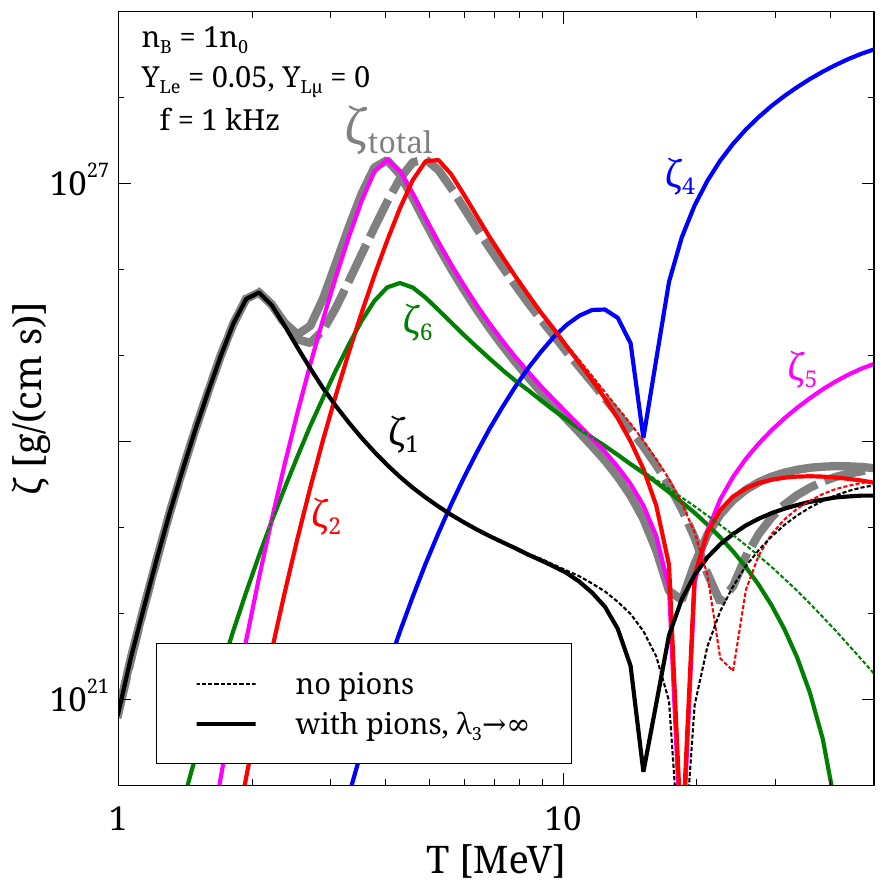}
\includegraphics[width=0.4\textwidth]{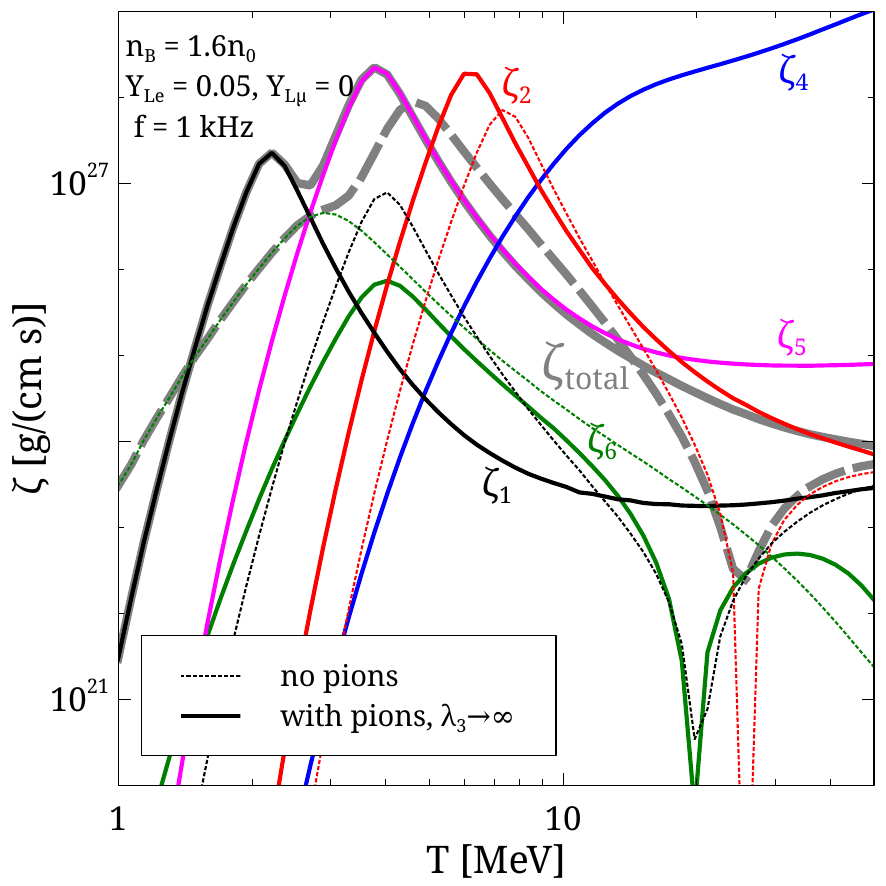}
\caption{Partial bulk viscosities in matter with $Y_{Le}=0.05$ and $Y_{L\mu} = 0$ at $1n_0$ (left column) and $1.6n_0$ (right column).  Results for pionless matter are in thin dotted lines.  The top row depicts the conventional definition of partial bulk viscosities, where $\lambda_3$ is set to zero.  The bottom row depicts the ``renormalized'' definition of partial bulk viscosities where $\lambda_3$ is sent to infinity.  The total bulk viscosity - \textit{not} the sum of the partial bulk viscosities (and discussed in Sec.~\ref{sec:results}) - is shown with a thick grey line. 
 For clarity, for this figure the x-axes are displayed with a log scale.}
\label{fig:partial_bulk_viscosities}
\end{figure*}

In Fig.~\ref{fig:partial_bulk_viscosities} we plot, just as an illustration, the partial bulk viscosities in the case of the pionless EoS ($\zeta_1^{\text{no } \pi}$, $\zeta_2^{\text{no } \pi}$, and $\zeta_6^{\text{no } \pi}$), for the EoS containing thermal pions where $\lambda_3$ is not sent to infinity (that is, the conventional definition of partial bulk viscosities in Appendix \ref{appendix:partialbv}), and for the EoS containing thermal pions but in the $\lambda_3\rightarrow\infty$ limit $\zeta_i^{\lambda_3\rightarrow\infty}$ ($i=1,2,4,5,6$).  The full result for the bulk viscosity is shown with a thick grey line - dashed for the EoS without pions and solid for the EoS with pions.  We see that most of the partial bulk viscosities display the predicted resonant structure, often with a conformal point at a temperature above that of the resonant maximum.  The quantities $\zeta_4$ and $\zeta_5$ increase with temperature, because of the strong temperature dependence of the relevant susceptibilities (this overrides the traditional resonance structure, which assumes that the susceptibilities do not depend strongly on temperature \cite{Alford:2010gw}).  The total bulk viscosity (with pions) seems to track the ``renormalized'' ($\lambda_3\rightarrow\infty$) partial bulk viscosities much better than the traditional ($\lambda_3=0$) versions.  We will refer back to this figure in the next section.  
\section{Results}\label{sec:results}
\subsection{Equilibration rates and bulk viscosity}
\begin{figure*}\centering
\includegraphics[width=0.4\textwidth]{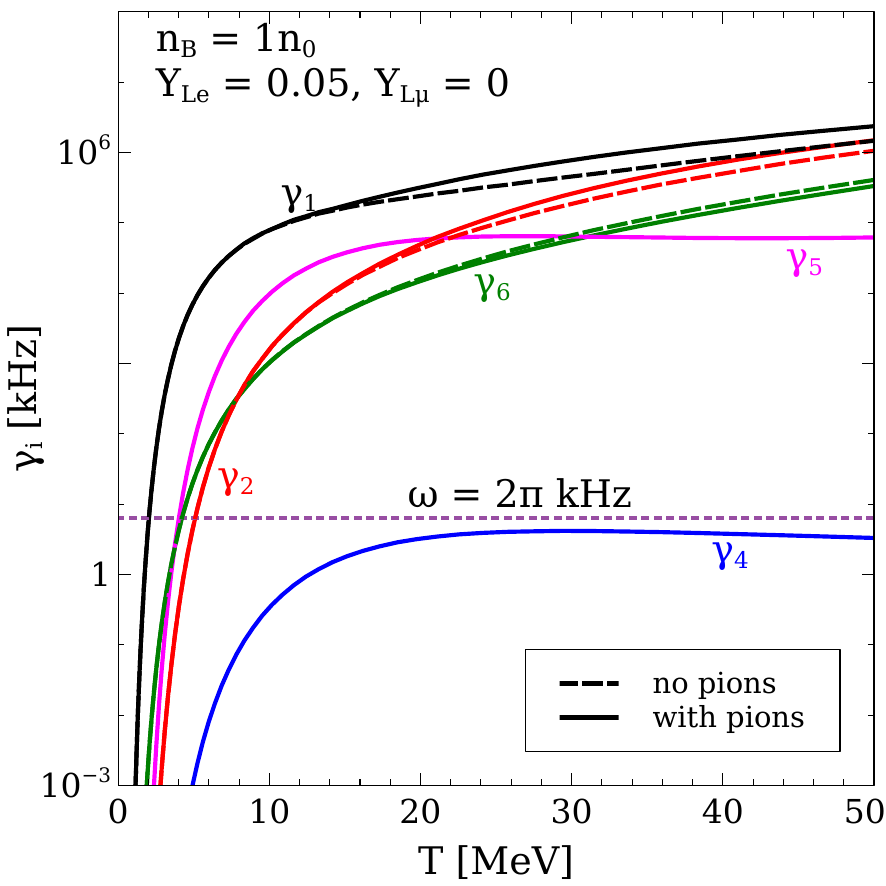} 
\includegraphics[width=0.4\textwidth]{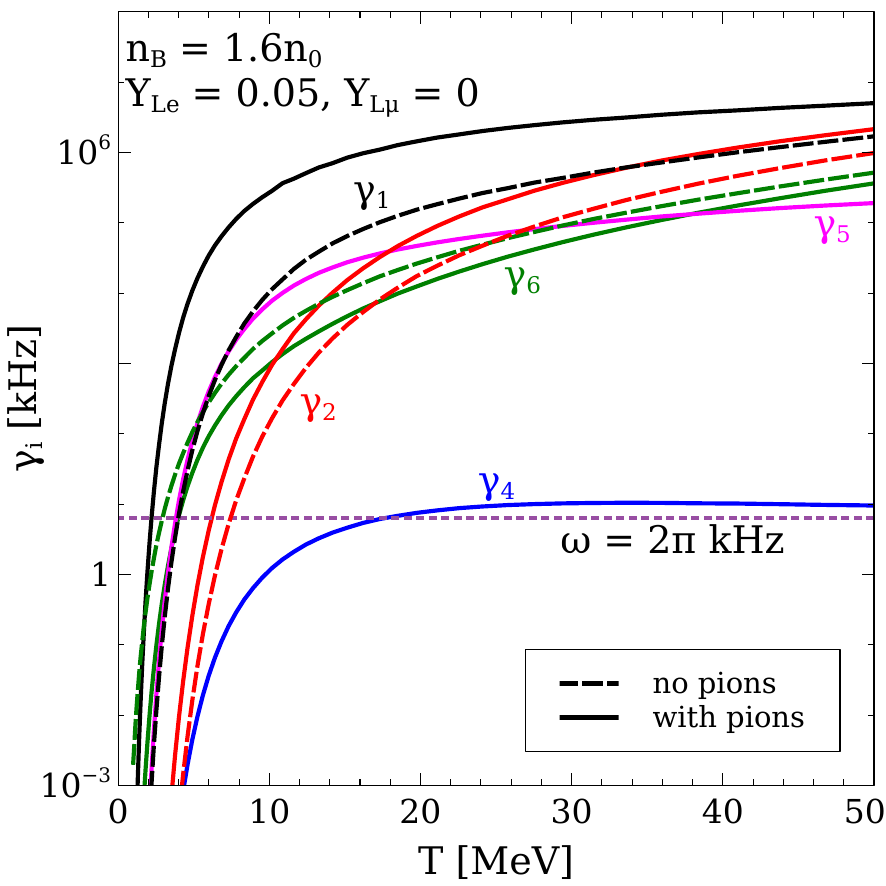}\\
\includegraphics[width=0.4\textwidth]{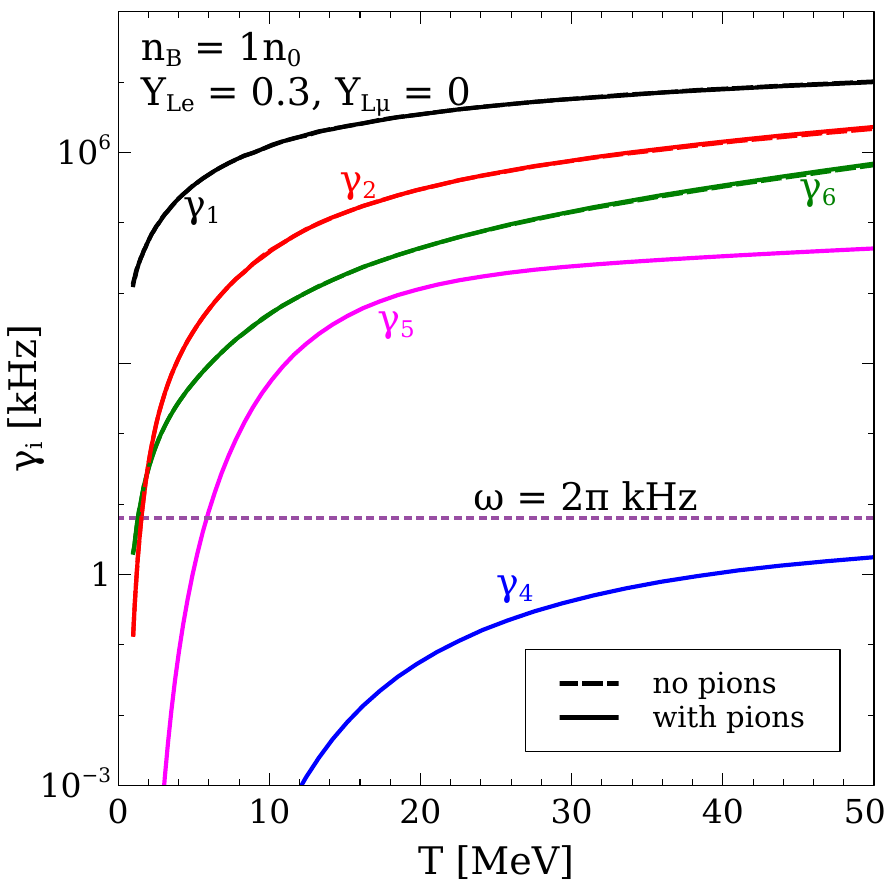}
\includegraphics[width=0.4\textwidth]{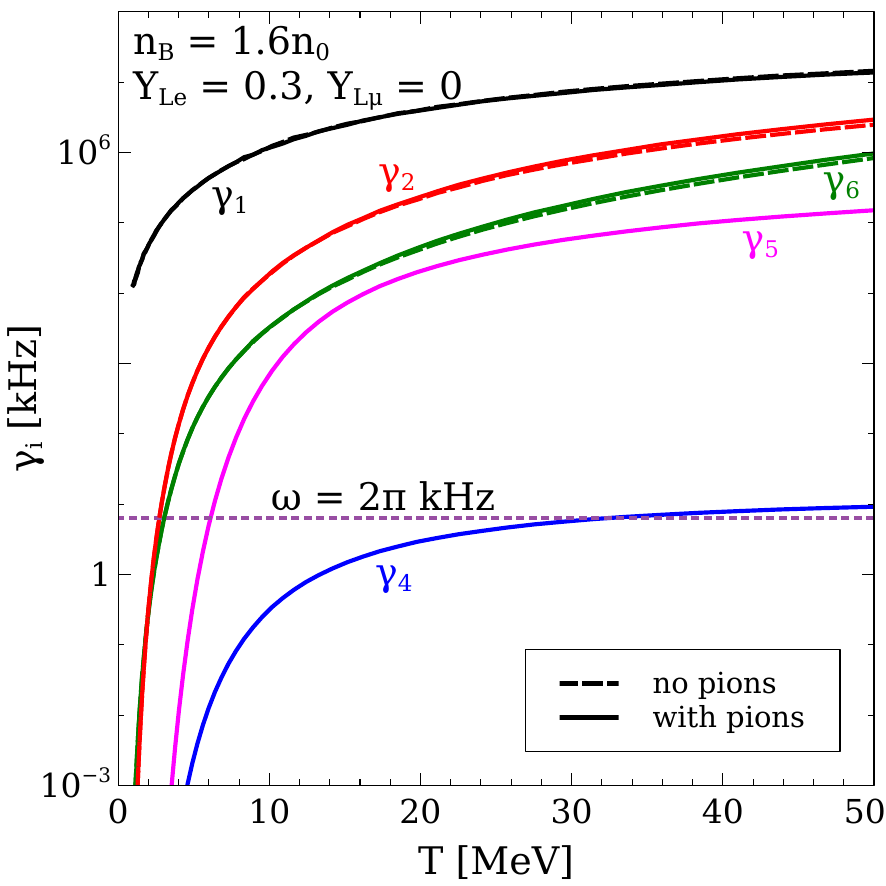}
\caption{Individual beta equilibration rates $\gamma_i^{\text{no }\pi}$ and $\gamma_i^{\lambda_3\rightarrow\infty}$ in neutron star merger conditions (top panels) and in supernovae conditions (bottom panels).  The dashed lines correspond to an EoS without pions, while the solid lines indicate an EoS with pions.  The rates $\gamma_4^{\lambda_3\rightarrow\infty}$ and $\gamma_5^{\lambda_3\rightarrow\infty}$ of course have no counterpart in matter without pions.  }
\label{fig:gamma}
\end{figure*}
\begin{figure*}\centering
\includegraphics[width=0.4\textwidth]{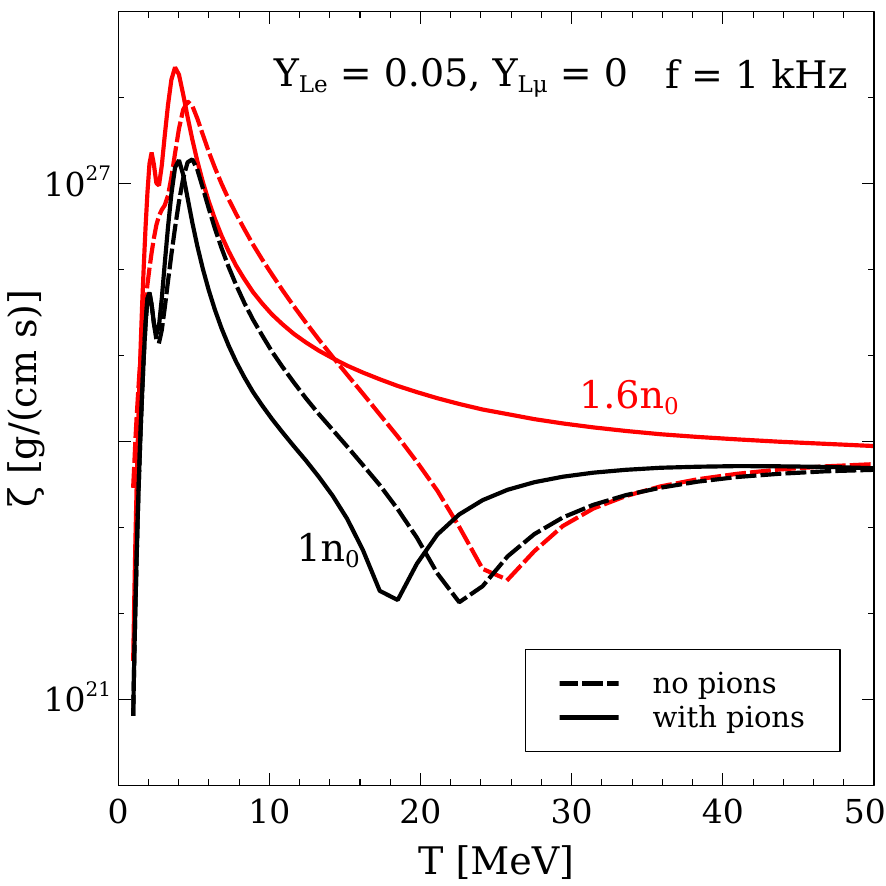}
\includegraphics[width=0.4\textwidth]{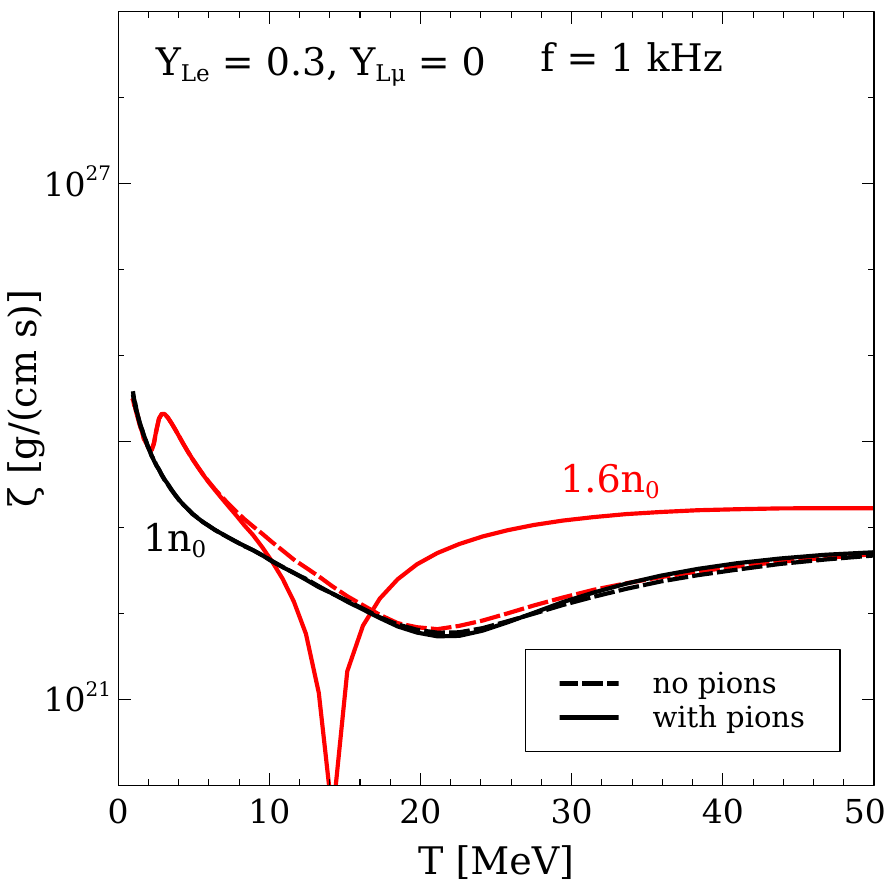}
\caption{Bulk viscosity in dense matter with and without pions, for a harmonic, small-amplitude, density oscillation with frequency 1 kHz.  The left panel corresponds to neutron star merger conditions and the right panel to supernovae conditions.}
\label{fig:bulk_visc}
\end{figure*}
In this section, we will use the EoS developed in Sec.~\ref{sec:EoS} and calculate the bulk viscosity using the formulas in Sec.~\ref{sec:bulkviscosity}, comparing the EoSs with and without pions.  The presence of pions both changes the EoS itself (the particle fractions, the susceptibilities, etc) as well as provides new beta equilibration pathways.  We examine both in this section.  Our results for the bulk viscosity rely on the susceptibilities (Fig.~\ref{fig:A_susc} and \ref{fig:BCD_susc}) and on the beta equilibration rates $\gamma_i$ (Fig.~\ref{fig:gamma}) and are themselves plotted in Figs.~\ref{fig:bulk_visc} and \ref{fig:bulk_visc_spectrum}, and in Appendix \ref{appendix:bv_finite_YLmu} for different lepton number conditions.

In matter with one equilibrating particle fraction, the behavior of the bulk viscosity resulting from a small amplitude, harmonic density oscillation of frequency $\omega$ is well known.  As long as the susceptibilities do not strongly depend on temperature, but the equilibration rate does, then the bulk viscosity has a resonant structure as a function of temperature, with a peak at the temperature at which the Urca rate matches the density oscillation frequency $\omega$ - see Fig.~7 in \cite{Alford:2010gw} or Fig.~2 in \cite{Alford:2019qtm} for examples.  The partial bulk viscosities plotted in Fig.~\ref{fig:partial_bulk_viscosities} match this form as well, though the susceptibilities in our model \textit{do} have a significant temperature dependence, modulating the baseline resonant structure. 

The matter studied in this paper, with three independent particle fractions $\{x_p,x_{\mu},x_{\pi}\}$ could have as many as three resonant peaks as a function of temperature (at fixed baryon density).  However, the rate $\lambda_3$ (or $\gamma_3$) is essentially infinitely fast in the temperature range considered.  That is, the resonant maximum of the corresponding partial bulk viscosity $\zeta_3$ (Eq.~\ref{eq:zeta3eqn}) would be at a temperature $T\lesssim 1\text{ MeV}$ and thus we would only expect two resonant peaks in the displayed temperature range.  The susceptibilities, as we see in Figs.~\ref{fig:A_susc} and \ref{fig:BCD_susc}, do have a significant temperature dependence, which will cause the overall bulk viscosity $\zeta(T)$ to have nontrivial behavior.  Now we explain the equilibration rates $\gamma$ (Fig.~\ref{fig:gamma}) and the bulk viscosity curves (Fig.~\ref{fig:bulk_visc}).

We consider neutron star merger conditions ($Y_{Le}=0.05, Y_{L\mu}=0$) first.  The top left panel of Fig.~\ref{fig:gamma} indicates that in the EoS without pions, all three rates $\gamma_1$, $\gamma_2$, and $\gamma_6$ rise above $f=1\text{ kHz}$ in the temperature range of a few MeV.  Indeed, one sees that the three partial bulk viscosities (Fig.~\ref{fig:partial_bulk_viscosities}) have resonant peaks in that temperature range.  The full calculation for the bulk viscosity in pionless matter (Fig.~\ref{fig:bulk_visc}) tracks the partial bulk viscosity $\zeta_1^{\text{no }\pi}$ for low temperature and through its resonant maximum at $T\approx 2$ MeV.  As temperature increases, both $\gamma_2$ and $\gamma_6$ cross resonance at very similar temperatures, leading to a second resonant peak in the total bulk viscosity, that doesn't quite overlap with any of the partial bulk viscosities.  There is no third resonant peak, because the pionless EoS only has two equilibrating quantities, $x_p$ and $x_{\mu}$, so the third, slowest, rate is redundant.  As temperature increases further, the bulk viscosity continues to decrease, since all particle fractions equilibrate very quickly compared to the millisecond timescale density oscillations. 
 At a temperature of about 20 MeV, the system becomes partly conformal, because $A_1$ and $A_2$ pass through zero, and the bulk viscosity dips down to a sharp minimum, but not to zero (unlike the partial bulk viscosities, which do dip to zero).  So, without pions, the neutrino-trapped $npe\mu$ bulk viscosity has two resonant peaks as a function of temperature, both at well below $T=10\text{ MeV}$, and then a partial conformal dip at a temperature around 20 MeV.    

When pions are added to the EoS and their reactions are considered, the existing rates $\gamma_1$, $\gamma_2$, $\gamma_6$ change very little\footnote{As can be seen in the top left panel of Fig.~\ref{fig:gamma}, including pions does change the rates at high temperature, where the pion population becomes substantial.  But at these high temperatures, the bulk viscosity is very far off of resonance because the rates are much faster than the 1 kHz density oscillation, and so the bulk viscosity is small.}.  However, three new equilibration rates involving pions, $\gamma_3$, $\gamma_4$, and $\gamma_5$ enter the picture.  The direct Urca (electron) rate $\gamma_1$ is still the fastest (besides $\gamma_3$), and leads to the first resonant peak in the bulk viscosity, just as in the pionless case (though, in reality, this is the second resonant peak, because $\gamma_3$ would reach resonance at some temperature below the 1 MeV minimum studied here).  The next resonant peak comes from $\gamma_5$, the pion decay to muons, which supersedes the reactions that dominate in the absence of pions.  In fact, the resonance shifts to slightly lower temperatures.  At temperatures just above that resonant peak ($T\gtrsim 4$ MeV), all three independent particle fractions are quick to equilibrate and thus the bulk viscosity decreases with temperature.  While Fig.~\ref{fig:gamma} shows other processes passing above the 1 kHz line (and thus Fig.~\ref{fig:partial_bulk_viscosities} shows other resonant peaks at higher temperature, for example, $\zeta_4$), those never manifest in the full bulk viscosity calculation because they are redundant.  The fastest three rates control the bulk viscosity in neutrino-trapped $npe\mu\pi^-$ matter.

At higher density ($1.6n_0$), without pions, now $\gamma_6$ (muon-electron conversion) is the fastest rate (see top right panel of Fig.~\ref{fig:gamma}), but it crosses 1 kHz at about the same temperature as $\gamma_1$ did at $1n_0$, so the resonant peak is still at $T\approx 2$ MeV.  Then, $\gamma_1$ crosses resonance at $T \approx 4\text{ MeV}$, and then $\gamma_2$ does at $T \approx 7\text{ MeV}$.  These two peaks seem to interfere with each other and what results appears to be an interference between the two resonances ($\gamma_1$ and $\gamma_2$), as the peak location and magnitude are between those of the two partial viscosities.  As temperature increases further, the bulk viscosity decreases until the dip indicating partial conformality.  

At this higher density, when pions are included in the EoS, their population is substantial for the entire temperature range discussed here.  Thus, the reaction rates $\gamma$ change substantially when pions are added to the EoS.  When the EoS includes pions, $\gamma_1$ is much faster than it is without pions present, and it becomes the first reaction to reach resonance as $T$ increases (besides $\gamma_3$, of course).  The total bulk viscosity thus tracks $\zeta_1^{\lambda_3\rightarrow\infty}$ in the range around its resonance.  The pion decay rate $\gamma_5$ reaches resonance at $T\approx 4$ MeV, and the total bulk viscosity tracks $\zeta_5^{\lambda_3\rightarrow\infty}$ in this region.  At higher temperatures, equilibration of the three independent particle fractions is faster than millisecond timescales, and thus the bulk viscosity decreases with temperature.  The partial conformal point at high temperature is smoothed out dramatically in the presence of pions.  

Overall, in merger conditions, the pions provide new equilibration pathways, which shift the temperature of the resonant peaks in the bulk viscosity, though not dramatically.  More importantly, the presence of pions in the EoS modify the existing susceptibilities and contribute new susceptibilities which alter the maximum value that bulk viscosity reaches. 
 While the bulk viscosity increases with density even without pions, the pions further increase the bulk viscosity at high density.  The pions naturally have more of an effect at higher densities where their population is higher.  

In supernovae conditions, namely $Y_{Le}=0.3, Y_{L\mu} = 0$, many of the equilibration rates (bottom left panel of Fig.~\ref{fig:gamma}) are already fast compared to millisecond time scales, even at low temperature.  This is due to the high conserved lepton fraction in the system.  Adding pions to the EoS does not change these rates much at all, except for introducing the new equilibration mechanisms $\gamma_3$, $\gamma_4$, and $\gamma_5$.  With or without pions, the bulk viscosity in the displayed temperature range is always on the downhill side of the resonance, decreasing with temperature, until the partial conformal points at temperatures above 20 MeV.  At higher density ($1.6n_0$), $\gamma_1$ is already faster than millisecond timescales even for the lowest temperatures shown.  But $\gamma_2$ and $\gamma_6$ both cross resonance at $T\approx 2$ MeV, leading to a resonant maximum in the bulk viscosity curve,  This feature is the same with and without pions - unlike in merger conditions, the rates containing pions do not substantially contribute to the beta equilibration.  The only effect the pions have at $1.6n_0$ is to create a partial conformal point at $T\approx 15$ MeV where none existed in the pionless case.  All partial bulk viscosities have conformal points near this temperature (but not all at the same temperature).  Since the partial conformal point temperatures do not quite coincide, then the bulk viscosity does not drop to zero (and even if it did, that would only be the chemical equilibration bulk viscosity - there would still be a small bulk viscosity from thermal equilibration \cite{Kolomeitsev:2014gfa}).

\begin{figure*}\centering
\includegraphics[width=0.4\textwidth]{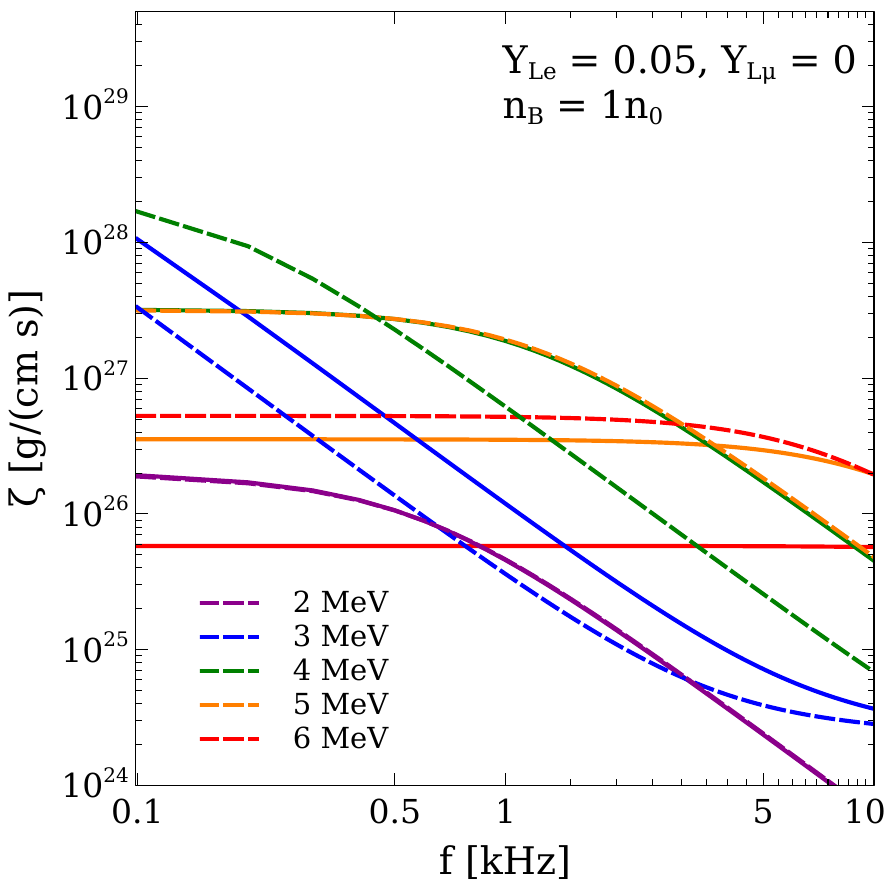}
\includegraphics[width=0.4\textwidth]{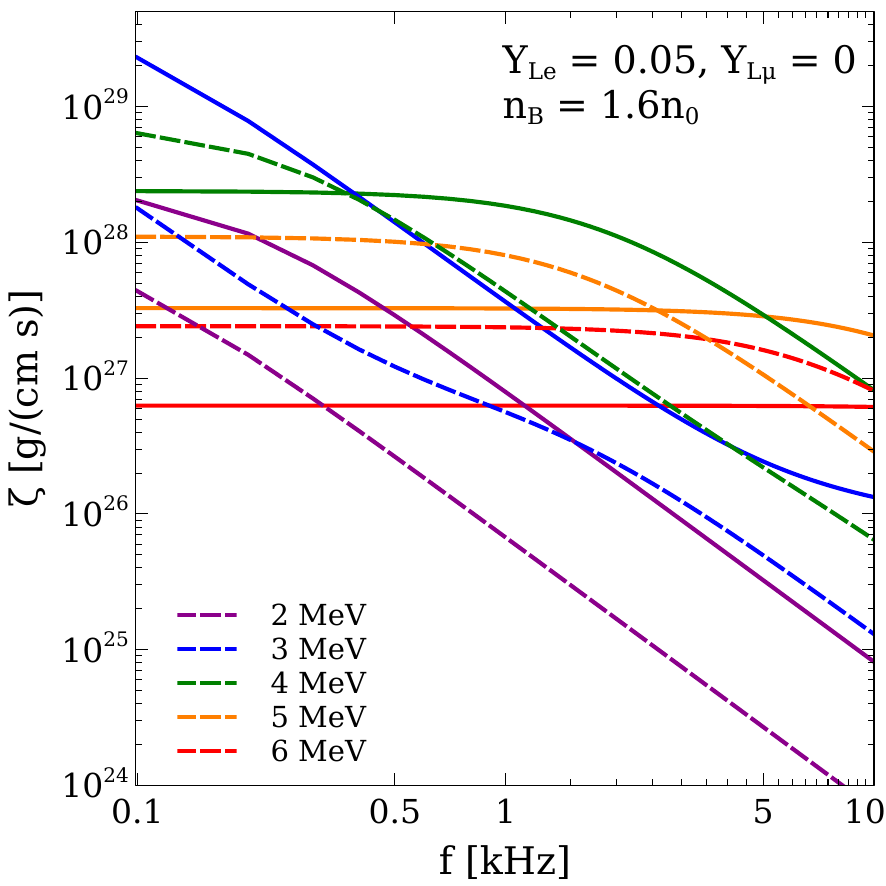}
\caption{Bulk viscosity as a function of density oscillation frequency $f=\omega/(2\pi)$ for neutrino-trapped matter with $Y_{Le}=0.05$ and $Y_{L\mu}=0$.  The left panel depicts matter at $1n_0$ and the right panel matter at $1.6n_0$.  Dashed lines correspond to matter without pions, and solid lines correspond to matter with a thermal pion population.}
\label{fig:bulk_visc_spectrum}
\end{figure*}

The (oscillation) frequency-dependence of the bulk viscosity in merger conditions is plotted in Fig.~\ref{fig:bulk_visc_spectrum}.  The spectrum of oscillations in the merger has a wide range of frequencies, with significant amplitude from hundreds of Hz up to 2 kHz (see the bottom panel of Fig.~4 of \cite{Alford:2017rxf}).  Likely an even wider range of frequencies is relevant.  The bulk viscosity is highest at low frequencies and drops monotonically with increasing density oscillation frequency.  There are no spectral peaks.  At high enough temperatures, the bulk viscosity becomes frequency independent.  Note that the energy dissipation rate has its own frequency dependence, so the higher values of bulk viscosity at low frequencies do not lead to quicker damping timescales.  In fact, for small $\omega$, the damping time $\tau\sim\omega^{-2}$, though in the limit $\omega\rightarrow\infty$, the damping time $\tau\sim\omega^0$. 

\subsection{Particle dynamics during an oscillation}
In the derivation of the bulk viscosity coefficient, we considered the evolution of the particle content throughout one oscillation.  This evolution ultimately gives rise to bulk viscous dissipation as the particle fraction evolution becomes out of phase with the density oscillation.  In this section, we plot the particle fraction as a function of time to better understand the dynamics that give rise to bulk viscous dissipation.  The density oscillation amplitude $\delta n_B$ is chosen to be $0.03n_0$, small enough to remain in the subthermal regime except at the lowest temperatures studied.  
\begin{figure*}\centering
\includegraphics[width=0.4\textwidth]{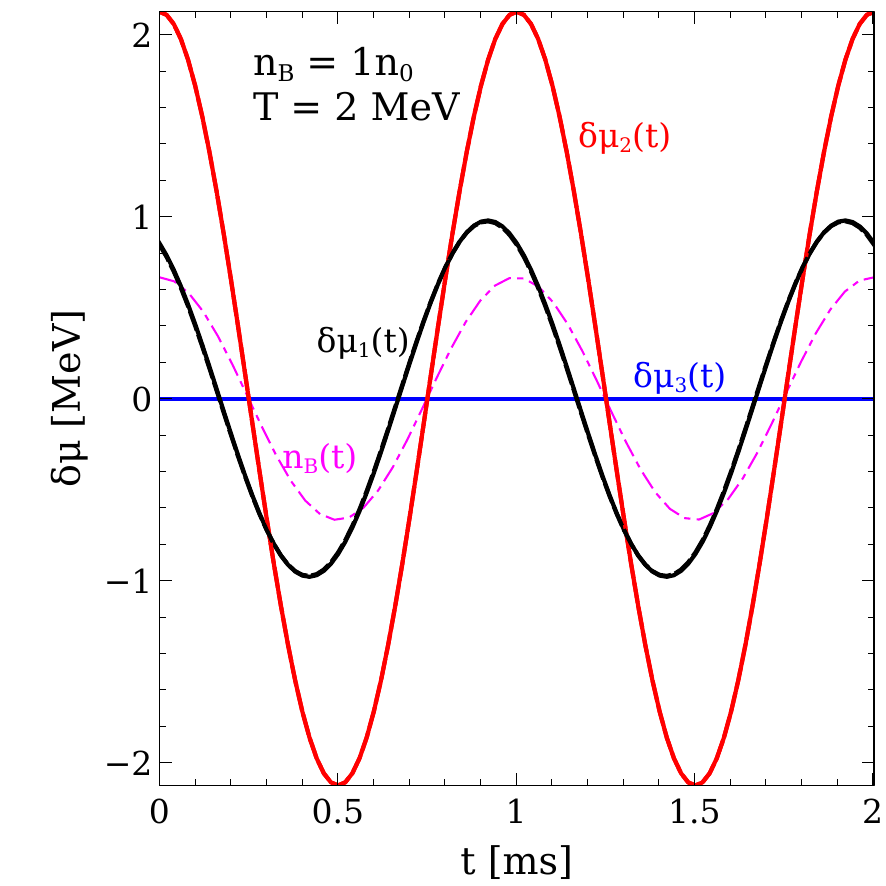}
\includegraphics[width=0.4\textwidth]{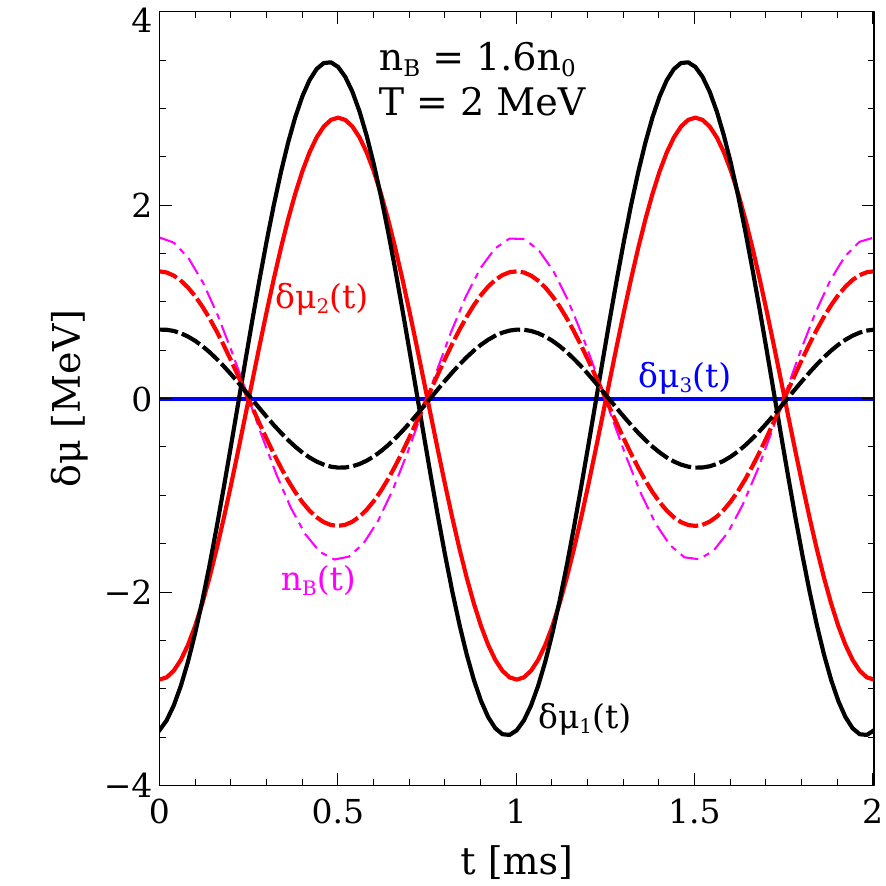}\\
\includegraphics[width=0.4\textwidth]{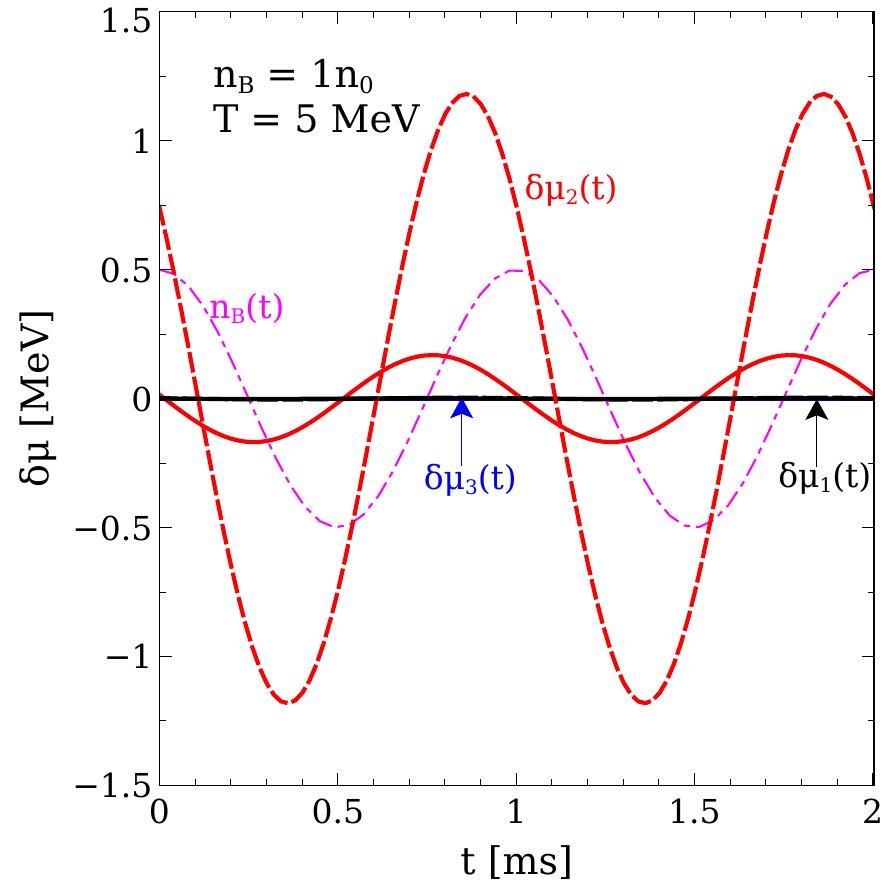}
\includegraphics[width=0.4\textwidth]{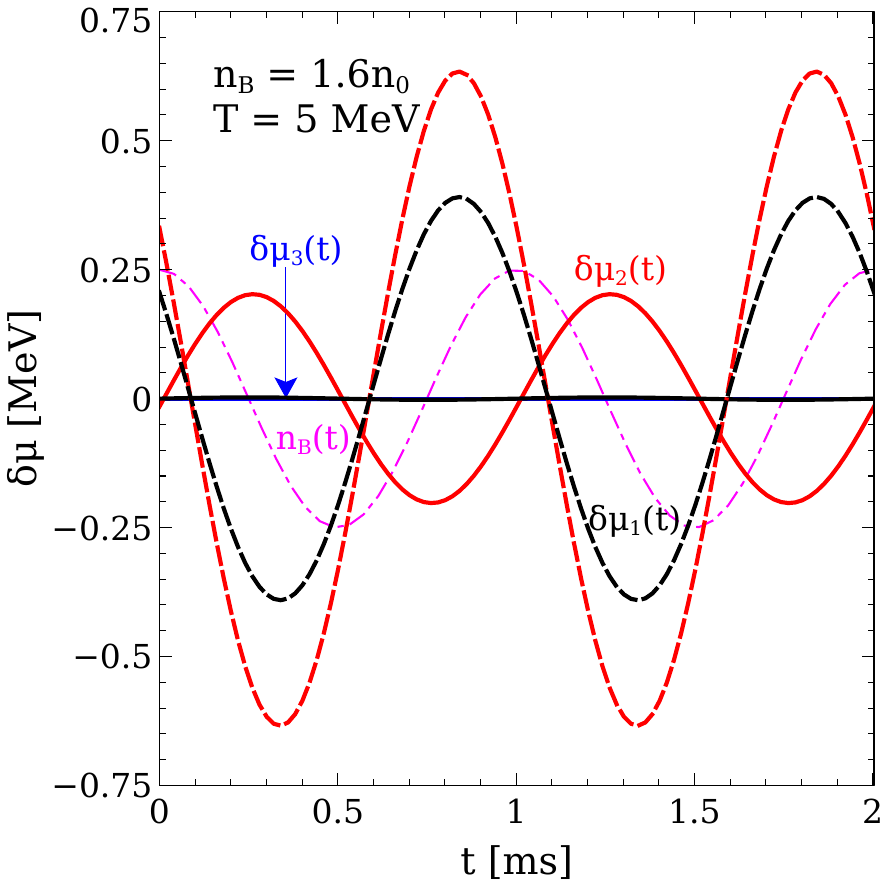}\\
\includegraphics[width=0.4\textwidth]{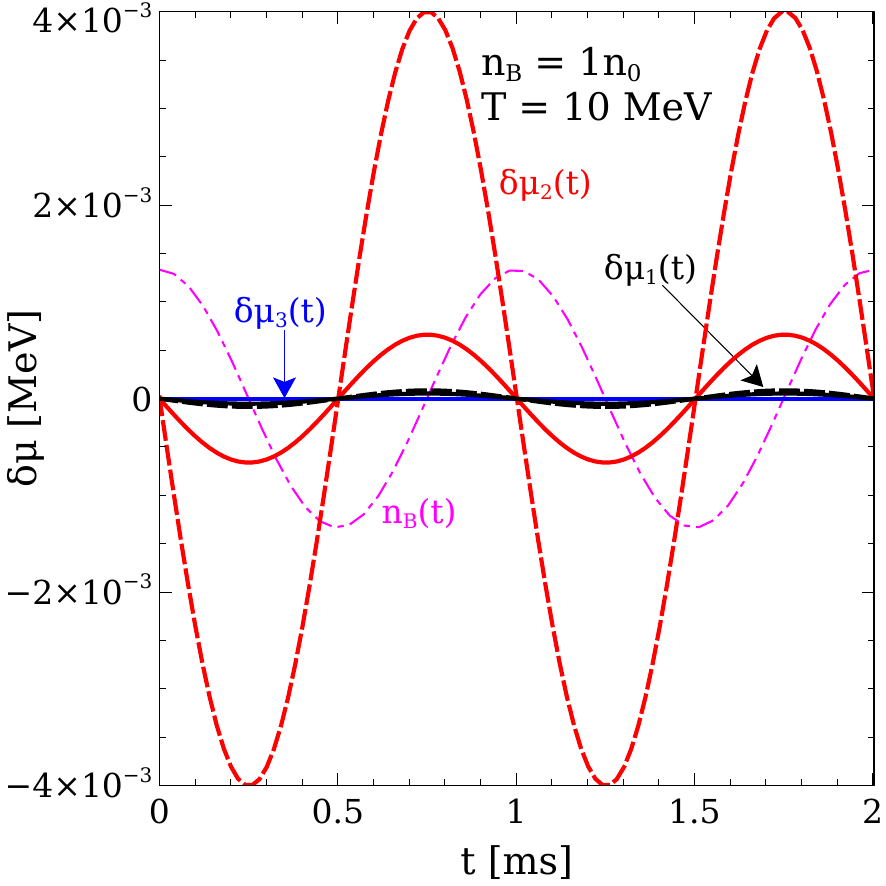}
\includegraphics[width=0.4\textwidth]{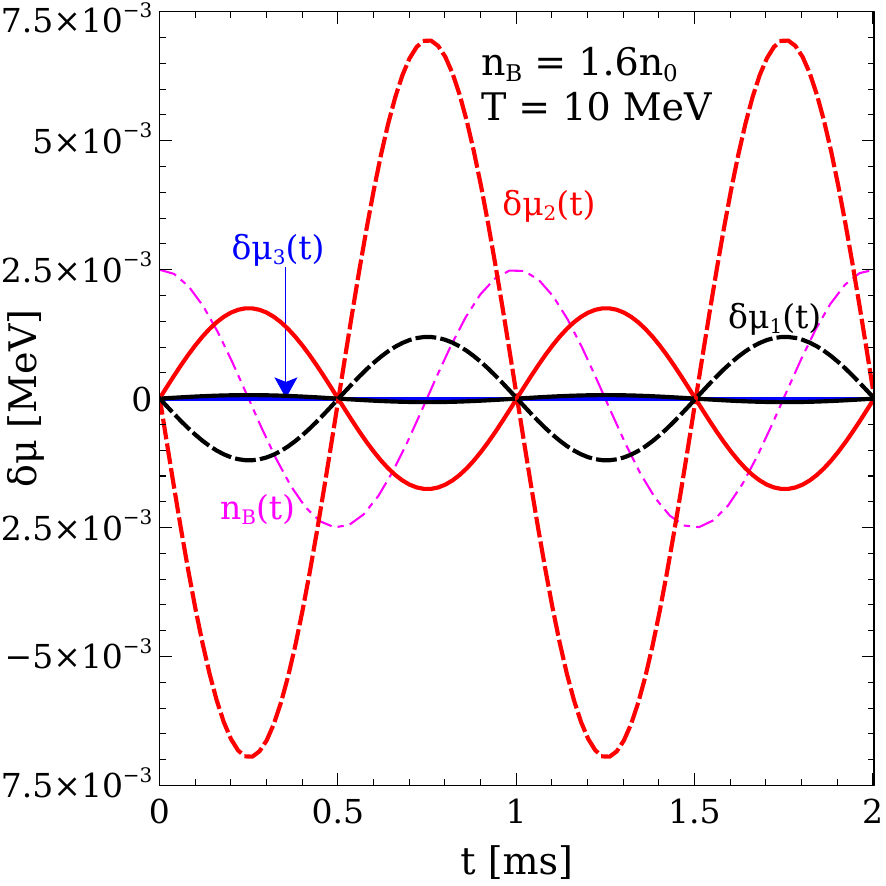}\\
\caption{In matter with $Y_{Le}=0.05, Y_{L\mu}=0$, the evolution of the three $\delta\mu$ quantities over the course of two oscillation periods (with oscillation amplitude $\delta n_B = 0.03n_0$).  Dashed lines denote matter without pions, solid lines denote matter including thermal pions.  Rows 1, 2, and 3 are at temperatures of 2, 5, and 10 MeV, respectively.  The left column is in matter at $1n_0$ and the right column is for matter at $1.6n_0$.  Note that the y axes do not have the same scale.  The density oscillation $n_B(t)$ is overlaid in pink so that one can tell the degree to which the $\delta\mu_i$ are in phase with the density oscillation.}
\label{fig:dmus}
\end{figure*}
\begin{figure*}\centering
\includegraphics[width=0.4\textwidth]{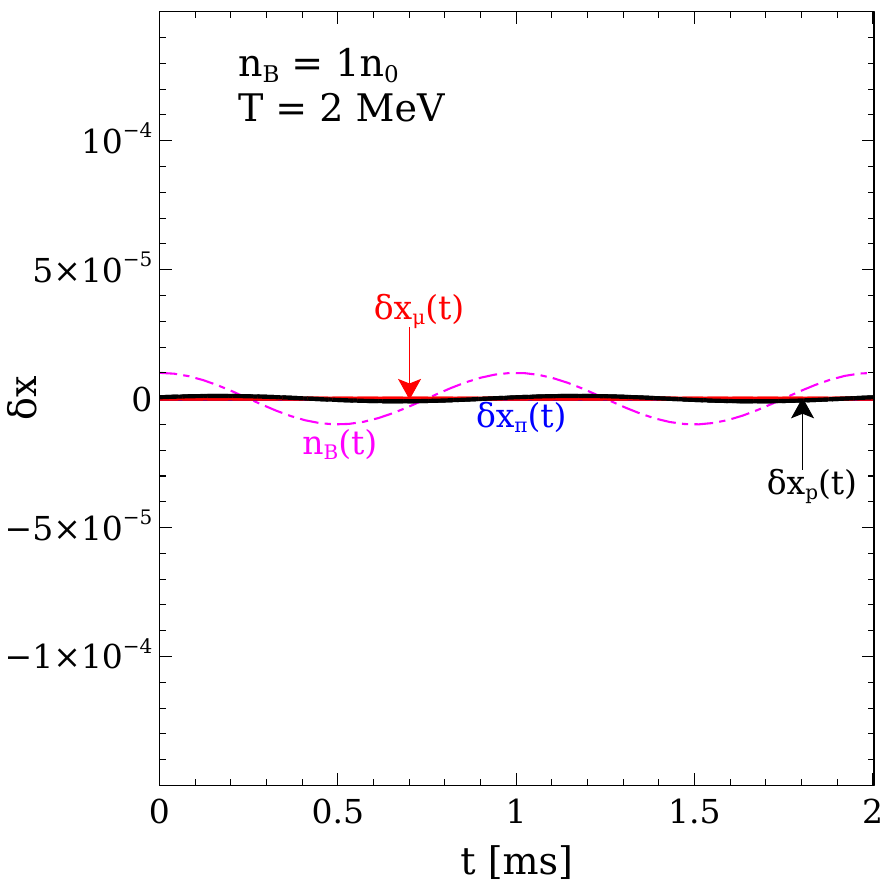}
\includegraphics[width=0.4\textwidth]{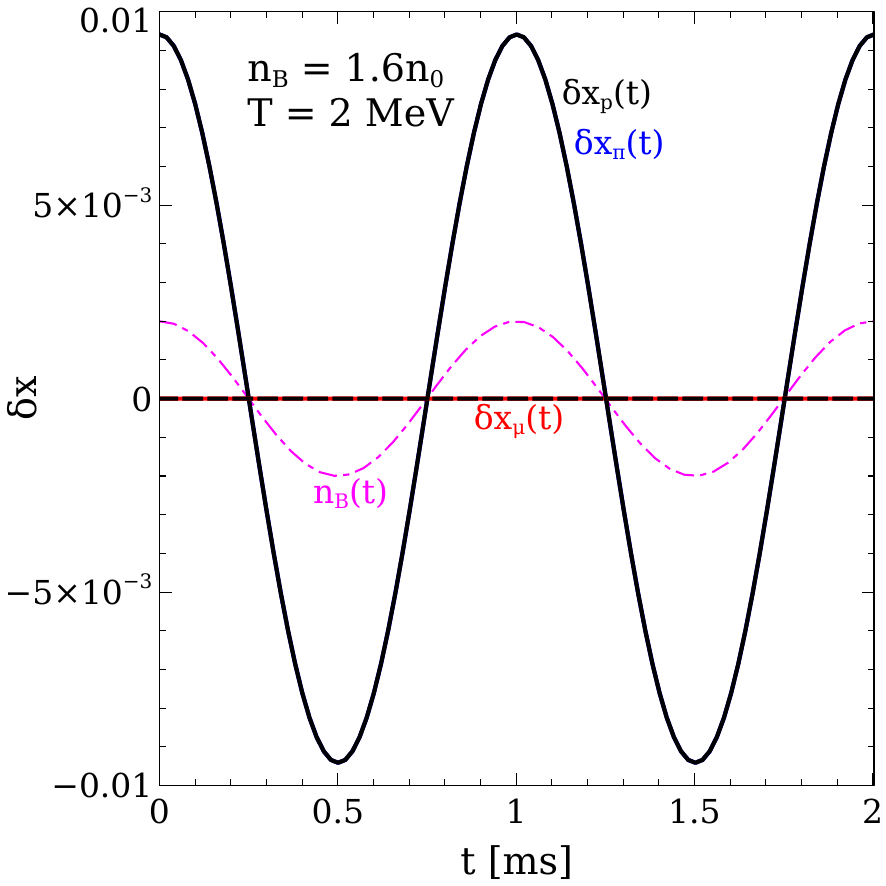}\\
\includegraphics[width=0.4\textwidth]{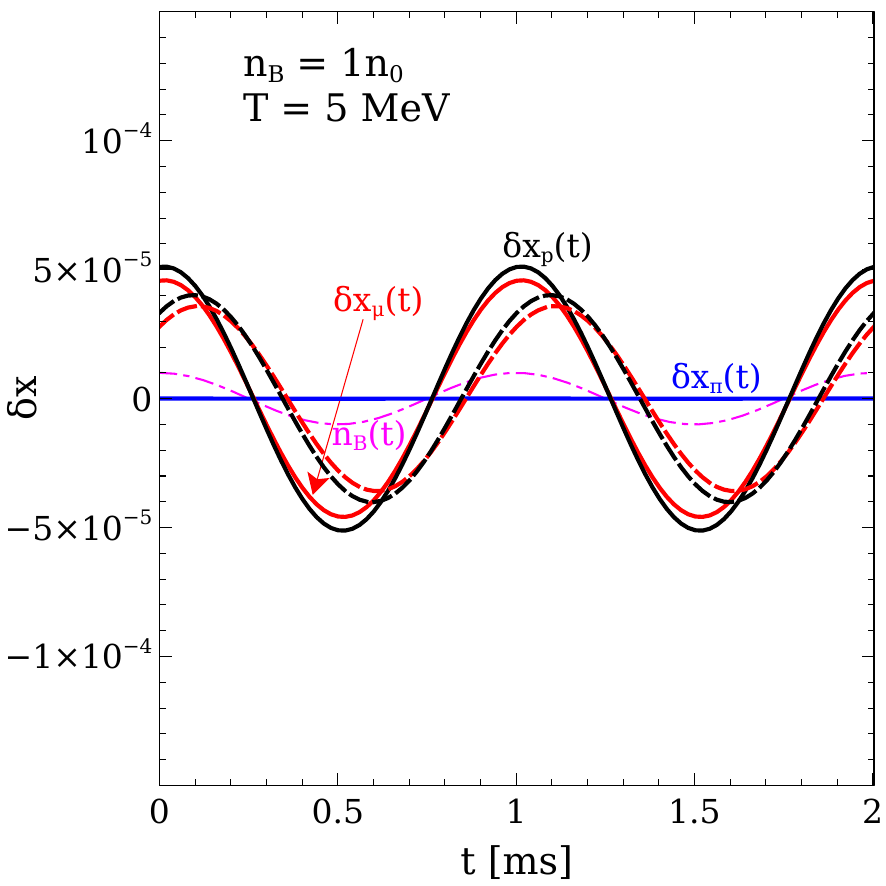}
\includegraphics[width=0.4\textwidth]{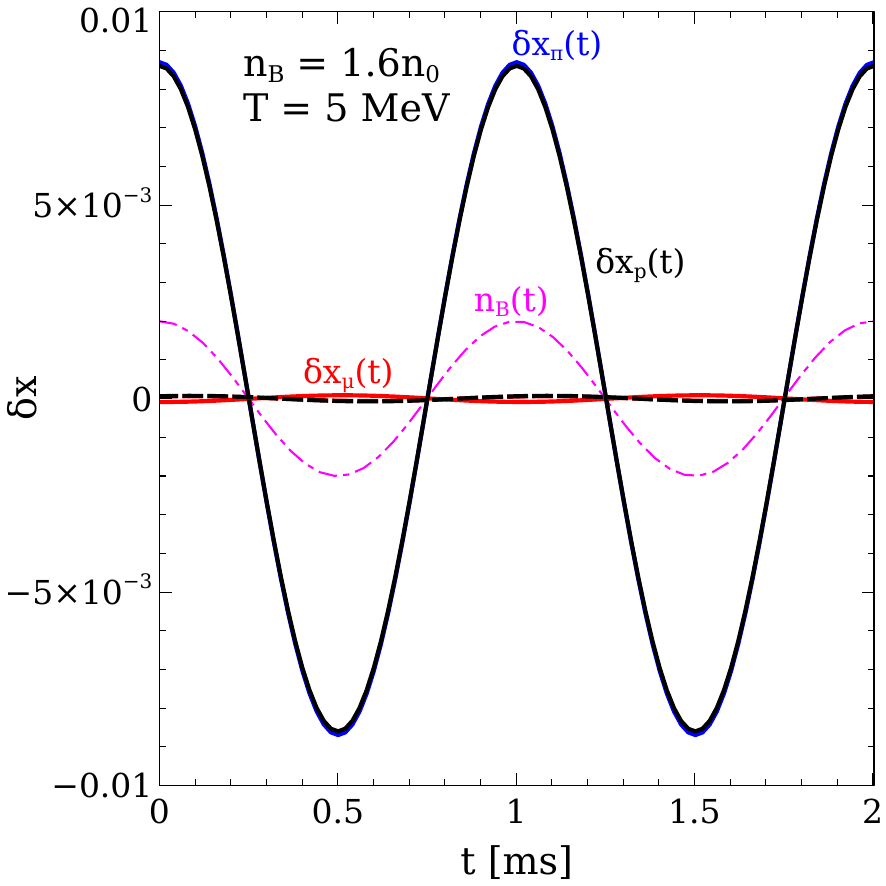}\\
\includegraphics[width=0.4\textwidth]{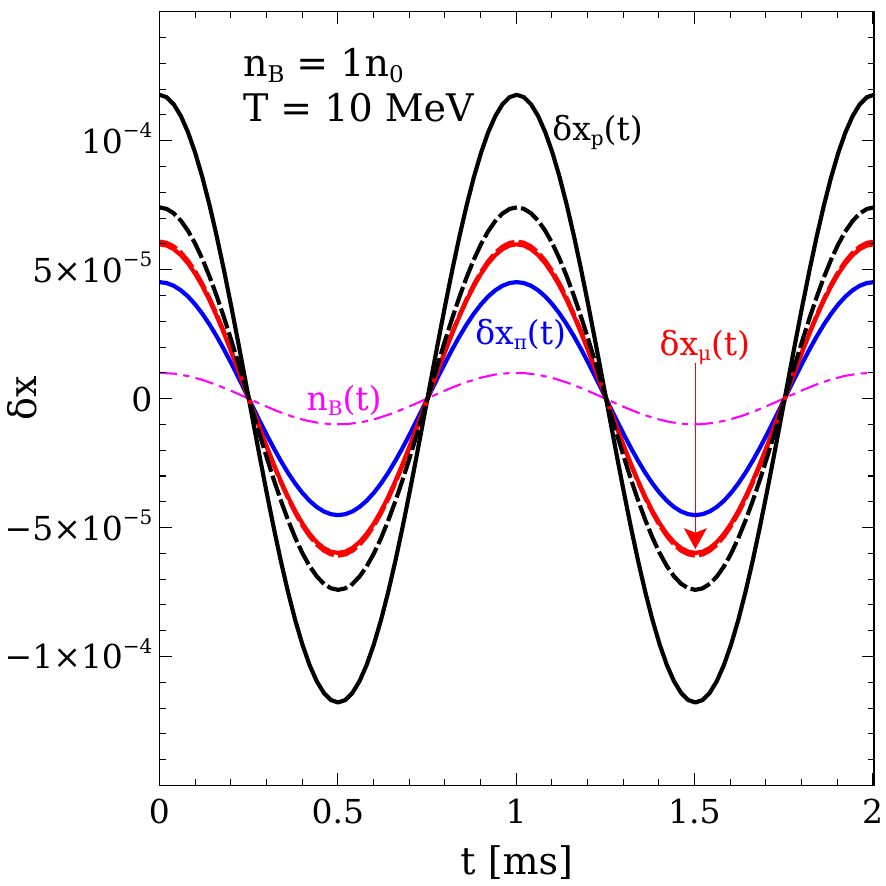}
\includegraphics[width=0.4\textwidth]{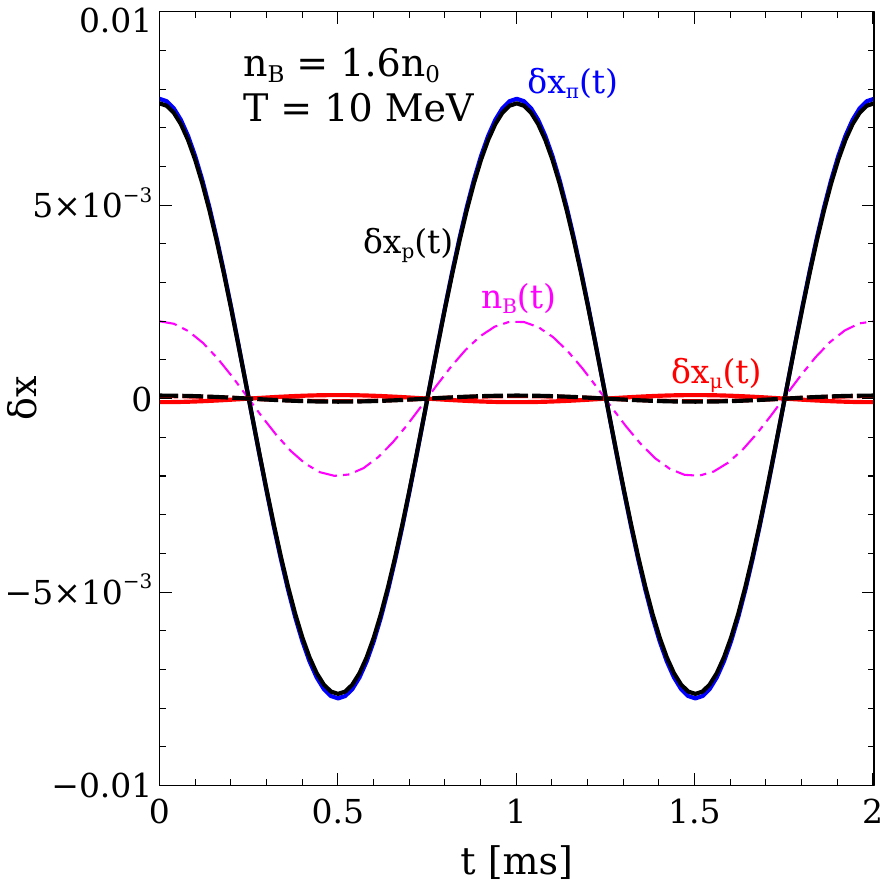}\\
\caption{In matter with $Y_{Le}=0.05, Y_{L\mu}=0$, the evolution of the three independent particle fractions $x_p$, $x_{\mu}$, and $x_{\pi}$ over the course of two oscillation periods (with oscillation amplitude $\delta n_B = 0.03n_0$).  Dashed lines denote matter without pions, solid lines denote matter including thermal pions.  Rows 1, 2, and 3 are at temperatures of 2, 5, and 10 MeV, respectively.  The left column depicts matter at $1n_0$ and the right column shows matter at $1.6n_0$.  The density oscillation $n_B(t)$ is overlaid in pink so that one can tell the degree to which the particle fractions are in phase with the density oscillation.}
\label{fig:dxs}
\end{figure*}

In Fig.~\ref{fig:dmus} we plot the evolution of $\delta\mu_1$, $\delta\mu_2$, and $\delta\mu_3$ over the course of the harmonic density oscillation.  These quantities are obtained from Eq.~\ref{eq:dmu_exp}, where the real and imaginary parts of $\delta x_i$ are obtained from the solution of the system of equations described after Eq.~\ref{eq:zeta_intermediate_expression}.  When any of the $\delta\mu_i$ departs from zero, the matter is no longer in chemical equilibrium.  In the limit $\lambda_3\rightarrow\infty$, $\delta\mu_3$ is forced to zero, so the reaction $n\leftrightarrow p+\pi^-$ is always infinitesimally close to equilibrium.  

At $1n_0$ and $T=2\text{ MeV}$ (top left panel of Fig.~\ref{fig:dmus}), both $\delta\mu_1(t)$ and $\delta\mu_2(t)$ oscillate up to a couple of MeV out of beta equilibrium.  The addition of pions into the system at this low temperature and density makes no difference because the pion number is very low.  At its maximum, the quantity $\delta\mu_2$ exceeds 2 MeV, and is no longer small compared to the temperature, which indicates that the subthermal approximation taken here loses validity - suprathermal corrections \cite{Alford:2010gw,Madsen:1992sx} should be taken into account at these low temperatures.  As the temperature rises to 5 MeV, the chemical reaction rates grow quickly and now the direct Urca process (with an electron) is fast enough to keep $\delta\mu_1$ very close to chemical equilibrium throughout the oscillation.  The quantity $\delta\mu_2$ is still pushed by about 1 MeV out of equilibrium, as the muon population is too low for the direct Urca processes with a muon ($\gamma_2$) (or muon-electron conversion $\gamma_6$) to equilibrate it efficiently.  When pions are included in the EoS, their population is sufficiently high at $T=5\text{ MeV}$ that $\delta\mu_2$ is kept closer to equilibrium than without the pions present.  Finally, at $T=10\text{ MeV}$, the system (with or without pions) remains extremely close to beta equilibrium (note the y-axis scale in the bottom left panel of Fig.~\ref{fig:dmus}) throughout the density oscillation, indicating that the bulk viscosity will be small at this high temperature.

At a higher density ($1.6n_0$), the pion density at all temperatures is high enough to impact the chemical equilibration pathways.  In an EoS without pions, at $T=2$ MeV, the fastest reaction is muon-electron conversion $\gamma_6$ which equilibrates $\delta\mu_1-\delta\mu_2$.  When pions are added, $\gamma_3$ equilibrates $\delta\mu_3$ very efficiently.  But $\gamma_6$ is reduced and $\gamma_1$ is enhanced due to the presence of the pions in the EoS.  In this situation, the addition of pions to the EoS makes the system depart further from chemical equilibrium than without pions (though, admittedly the system departs from the subthermal regime at this low temperature).  At the higher temperatures of 5 or 10 MeV, the same trends are present as at their $1n_0$ counterparts.  The speed-up in the equilibration rates at this higher density keeps $\delta\mu$ and $\delta\mu_2$ much closer to zero, and the pions help to keep the system closer to beta equilibrium.

The evolution of the particle fractions $x_p(t)$, $x_{\mu}(t)$, and $x_{\pi}(t)$ are plotted in Fig.~\ref{fig:dxs} at different temperatures and densities, with and without pions in the EoS.  These quantities are obtained from Eq.~\ref{eq:xi_exp}, where the real and imaginary parts of $\delta x_i$ are the solutions of the system of equations described after Eq.~\ref{eq:zeta_intermediate_expression}.  Note that we plot $\delta x_i(t)$, which indicates how far the particle fraction of species $i$ is from its beta equilibrium value at the \textit{background} density, not the concurrent value of the density $n_B(t)$.

In studying these plots, both the height and the phase of the sinusoidal variations in the particle fractions are informative.  Our intuition for the behavior of $\delta x_i(t)$ comes from the $npe$ matter case.  For example, Eq.~34 in the arXiv version of \cite{Harris:2024evy} indicates that in cold matter, where the beta equilibration is slow, $\delta x_p(t) \sim (\delta n_B/n_B)\left[A/(\vert B\vert n_B)\right](\gamma/\omega)\sin{(\omega t)}$, which\footnote{Here, $A$ and $B$ correspond to $A_1$ and $B_1$ in this text (removing, of course, the designation of fixed muon and pion fractions as the expression in Ref.~\cite{Harris:2024evy} applies to $npe$ matter).} is out of phase with the density oscillation, but also vanishes in the slowly equilibrating limit (that is, when the Urca rate is too slow to chemically equilibrate the matter, the particle fraction does not change from the beta equilibrium value at the background density).  In the opposite limit, where the beta equilibration is very fast, $\delta x_p(t)\sim (\delta n_B/n_B)\left[A/(\vert B\vert n_B)\right]\cos{(\omega t)}$.  That is, the particle fraction oscillation is in phase with the density oscillation and, at the maximum in the $x_p(t)$ oscillation cycle, has a magnitude that is independent of the equilibration rate.  The particle fraction just tracks the beta equilibrium curve $x_i^{\beta-eq.}[n_B(t)]$ throughout the density oscillation.  

While the equilibration rate $\gamma$ is a strong function of $T$, it is not necessarily the case that the above description of $\delta x_p(t)$ at low and high $\gamma$ can be mapped to low and high temperature, as the susceptibility prefactor in $\delta x_p(t)$ depends on temperature too.  For low temperature $npe$ matter, the susceptibilities are relatively temperature-independent.  But for a system including thermal pions, the susceptibilites are not independent of temperature, as we have seen (c.f.~Figs.~\ref{fig:A_susc} and \ref{fig:BCD_susc}).  

Without pions (dashed lines in Fig.~\ref{fig:dxs}), we see the expected behavior.  The fractions $x_p(t)$ and $x_{\mu}(t)$ deviate little from their beta equilibrium values at low temperature, and as temperature increases, they deviate further, as the equilibration is fast enough to force the particle fractions to trace the beta equilibrium curve $x_i^{\beta-eq.}[n_B(t)]$.  The particle fraction oscillations also become more in-phase with the density oscillation with increasing temperature.

When pions are included in the EoS, the expected behavior (increasing deviation $\delta x_i(t)$ with increasing temperature, up to some maximum deviation) still occurs at $1n_0$, but at $1.6n_0$, the trend reverses for some of the particle species.  Indeed, the presence of the pions (through their contribution to the susceptibilities) causes the particle fractions $x_p(t)$ and $x_{\pi}(t)$ to oscillate less dramatically as temperature increases, while the muon fraction $x_{\mu}(t)$ oscillation amplitude increases in the expected way with increasing temperature.   
\section{Conclusions}
We studied the bulk viscosity stemming from flavor-changing interactions in dense matter, including for the first time a population of thermal pions.  To include the pions in the EoS, we follow the virial expansion approach developed in \cite{Fore:2019wib}.  We include reactions involving pions in the set of reactions that give rise to bulk viscosity.  This work focuses on neutrino-trapped nuclear matter, which is the bulk of the matter in a neutron star merger remnant or supernovae environment.  Our goal was to understand how the presence of thermal pions modifies the beta equilibration and bulk viscosity of dense matter.  

The bulk viscosity is a function of the reaction rates in the system and compositional properties of the EoS, such as the susceptibilities (which we demonstrated here can be related to multicomponent generalizations of the compressibility of nuclear matter).  We calculated these quantities and the resultant bulk viscosity in thermodynamic conditions encountered in neutron star mergers and supernovae, and compared the results with and without thermal pions in the EoS.  We also found relationships between the local maxima in the bulk viscosity as a function of temperature and various compressibilities of the EoS, making apparent the possibility of using the measurement of the bulk viscosity to learn about the nature of dense matter.  Finally, we investigated how the composition of the dense matter evolves over the course of a small-amplitude density oscillation. 

Our main finding is that thermal pions can significantly alter the density and temperature dependence of the bulk viscosity of hot dense matter.  The pion population rapidly increases with density and temperature, and so the bulk viscosity deviates most strongly from the pionless case at higher densities and temperatures.  The addition of pions to the EoS modifies the existing nuclear matter susceptibilities to a relatively small extent (Figs.~\ref{fig:A_susc} and \ref{fig:BCD_susc}) but also introduces new susceptibilities that involve the pion degree of freedom and therefore do not exist without pions.  These new susceptibilities have a significant effect on the bulk viscosity.  In addition, pion reaction rates have two different effects.  The (essentially) infinitely fast strong interaction $n+n\rightarrow n+p+\pi^-$ does not produce a new resonant peak in the bulk viscosity in the temperature range studied, but does (as one can see from Fig.~\ref{fig:partial_bulk_viscosities}) modify the maximum value of the bulk viscosity and can move or even eliminate or introduce partial conformal points in the bulk viscosity at fixed density, $\zeta(T)$.  The slower pion decays ($\pi^-\rightarrow \mu^-+\bar{\nu}_{\mu}$) in some conditions can overtake the existing nucleonic or leptonic equilibration processes and can dominate beta equilibration.  

In the specific thermodynamic conditions that we studied, the addition of thermal pions to the EoS and the inclusion of their flavor-changing reactions into the bulk viscosity calculation enhances the maximum value of bulk viscosity in NS merger conditions by a factor of a few at higher density, while shifting the second peak in the bulk viscosity to a slightly lower temperature.  In supernovae conditions, the inclusion of pions introduces a partial conformal point, where the bulk viscosity drops precipitously, but not all the way to zero.  In merger conditions, the maximum bulk viscosity [in excess of $10^{27}\text{ g/(cm s)}$] occurs at temperatures of a few MeV.  Unfortunately, at these temperatures, the neutrino MFP is rather long, and it is unlikely that our assumption of a thermally equilibrated Fermi sea of neutrinos is valid.  At higher temperatures ($T\gtrsim$ 5-10 MeV) where our calculation is likely valid, the bulk viscosity predicted is quite small because the reaction rates are much faster than the millisecond hydrodynamical timescales.  So, while pions enhance the bulk viscosity in neutrino-trapped matter, the bulk viscosity is likely still too small to impact neutron star mergers, unless density oscillations persist in long-lived merger remnants. 

This work is only the beginning of the study of the role of pions in transport in hot, dense matter environments.  This calculation used a virial EoS to model the pions, an approach that is valid at the high temperatures and low densities considered here and captures the qualitative trends associated with attractive p-wave interactions and repulsive s-wave interactions.  However, a microscopic treatment of the pion-nucleon system at finite temperature and density using Chiral Perturbation Theory (ChiPT) or other effective Hamiltonians that are constrained by pion-nucleon and nucleon-nucleon scattering data is needed to make more reliable quantitative predictions and assess the role of pions over the wide range of densities and temperatures of interest to our study. Recent work has shown that ChiPT provides useful guidance for estimating the uncertainties associated with the calculation of the charged pion masses in neutron-rich matter \cite{Fore:2023gwv}. However, more work is necessary to address the role of the attractive p-wave interactions that play a critical role in enhancing the thermal population of negatively charged pions at finite temperatures.      

Furthermore, the bulk viscosity should be calculated in the neutrino-transparent regime, and the work in this paper should be extended to the suprathermal regime.  As a first step to include pions in neutron star merger simulations, a nucleonic EoS where pions were included in a makeshift way was implemented recently in a merger simulation \cite{Vijayan:2023qrt}.  In the future, EoSs that explicitly include pions as degrees of freedom (and are valid for wide ranges of density and temperatures) should be developed and used in simulations.  Additionally, just as Urca rates are beginning to be included in merger simulations, rates involving pions should be included in simulations with EoSs that include pions, allowing for the effects of pions on energy transport and dissipation to be studied.
\begin{acknowledgments}
SPH thanks Mark Alford for collaboration in the early stages of the project, Albino Perego and Eleonora Loffredo for information about the conserved lepton fractions in neutron star mergers, and Alex Haber for comments on the draft of this paper.  The work of SPH was supported by the U.S. Department of Energy grant DE-FG02-00ER41132 as well as the National Science Foundation grants PHY-1430152 (JINA Center for the Evolution of the Elements) and PHY 21-16686. 
BF is supported by the U.S. Department of Energy, Office of Science, Office of Nuclear Physics, under contracts DE-AC02-06CH11357, by the 2020 DOE Early Career Award number ANL PRJ1008597, by the NUCLEI SciDAC program, and Argonne LDRD awards.
The work of SR was supported by the U.S. Department of Energy grant DE-FG02-00ER41132. SR also thanks the members of the N3AS Physics Frontier Center, funded by the NSF Grant No. PHY-2020275 for useful conversations.
\end{acknowledgments}
\appendix
\section{Bulk viscosity of matter with net muon lepton fraction $Y_{L\mu}$}\label{appendix:bv_finite_YLmu}
In the main text of the paper, we focused on the more likely case at densities near $n_0$, where muonless, neutrino-transparent matter is heated up and becomes neutrino trapped, but without any net conserved muon number $Y_{L\mu}.$  We were able to discuss essentially all aspects of the bulk viscosity of this neutrino-trapped $npe\mu\pi$ matter.  However, it is conceivable that in some cases, there would be a net $Y_{L\mu}$, and thus in this appendix we calculate the bulk viscosity of matter in two scenarios where $Y_{Le}=Y_{L\mu}$.

\begin{figure*}\centering
\includegraphics[width=0.4\textwidth]{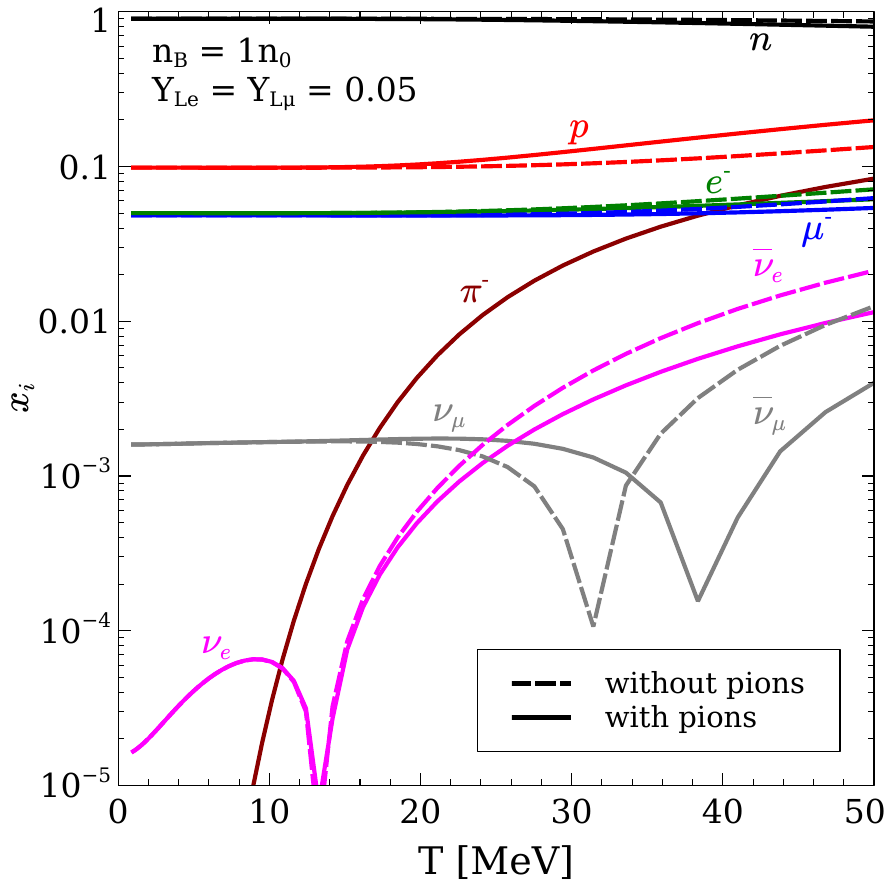}
\includegraphics[width=0.4\textwidth]{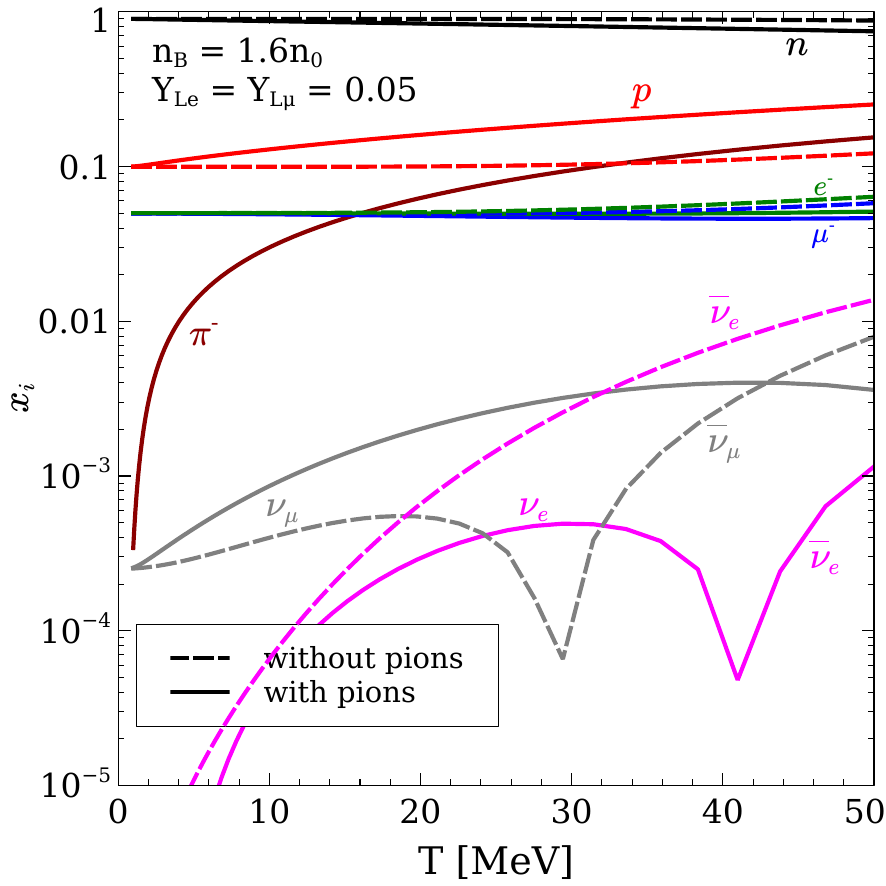}
\includegraphics[width=0.4\textwidth]{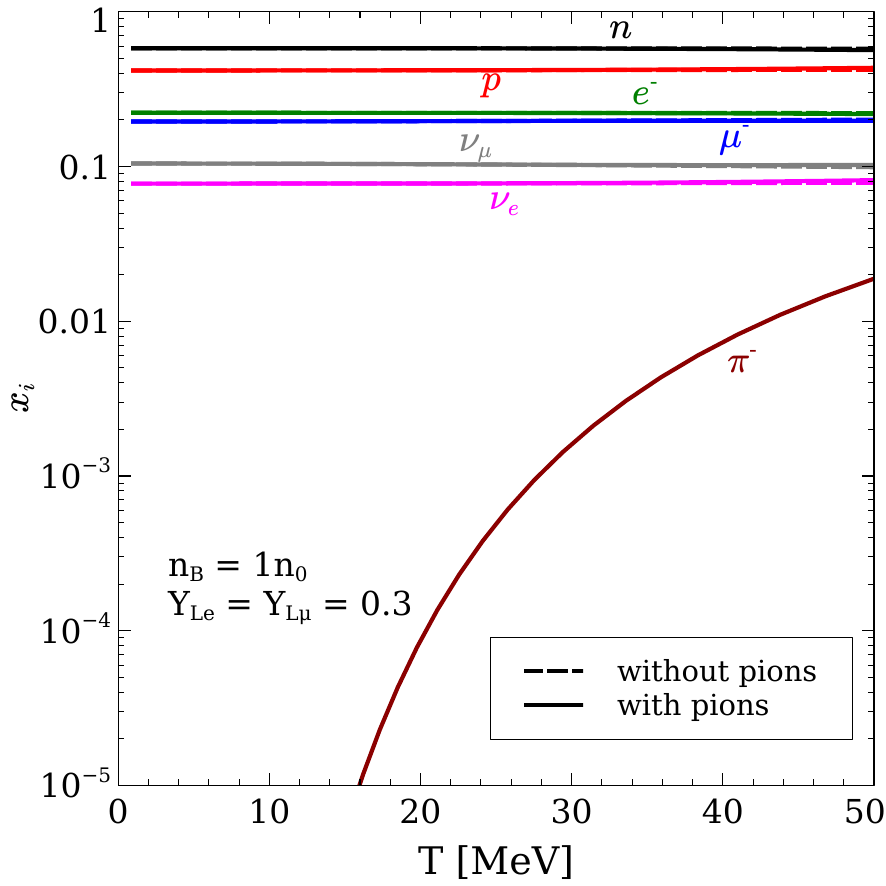}
\includegraphics[width=0.4\textwidth]{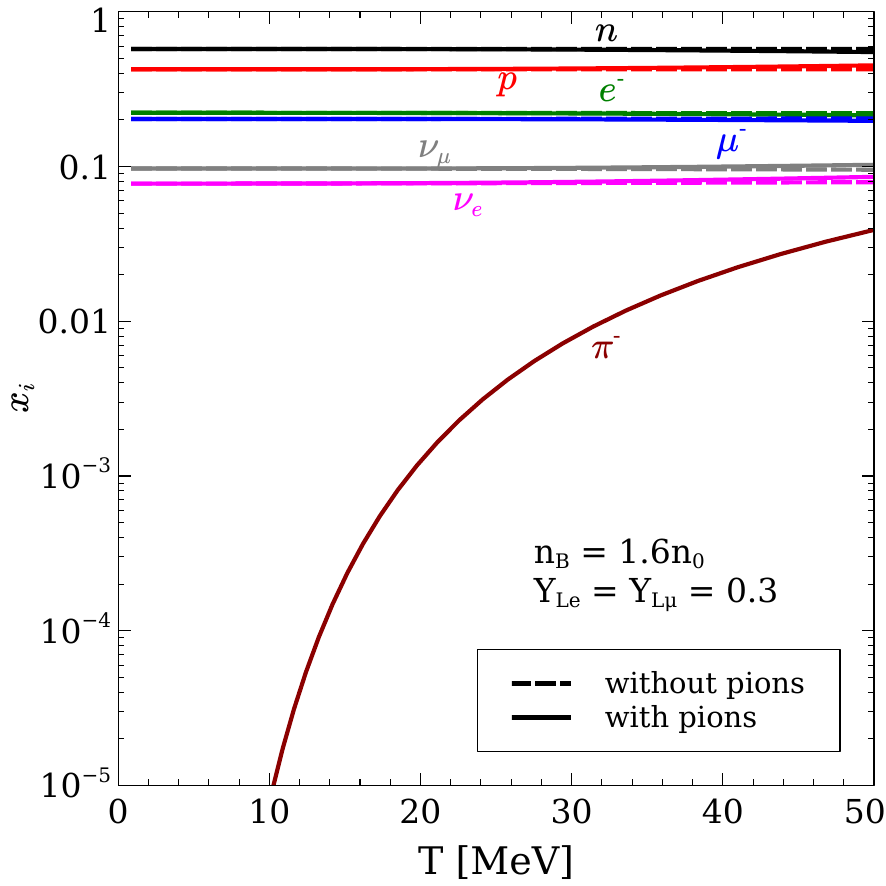}
\caption{Particle content in beta equilibrium in various thermodynamic conditions expected to exist in neutron star mergers (top two panels) and in supernovae (bottom two panels), but with equal conserved lepton fractions.  Same conventions as Fig.~\ref{fig:xi}.}
\label{fig:xi_appendix}
\end{figure*}

The particle fractions in matter with net conserved muon number are plotted in Fig.~\ref{fig:xi_appendix}.  When $Y_{Le}=Y_{L\mu}=0.05$, the electron and muon populations are roughly equal.  At $n_0$, the proton fraction is 10\% instead of around 5\% (in the $Y_{L\mu}=0$ case).  When pions are added to the system, their population exponentially rises with temperature, becoming appreciable at $T\approx 20\text{ MeV}$.  When the pion population becomes appreciable, the proton fraction is raised and the lepton fractions are slightly lowered.  At $1.6n_0$, the pions content becomes appreciable at very low temperatures, just a couple MeV.  Thus the proton fraction is larger than without pions for the basically the entire temperature range plotted.  

In the high lepton number case ($Y_{Le}=Y_{L\mu}=0.3$) potentially relevant for supernovae, the muon content is much closer to the electron density, but the pion population is still strongly suppressed, and does not affect the other particle fractions substantially.  

\begin{figure*}\centering
\includegraphics[width=0.4\textwidth]{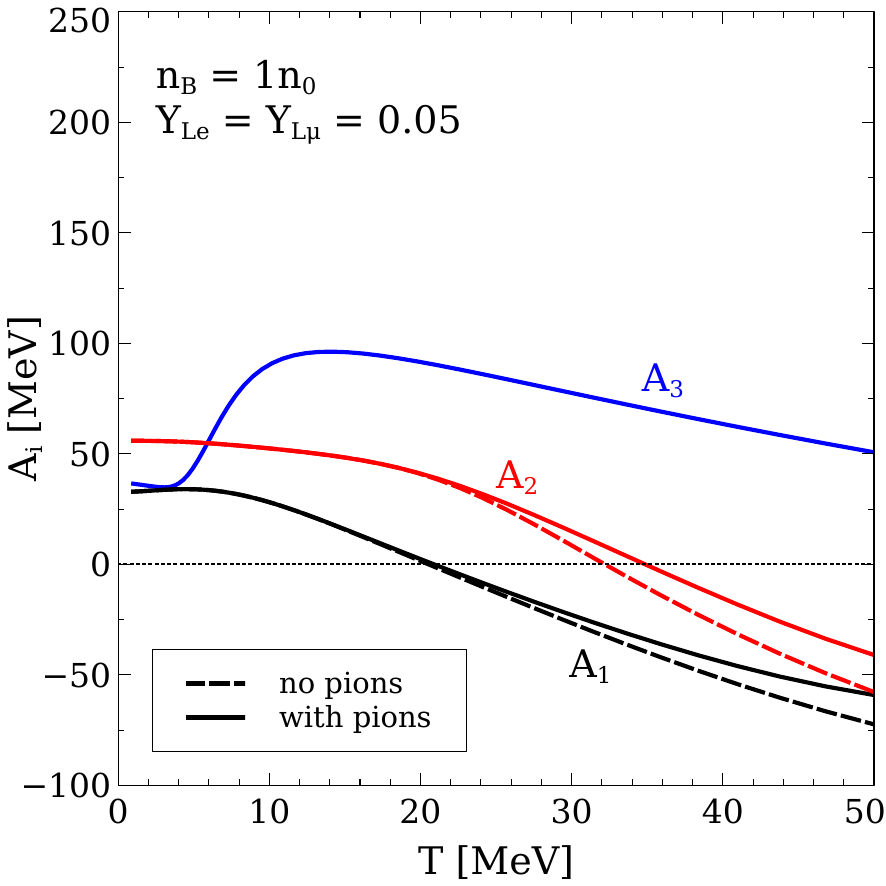} 
\includegraphics[width=0.4\textwidth]{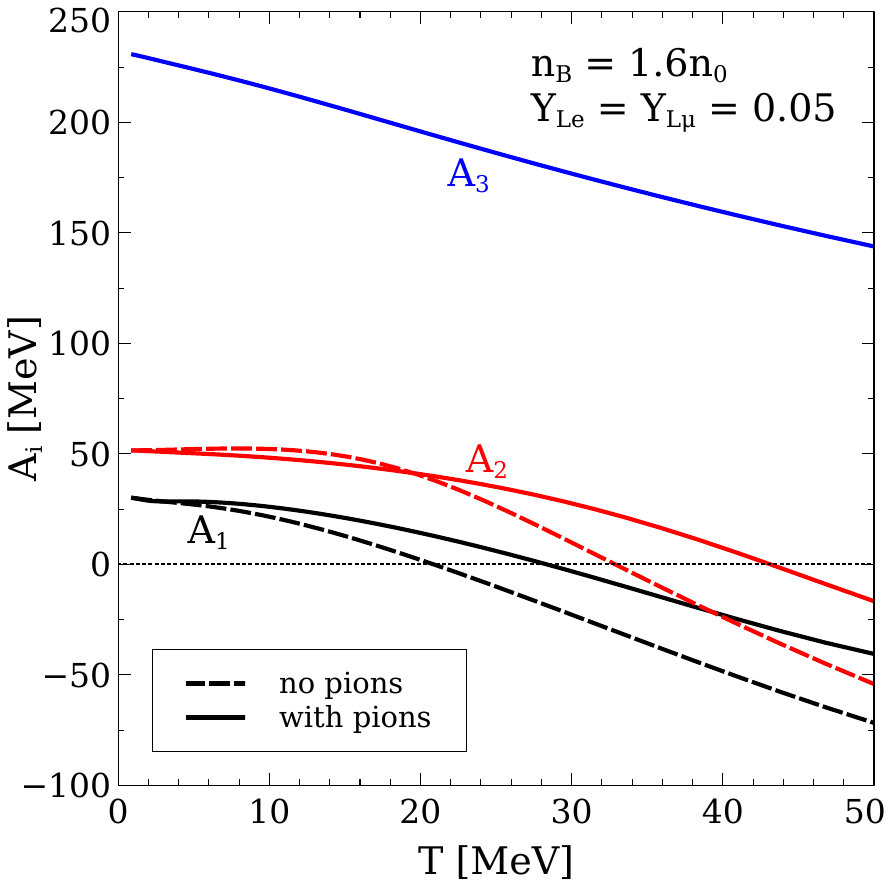}\\
\includegraphics[width=0.4\textwidth]{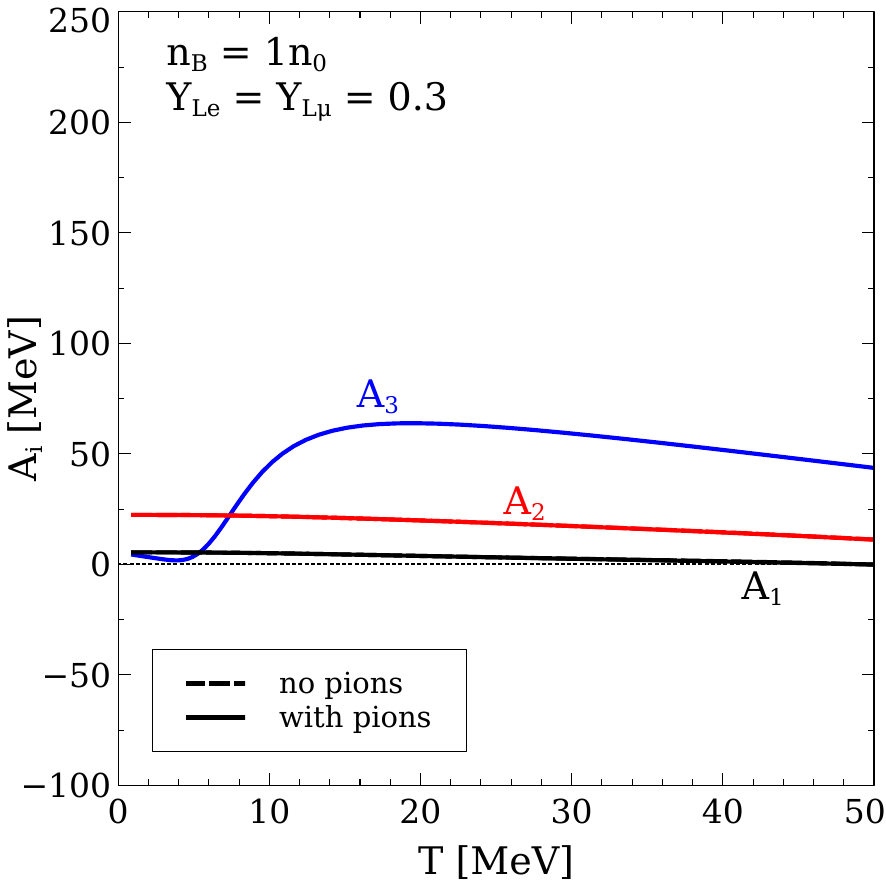} 
\includegraphics[width=0.4\textwidth]{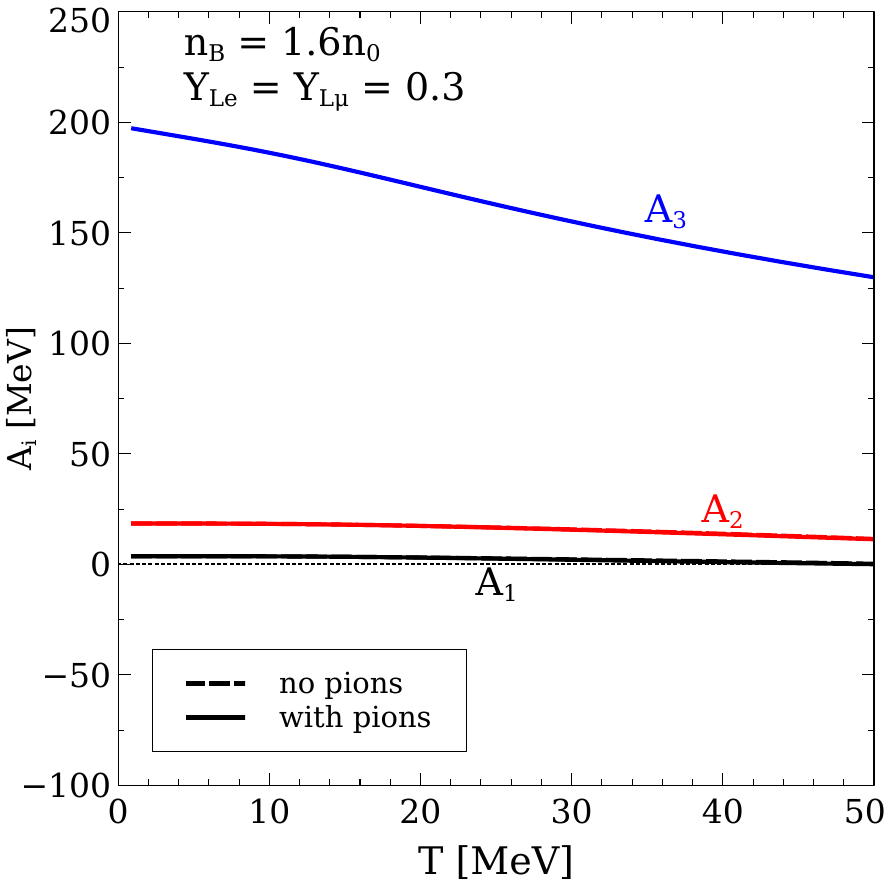}
\caption{Susceptibilities of the $A$ type from the EoS described in Sec.~\ref{sec:EoS} plotted in merger conditions (top panels) and in supernovae conditions (bottom panels), but with equal conserved lepton fractions.  The dashed lines correspond to an EoS without pions, while the solid lines indicate an EoS with pions.  $A_3$ has no counterpart in a system without pions.}
\label{fig:A_susc_appendix}
\end{figure*}

\begin{figure*}\centering
\includegraphics[width=0.4\textwidth]{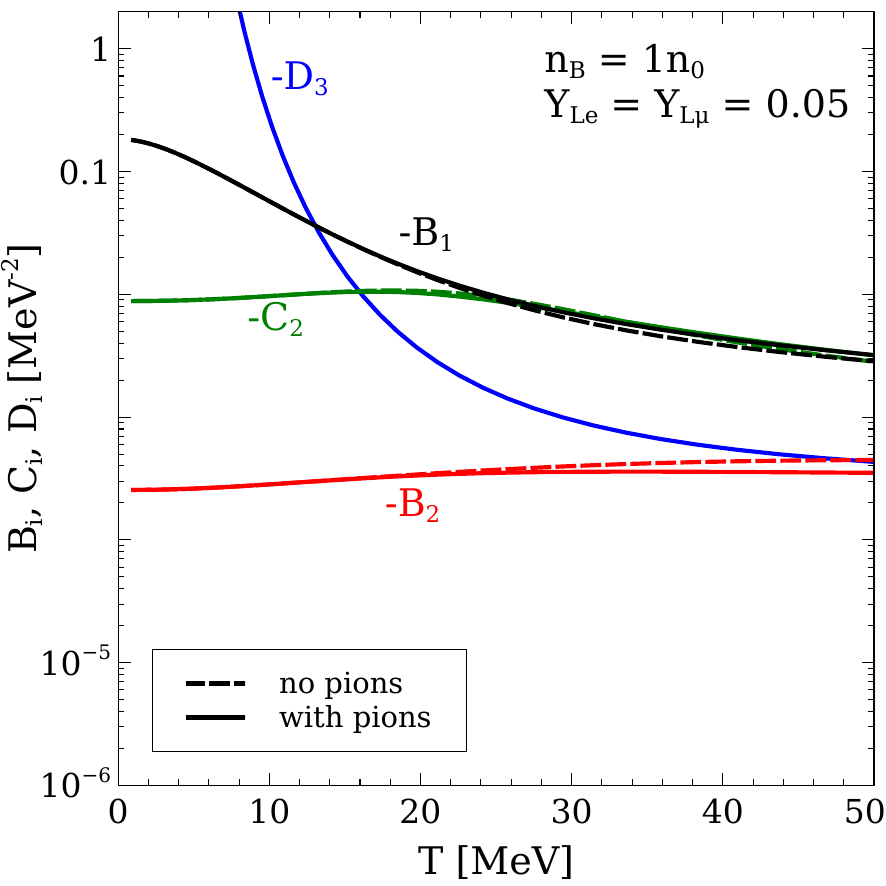}
\includegraphics[width=0.4\textwidth]{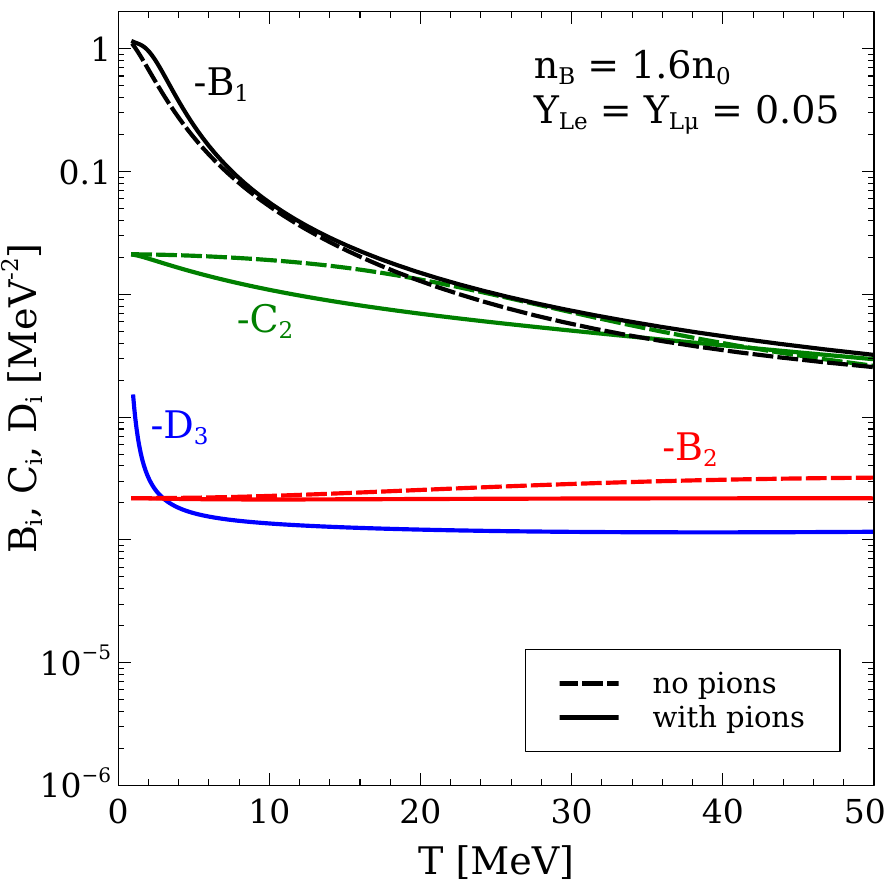}\\
\includegraphics[width=0.4\textwidth]{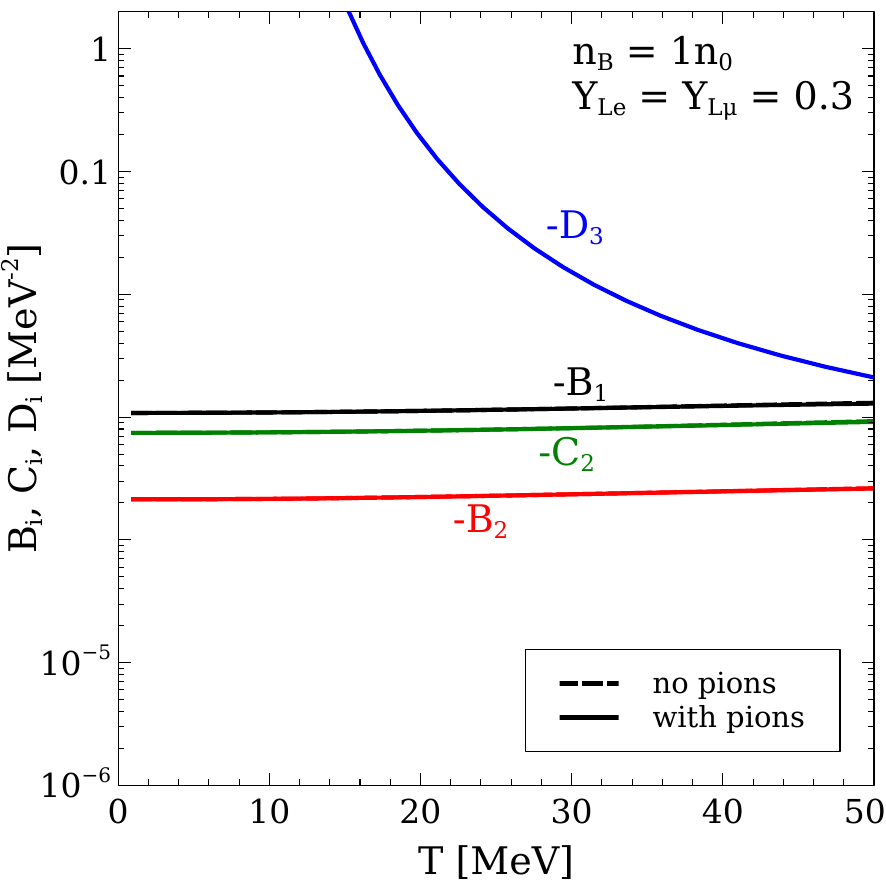}
\includegraphics[width=0.4\textwidth]{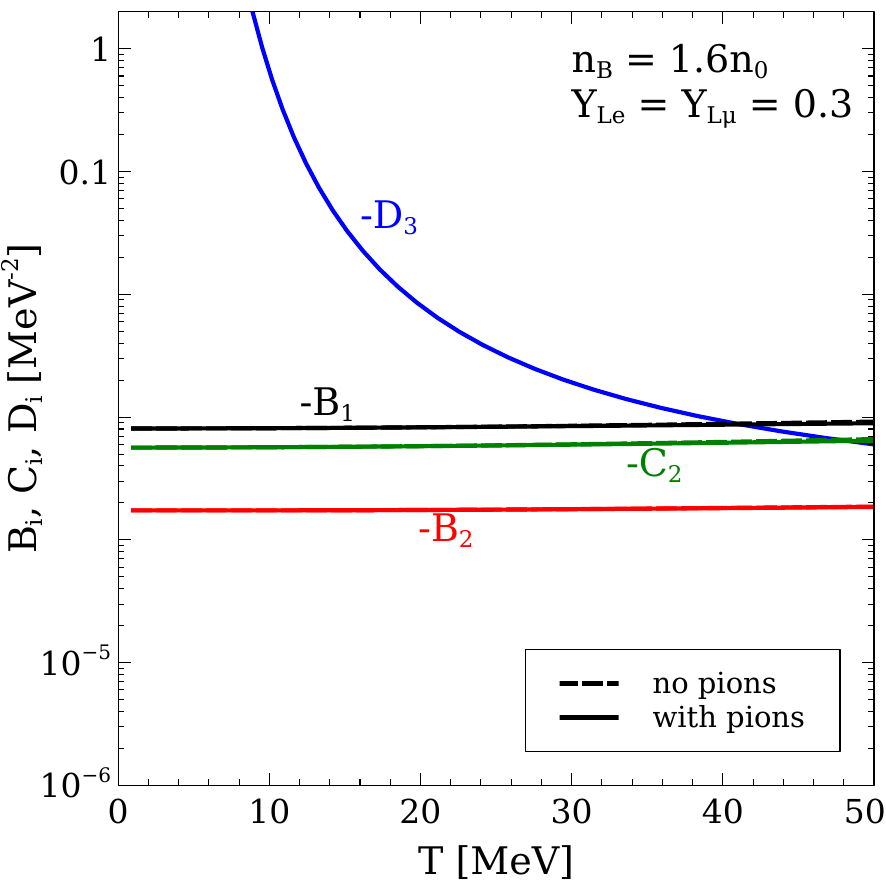}
\caption{Susceptibilities of the $B$, $C$, and $D$ types from the EoS described in Sec.~\ref{sec:EoS} plotted in merger conditions (top panels) and in supernovae conditions (bottom panels), but with equal conserved lepton fractions.  The dashed lines correspond to an EoS without pions, while the solid lines indicate an EoS with pions.  $D_3$ has no counterpart in a system without pions.}
\label{fig:BCD_susc_appendix}
\end{figure*}

The susceptibilities $A_i$, plotted in Fig.~\ref{fig:A_susc_appendix}, are essentially the same as the ones plotted in the main text (Fig.~\ref{fig:A_susc}).  However, $A_2$ does not cross zero in the displayed temperature range when $Y_{Le}=Y_{L\mu}=0.3$, but it does in the more physical realistic scenario $Y_{Le}=0.3, Y_{L\mu}=0$.

The susceptibilities $B_i$, $C_i$, and $D_i$ are plotted in Fig.~\ref{fig:BCD_susc_appendix}.  They are relatively similar to their counterparts in $Y_{L\mu}=0$ matter (Fig.~\ref{fig:BCD_susc}).  

\begin{figure*}\centering
\includegraphics[width=0.4\textwidth]{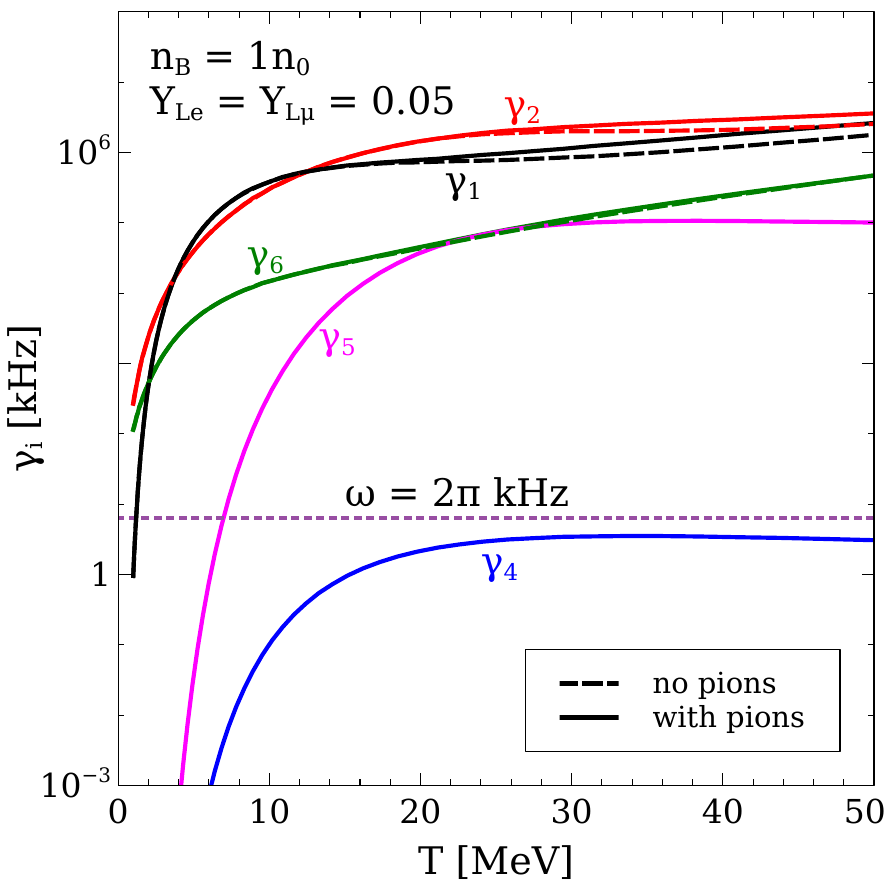} 
\includegraphics[width=0.4\textwidth]{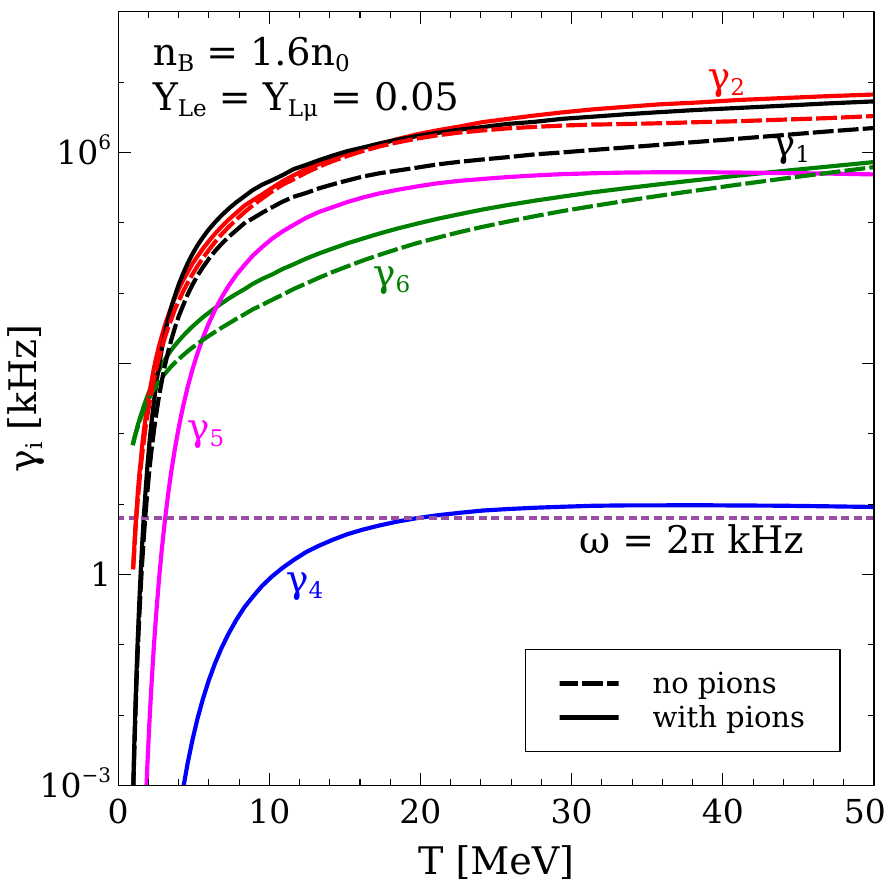}\\
\includegraphics[width=0.4\textwidth]{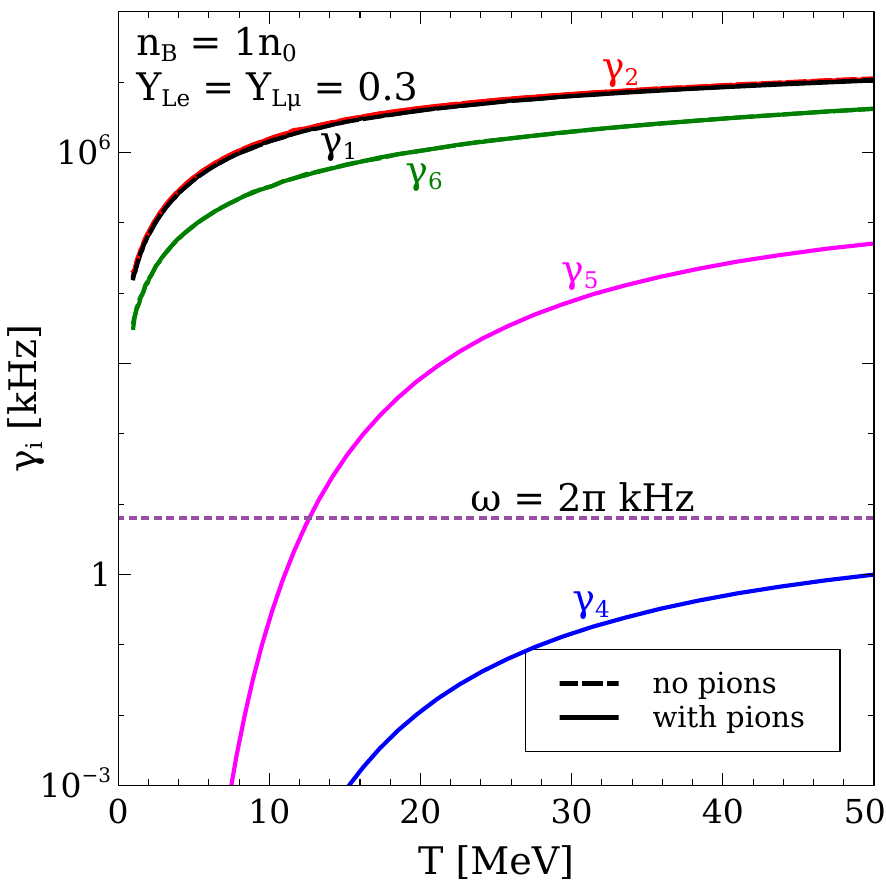}
\includegraphics[width=0.4\textwidth]{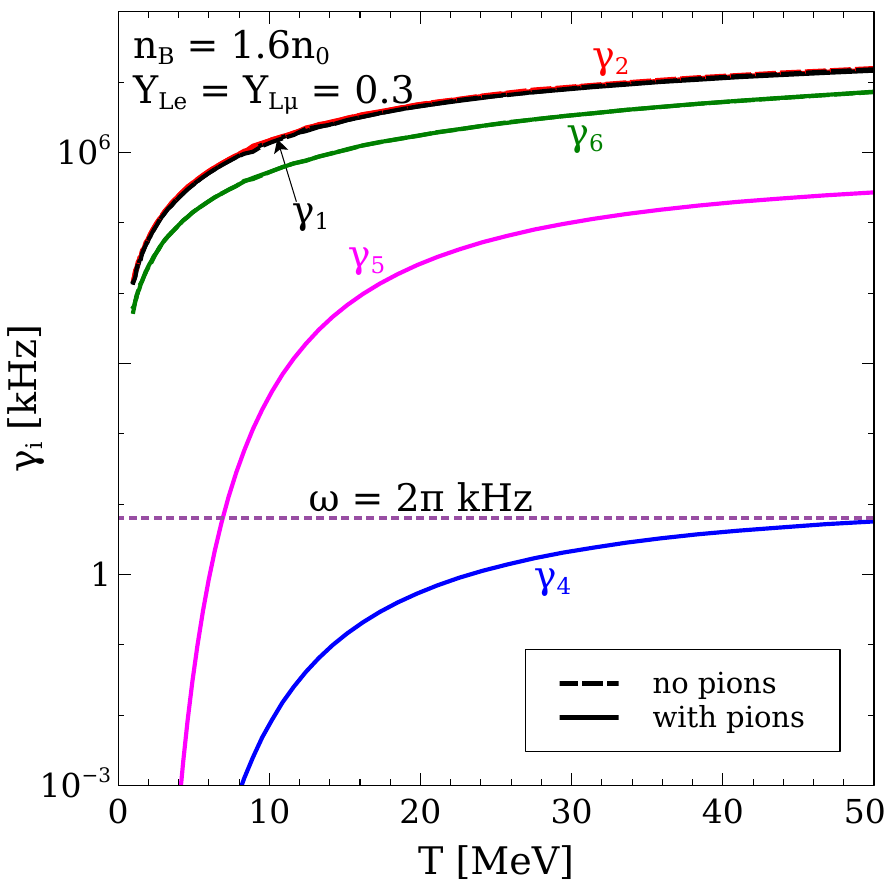}
\caption{Individual beta equilibration rates $\gamma_i^{\text{no }\pi}$ and $\gamma_i^{\lambda_3\rightarrow\infty}$ in neutron star merger conditions (top panels) and in supernovae conditions (bottom panels), but with equal conserved lepton fractions.  The dashed lines correspond to an EoS without pions, while the solid lines indicate an EoS with pions.  The rates $\gamma_4^{\lambda_3\rightarrow\infty}$ and $\gamma_5^{\lambda_3\rightarrow\infty}$ of course have no counterpart in matter without pions.  }
\label{fig:gamma_appendix}
\end{figure*}
\begin{figure*}\centering
\includegraphics[width=0.4\textwidth]{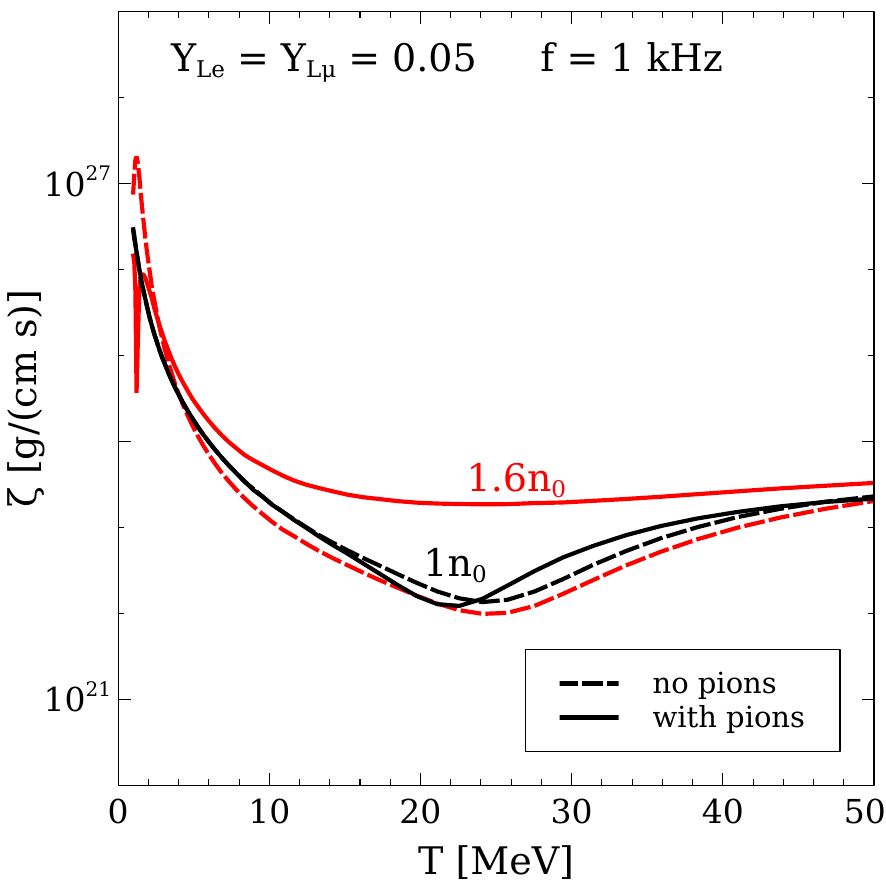}
\includegraphics[width=0.4\textwidth]{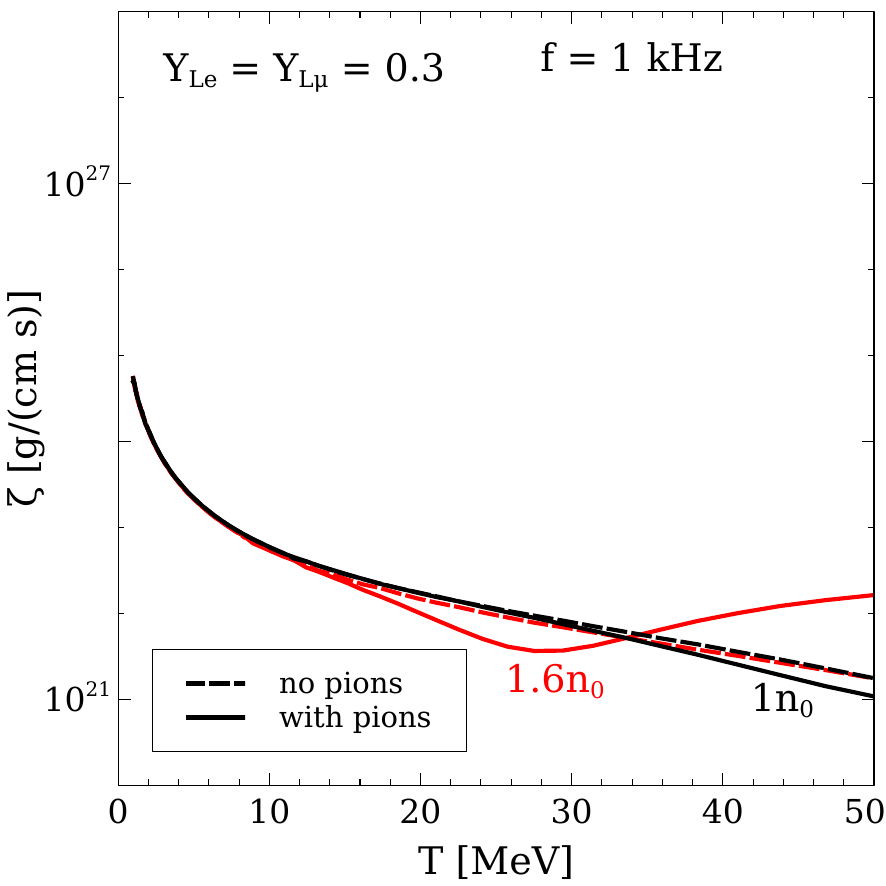}
\caption{Bulk viscosity in dense matter with and without pions, for a harmonic, small-amplitude, density oscillation with frequency 1 kHz.}
\label{fig:bulk_visc_appendix}
\end{figure*}

The equilibration rates $\gamma_i$ are plotted in Fig.~\ref{fig:gamma_appendix}.  In merger conditions, the rates are much faster when $Y_{L\mu}=0.05$ than when $Y_{L\mu}=0$. 
 Adding pions to the EoS in these equal lepton fraction conditions modifies the beta equilibration rates slightly, especially as temperature rises.  In supernovae conditions, the rates are hardly modified at all, because the pion content is very low.

The bulk viscosity in this equal-lepton-fraction matter is plotted in Fig.~\ref{fig:bulk_visc_appendix}.  In matter with $Y_{Le}=Y_{L\mu}=0.05$, the bulk viscosity is always on the ``downhill side'' of the resonance at $n_0$ (that is, all equilibration channels are equilibrated faster than the millisecond oscillation timescale).  This can be seen from the top left panel of Fig.~\ref{fig:gamma_appendix} as well.  Adding pions to the EoS at $1n_0$ merely shifts the dip in the bulk viscosity at $T\approx 25$ MeV.  At $1.6n_0$, the rates are actually a bit slower, and, without pions, the rate $\gamma_2$ passes through resonance, creating a peak in the bulk viscosity at $T\approx 1.2 \text{ MeV}$.  With pions, the infinitely fast $\gamma_3$ rate leads to a dip in $\zeta_2^{\gamma_3\rightarrow\infty}$, causing the sharp drop in the bulk viscosity at $T$ just above 1 MeV.  The pions smooth out the dip in the bulk viscosity at higher temperatures too.  

In matter with $Y_{Le}=Y_{L\mu}=0.3$, again all of the particle fractions are equilibrated quickly compared to millisecond timescales.  The bulk viscosity is, even at $T=1\text{ MeV}$, quite small and decreases further with temperature.  At $1.6n_0$, the quickly equilibrating pions cause a dip in the bulk viscosity.  
\begin{widetext}
\section{Weak interaction rate phase space integrals}\label{sec:rates}
In this section, we write down the expressions for the phase space integrals for the rate calculations in this paper and give a few details, but mostly we refer the reader to previous works which give a complete description of the calculations.
\subsection{Direct Urca (with electrons)}
The rate (per volume) of the neutron decay process $n\rightarrow p + e^- + \bar{\nu}_e$ is 
\begin{equation}
    \Gamma = \int \dfrac{\mathop{d^3p_n}}{(2\pi)^3}\dfrac{\mathop{d^3p_p}}{(2\pi)^3}\dfrac{\mathop{d^3p_e}}{(2\pi)^3}\dfrac{\mathop{d^3p_{\bar{\nu}_e}}}{(2\pi)^3}(2\pi)^4\delta^4(p_n-p_p-p_e-p_{\bar{\nu}_e})\dfrac{\sum_{\text{spins}}\vert\mathcal{M}\vert^2}{2^4E_nE_pE_eE_{\bar{\nu}_e}}f_n(1-f_p)(1-f_e)(1-f_{\bar{\nu}_e}),
\end{equation}
where $f$ denotes the Fermi-Dirac distribution.  The matrix element is given by 
\begin{equation}
    \sum_{\text{spins}}\vert \mathcal{M}\vert^2 = 32G_F^2\cos^2{\theta_c}\left[(1+g_A)^2(p_p\cdot p_e)(p_n\cdot p_{\nu})+(g_A^2-1)m^2(p_e\cdot p_{\nu})+(g_A-1)^2(p_n\cdot p_e)(p_p\cdot p_{\nu})\right],\label{eq:durca_matrix_element}
\end{equation}
but we will take the nucleons to be nonrelativistic and eliminate the (small) term that is proportional to $(1-g_A^2)$, which reduces the matrix element down to the momentum-independent quantity (see Appendix C in \cite{Harris:2020rus})
\begin{equation}
    \dfrac{\sum_{\text{spins}}\vert\mathcal{M}\vert^2}{2^4E_nE_pE_eE_{\nu}} = 2G_F^2\cos^2{\theta_c}(1+3g_A^2).\label{eq:durca_matrix_element_NR}
\end{equation}
The phase space integral reduces from 12 to 8 dimensions after integrating over the delta functions.  Then a spherical coordinate system can be chosen in such a way to render three additional angular integrals trivial, leaving a 5-dimensional integral.  Two more integrals can be done analytically, leaving a 3-dimensional integral to do numerically.  This integral has been done with the full matrix element (Eq.~\ref{eq:durca_matrix_element}) with $g_A=1$ in \cite{Alford:2021lpp}, with the non-relativistic matrix element \ref{eq:durca_matrix_element_NR} in \cite{Alford:2019kdw}, and in the neutrino-transparent case in \cite{Alford:2021ogv,Alford:2018lhf}.

The rate of the inverse electron capture process $n + \nu_e\rightarrow e^- + p$ is 
\begin{equation}
    \Gamma = \int \dfrac{\mathop{d^3p_n}}{(2\pi)^3}\dfrac{\mathop{d^3p_p}}{(2\pi)^3}\dfrac{\mathop{d^3p_e}}{(2\pi)^3}\dfrac{\mathop{d^3p_{\nu_e}}}{(2\pi)^3}(2\pi)^4\delta^4(p_n-p_p-p_e+p_{\nu_e})\dfrac{\sum_{\text{spins}}\vert\mathcal{M}\vert^2}{2^4E_nE_pE_eE_{\nu_e}}f_n(1-f_p)(1-f_e)f_{\nu_e}.
\end{equation}
The matrix element for $n + \nu_e\rightarrow e^- + p$ is also given by Eq.~\ref{eq:durca_matrix_element_NR} (or Eq.~\ref{eq:durca_matrix_element}) due to crossing symmetry, and the phase space integration is done almost identically to the neutron decay case.

The direct Urca processes involving a muon are obtained by replacing the electron mass and chemical potential with the muon mass and chemical potential.  While electrons can be treated as ultrarelativistic particles (though we do not assume this in our calculation), muons cannot be.  The electron neutrino chemical potential is also changed to the muon neutrino chemical potential.  

\subsection{Pion decay to electron}
The rate of the pion decay process $\pi^-\rightarrow e^- + \bar{\nu}_e$ is 
\begin{equation}
    \Gamma = \int \dfrac{\mathop{d^3p_{\pi}}}{(2\pi)^3}\dfrac{\mathop{d^3p_e}}{(2\pi)^3}\dfrac{\mathop{d^3p_{\bar{\nu}_e}}}{(2\pi)^3}(2\pi)^4\delta^4(p_{\pi}-p_e-p_{\bar{\nu}_e})\dfrac{\sum_{\text{spins}}\vert\mathcal{M}\vert^2}{2^3E_{\pi}E_eE_{\bar{\nu}_e}}g_{\pi}(1-f_e)(1-f_{\bar{\nu}_e}), \label{eq:pi_e_nu_rate}
\end{equation}
where $g$ denotes the Bose-Einstein distribution.  The matrix element is given by 
\begin{align}
    \sum_{\text{spins}}\vert \mathcal{M}\vert^2 &= 4f_{\pi}^2G_F^2\left[2(p_{\pi}\cdot p_{\nu})(p_{\pi}\cdot p_e)-(p_{\pi}\cdot p_{\pi})(p_e\cdot p_{\nu})\right]\label{eq:pi_e_nu_matrix_element}\\
    &= 4f_{\pi}^2G_F^2m_e^2(p_e\cdot p_{\nu}),\nonumber
\end{align}
where the second line was obtained by using 4-momentum conservation.  The four-vector dot product $p_{\pi}\cdot p_{\pi}\neq m_{\pi}^2$ due to the pion self-energy.  

The rate of the process $\pi^- \rightarrow \mu^- + \bar{\nu}_{\mu}$ can be obtained by taking the rate integral \ref{eq:pi_e_nu_rate} with matrix element \ref{eq:pi_e_nu_matrix_element} and replacing the electron mass and chemical potential with those of the muon, as well as replacing the electron neutrino chemical potential with the muon neutrino chemical potential.
\subsection{Electron-muon conversion}
The rate of $\mu^-\rightarrow e^- + \bar{\nu}_e+\nu_{\mu}$ is
\begin{equation}
    \Gamma = \int \dfrac{\mathop{d^3p_{\mu}}}{(2\pi)^3}\dfrac{\mathop{d^3p_e}}{(2\pi)^3}\dfrac{\mathop{d^3p_{\bar{\nu}_e}}}{(2\pi)^3}\dfrac{\mathop{d^3p_{\nu_{\mu}}}}{(2\pi)^3}(2\pi)^4\delta^4(p_{\mu}-p_e-p_{\bar{\nu}_e}-p_{\nu_{\mu}})\dfrac{\sum_{\text{spins}}\vert\mathcal{M}\vert^2}{2^4E_{\mu}E_eE_{\bar{\nu}_e}E_{\nu_{\mu}}}f_{\mu}(1-f_e)(1-f_{\bar{\nu}_e})(1-f_{\nu_{\mu}}).
\end{equation}
The matrix element is given by 
\begin{equation}
    \sum_{\text{spins}}\vert \mathcal{M}\vert^2 = 128 G_F^2(p_{\mu}\cdot p_{\nu_e})(p_{\nu_{\mu}}\cdot p_e).\label{eq:muon_decay_matrix_element}
\end{equation}
The rate of $\mu^-+\bar{\nu}_{\mu}\rightarrow e^- + \bar{\nu}_e$ is
\begin{equation}
    \Gamma = \int \dfrac{\mathop{d^3p_{\mu}}}{(2\pi)^3}\dfrac{\mathop{d^3p_e}}{(2\pi)^3}\dfrac{\mathop{d^3p_{\bar{\nu}_e}}}{(2\pi)^3}\dfrac{\mathop{d^3p_{\bar{\nu}_{\mu}}}}{(2\pi)^3}(2\pi)^4\delta^4(p_{\mu}-p_e-p_{\bar{\nu}_e}-p_{\bar{\nu}_{\mu}})\dfrac{\sum_{\text{spins}}\vert\mathcal{M}\vert^2}{2^4E_{\mu}E_eE_{\bar{\nu}_e}E_{\bar{\nu}_{\mu}}}f_{\mu}(1-f_e)(1-f_{\bar{\nu}_e})(1-f_{\bar{\nu}_{\mu}}).
\end{equation}
The rate of $\mu^-+\nu_e \rightarrow e^- + \nu_{\mu}$ is 
\begin{equation}
    \Gamma = \int \dfrac{\mathop{d^3p_{\mu}}}{(2\pi)^3}\dfrac{\mathop{d^3p_e}}{(2\pi)^3}\dfrac{\mathop{d^3p_{\nu_e}}}{(2\pi)^3}\dfrac{\mathop{d^3p_{\nu_{\mu}}}}{(2\pi)^3}(2\pi)^4\delta^4(p_{\mu}-p_e-p_{\nu_e}-p_{\nu_{\mu}})\dfrac{\sum_{\text{spins}}\vert\mathcal{M}\vert^2}{2^4E_{\mu}E_eE_{\nu_e}E_{\nu_{\mu}}}f_{\mu}(1-f_e)(1-f_{\nu_e})(1-f_{\nu_{\mu}}).
\end{equation}
The rate of $\mu^- + \nu_e + \bar{\nu}_{\mu}\rightarrow e^-$ is given by
\begin{equation}
    \Gamma = \int \dfrac{\mathop{d^3p_{\mu}}}{(2\pi)^3}\dfrac{\mathop{d^3p_e}}{(2\pi)^3}\dfrac{\mathop{d^3p_{\nu_e}}}{(2\pi)^3}\dfrac{\mathop{d^3p_{\bar{\nu}_{\mu}}}}{(2\pi)^3}(2\pi)^4\delta^4(p_{\mu}-p_e-p_{\nu_e}-p_{\bar{\nu}_{\mu}})\dfrac{\sum_{\text{spins}}\vert\mathcal{M}\vert^2}{2^4E_{\mu}E_eE_{\nu_e}E_{\bar{\nu}_{\mu}}}f_{\mu}(1-f_e)(1-f_{\nu_e})(1-f_{\bar{\nu}_{\mu}}).
\end{equation}
All four reactions in this section have the same matrix element (Eq.~\ref{eq:muon_decay_matrix_element}) due to crossing symmetry.  The phase space integration proceeds the same way as for the Urca processes, and is (briefly) described in \cite{Alford:2021lpp}.
\section{Maxwell Relations}\label{sec:maxwell}
The first law of thermodynamics in neutrino-trapped $npe\mu\pi^-$ matter can be written
\begin{equation}
    \mathop{dE} = -P\mathop{dV}+T\mathop{dS}+\mu_n\mathop{dN_n}+\mu_p\mathop{dN_p}+\mu_e\mathop{dN_e} +\mu_{\mu}\mathop{dN_{\mu}}+\mu_{\nu_e}\mathop{dN_{\nu_e}} + \mu_{\nu_{\mu}}\mathop{dN_{\nu_{\mu}}}+ \mu_{\pi}\mathop{dN_{\pi}}.
\end{equation}
We normalize all quantities by baryon number $N_B$
\begin{equation}
    \mathop{d\left(\dfrac{\varepsilon}{n_B}\right)} = \dfrac{P}{n_B^2}\mathop{dn_B} + T\mathop{d\left(\dfrac{s}{n_B}\right)}+\mu_n\mathop{dx_n}+\mu_p\mathop{dx_p}+\mu_e\mathop{dx_e}+\mu_{\mu}\mathop{dx_{\mu}}+\mu_{\nu_e}\mathop{dx_{\nu_e}}+\mu_{\nu_{\mu}}\mathop{dx_{\nu_{\mu}}}+\mu_{\pi}\mathop{dx_{\pi}}.
\end{equation}
We know from the allowed chemical reactions that $\mathop{dx_n} = -\mathop{dx_p}$, $\mathop{dx_e}=\mathop{dx_p}-\mathop{dx_{\mu}}-\mathop{dx_{\pi}}$, $\mathop{dx_{\nu_e}} = \mathop{dx_{\pi}}+\mathop{dx_{\mu}}-\mathop{dx_p}$, and $\mathop{dx_{\nu_{\mu}}}=-\mathop{dx_{\mu}}$ and so all particle fractions can be expressed in terms of the proton, muon, and pion fractions.  We Legendre transform the temperature/entropy term \cite{swendsen2020introduction}, getting the final expression
\begin{equation}
    \mathop{d\left(\dfrac{\varepsilon-sT}{n_B}\right)} = \dfrac{P}{n_B^2}\mathop{dn_B}-\dfrac{s}{n_B}\mathop{dT}-\delta\mu_1\mathop{dx_p}+(\delta\mu_1-\delta\mu_2)\mathop{dx_{\mu}}+(\delta\mu_1-\delta\mu_3)\mathop{dx_{\pi}}.
\end{equation}
From this 1st law of thermodynamics, we can derive six Maxwell relations (if we only consider derivatives with respect to $n_B$ or $x_i$) \cite{swendsen2020introduction}.  The three that are relevant are written in Eqs.~\ref{eq:maxwell1}, \ref{eq:maxwell2}, and \ref{eq:maxwell3}.

\section{Full expression for bulk viscosity}\label{sec:full_expression_for_bulk_viscosity}
In this section, we give the definitions of $F,G,H,J,K,L$ in the expression for the bulk viscosity (Eq.~\ref{eq:full_bulk_viscosity}).  We keep all of the susceptibilities in these expressions, even though three are redundant (due to Maxwell relations) and $D_2 = C_3 = 0$ for our EoS.  Even if we were to take advantage of these simplifications, the resultant expression for the bulk viscosity would still be quite complicated.

First, we define the variables
\begin{subequations}
\begin{align}
    a &\equiv A_1\lambda_1+A_2\lambda_2+A_3\lambda_3\\
    b &\equiv B_1\lambda_1 + B_2\lambda_2 + B_3\lambda_3\\
    c &\equiv C_1\lambda_1 + C_2\lambda_2 + C_3\lambda_3\\
    d &\equiv D_1\lambda_1 + D_2\lambda_2 + D_3\lambda_3\\
    e &\equiv A_1 \lambda_6 - A_2(\lambda_2+\lambda_5+\lambda_6)+A_3\lambda_5\\
    f &\equiv B_1 \lambda_6 - B_2(\lambda_2+\lambda_5+\lambda_6)+B_3\lambda_5\\
    g &\equiv C_1 \lambda_6 - C_2(\lambda_2+\lambda_5+\lambda_6)+C_3\lambda_5\\
    h &\equiv D_1 \lambda_6 - D_2(\lambda_2+\lambda_5+\lambda_6)+D_3\lambda_5\\
    i &\equiv A_1\lambda_4+A_2\lambda_5-A_3(\lambda_3+\lambda_4+\lambda_5)\\
    j &\equiv B_1\lambda_4+B_2\lambda_5-B_3(\lambda_3+\lambda_4+\lambda_5)\\
    k &\equiv C_1\lambda_4+C_2\lambda_5-C_3(\lambda_3+\lambda_4+\lambda_5)\\
    l &\equiv D_1\lambda_4+D_2\lambda_5-D_3(\lambda_3+\lambda_4+\lambda_5).
\end{align}
\end{subequations}
Then,
\begin{subequations}
\begin{align}
    F &= (A_1-A_3)c^2f^2i - (A_1-A_2)cdf^2i-2(A_1-A_3)bcfgi+(A_1-A_2)bdfgi-A_1cdfgi+(A_1-A_3)b^2g^2i\\
    &+A_1bdg^2i+(A_1-A_2)bcfhi+A_1c^2fhi-(A_1-A_2)b^2ghi-A_1bcghi-(A_1-A_3)c^2efj+(A_1-A_2)cdefj\nonumber\\
    &+(A_1-A_3)bcegj+A_1cdegj+(A_1-A_3)acfgj-(A_1-A_2)adfgj-(A_1-A_3)abg^2j-A_1adg^2j\nonumber\\
    &-(A_1-A_2)bcehj-A_1c^2ehj+(A_1-A_2)abghj+A_1acghj+(A_1-A_3)cdfij-(A_1-A_2)d^2fij\nonumber\\
    &-(A_1-A_3)dg^2ij-(A_1-A_3)bchij+(A_1-A_2)bdhij+(A_1-A_3)cghij+(A_1-A_2)dghij-(A_1-A_2)ch^2ij\nonumber\\
    &-(A_1-A_3)cdej^2+(A_1-A_2)d^2ej^2+(A_1-A_3)achj^2-(A_1-A_2)adhj^2+(A_1-A_3)bcefk-(A_1-A_2)bdefk\nonumber\\
    &-(A_1-A_3)acf^2k+(A_1-A_2)adf^2k-(A_1-A_3)b^2egk-A_1bdegk+(A_1-A_3)abfgk+A_1adfgk\nonumber\\
    &+(A_1-A_2)b^2ehk+A_1bcehk-(A_1-A_2)abfhk-A_1acfhk-(A_1-A_3)bdfik-A_1d^2fik+(A_1-A_3)dfgik\nonumber\\
    &+(A_1-A_3)b^2hik+A_1bdhik-(A_1-A_3)cfhik+A_1dghik-A_1ch^2ik+(A_1-A_3)bdejk+A_1d^2ejk\nonumber\\
    &+(A_1-A_3)degjk-(A_1-A_3)abhjk-A_1adhjk-(A_1-A_2)dehjk-(A_1-A_3)aghjk+(A_1-A_2)ah^2jk\nonumber\\
    &-(A_1-A_3)defk^2-A_1dehk^2+(A_1-A_3)afhk^2+A_1ah^2k^2+(A_1-A_2)bdfil+A_1cdfil-(A_1-A_2)dfgil\nonumber\\
    &-A_1dg^2il-(A_1-A_2)b^2hil-A_1bchil+(A_1-A_2)cfhil+A_1cghil+(A_1-A_3)bcejl-2(A_1-A_2)bdejl\nonumber\\
    &-A_1cdejl-(A_1-A_3)acfjl+(A_1-A_2)adfjl-(A_1-A_3)cegjl+(A_1-A_3)ag^2jl+(A_1-A_2)abhjl\nonumber\\
    &+A_1achjl+(A_1-A_2)cehjl-(A_1-A_2)aghjl-(A_1-A_3)b^2ekl-A_1bdekl+(A_1-A_3)abfkl+A_1adfkl\nonumber\\
    &+(A_1-A_3)cefkl+(A_1-A_2)defkl+A_1degkl-(A_1-A_3)afgkl+A_1cehkl-(A_1-A_2)afhkl-2A_1aghkl\nonumber\\
    &+(A_1-A_2)b^2el^2+A_1bcel^2-(A_1-A_2)abfl^2-A_1acfl^2-(A_1-A_2)cefl^2-A_1cegl^2+(A_1-A_2)afgl^2\nonumber\\
    &+A_1ag^2l^2\nonumber\\
    G &= (A_1-A_2)b^2e+A_1bce-(A_1-A_2)abf-A_1acf-(A_1-A_2)cef-A_1ceg+(A_1-A_2)afg+A_1ag^2\\
    &+(A_1-A_3)b^2i+A_1bdi-2(A_1-A_3)cfi+(A_1-A_2)dfi+(A_1-A_3)g^2i-A_1chi-(A_1-A_2)ghi\nonumber\\
    &-(A_1-A_3)abj-A_1adj+(A_1-A_3)cej-2(A_1-A_2)dej+(A_1-A_2)ahj-(A_1-A_3)dij-A_1dek\nonumber\\
    &+(A_1-A_3)afk-(A_1-A_3)egk+2A_1ahk+(A_1-A_2)ehk+(A_1-A_3)hik-A_1dil-(A_1-A_2)hil\nonumber\\
    &+(A_1-A_3)ajl-(A_1-A_3)ekl+A_1al^2+(A_1-A_2)el^2\nonumber\\
    H &= A_1a+(A_1-A_2)e+(A_1-A_3)i\\
    J &= (dgj-chj-dfk+bhk+cfl-bgl)^2\\
    K &= c^2f^2-2bcfg+b^2g^2+2cdfj-2dg^2j-2bchj+2cghj+d^2j^2-2bdfk+2dfgk+2b^2hk-2cfhk\\
    &-2dhjk+h^2k^2-2bdjl+2chjl+2dfkl-2ghkl+b^2l^2-2cfl^2+g^2l^2\nonumber\\
    L &= b^2-2cf+g^2-2dj+2hk+l^2.
\end{align}
\end{subequations}
\section{Bulk viscosity without pions}\label{sec:bv_without_pions_appendix}
In this section, we give the definitions of  $G'$, $H'$, $K'$, and $L'$, in the bulk viscosity expression Eq.~\ref{eq:bv_without_pions}
\begin{align}
    G' &= [\lambda_2\lambda_6+\lambda_1(\lambda_2+\lambda_6)]\left\{(A_1B_2-A_2B_1)^2\lambda_1+\left[(A_1-A_2)B_2+A_1C_2\right]^2\lambda_2+\left[A_2(B_1-B_2)+A_1C_2\right]^2\lambda_6\right\}\\
    H' &= A_1^2\lambda_1+A_2^2\lambda_2+(A_1-A_2)^2\lambda_6\\
    K' &= [\lambda_2\lambda_6+\lambda_1(\lambda_2+\lambda_6)]^2\left[B_2^2-B_1(B_2+C_2)\right]^2\\
    L' &= B_1^2\lambda_1^2+2B_2^2\lambda_1\lambda_2+(B_2+C_2)^2\lambda_2^2+2(B_1-B_2)^2\lambda_1\lambda_6+2C_2^2\lambda_2\lambda_6+(B_2-B_1-C_2)^2\lambda_6^2
\end{align}
Ref.~\cite{Alford:2021lpp} studied bulk viscosity in neutrino-trapped $npe\mu$ matter in the cases where $\lambda_6\rightarrow 0$ and $\lambda_6\rightarrow \infty$.  In the limit $\lambda_6\rightarrow 0$, our expression simplifies to
\begin{equation}
    \zeta = \dfrac{\lambda_1\lambda_2\left\{(A_1B_2-A_2B_1)^2\lambda_1+\left[(A_1-A_2)B_2+A_1C_2\right]^2\lambda_2\right\}+(A_1^2\lambda_1+A_2^2\lambda_2)\omega^2}{\lambda_1^2\lambda_2^2\left[B_2^2-B_1(B_2+C_2)\right]^2+\left[B_1^2\lambda_1^2+2B_2^2\lambda_1\lambda_2+(B_2+C_2)^2\lambda_2^2\right]\omega^2+\omega^4}
\end{equation}
In the limit where $\lambda_6\rightarrow\infty$,
\begin{equation}
    \zeta = \dfrac{\left[A_2(B_1-B_2)+A_1C_2\right]^2(\lambda_1+\lambda_2)}{\left[B_2^2-B_1(B_2+C_2)\right]^2(\lambda_1+\lambda_2)^2+(B_2-B_1-C_2)^2\omega^2}.
\end{equation}
\section{Bulk viscosity $\lambda_3\rightarrow \infty$}\label{appendix:bv_l3inf}
In this section, we give the definitions of $Q,R,U$, and $W$ in the expression for the bulk viscosity (Eq.~\ref{eq:bulk_viscosity_l3infinity}).  We set $D_2 = C_3=0$, consistent with our EoS.
\begin{subequations}
\begin{align}
    Q &= \left[(\lambda_1+\lambda_4)(\lambda_2+\lambda_5)+(\lambda_1+\lambda_2+\lambda_4+\lambda_5)\lambda_6\right]\\
    &\times \Big\{\big(\lambda_1+\lambda_4\big)\bigg[A_2B_1-\dfrac{B_2\left[A_3(B_1-B_2)+A_2B_2+A_1D_3\right]}{B_2+D_3}\bigg]^2+\big(\lambda_2+\lambda_5\big)\bigg[A_1C_2+\dfrac{B_2\left[(A_1-A_2)D_3-A_3C_2\right]}{B_2+D_3}\bigg]^2\nonumber\\
    &+\lambda_6\bigg[A_2(B_1-B_2)+A_1C_2-\dfrac{A_3B_2(B_1-B_2+C_2)}{B_2+D_3}\bigg]^2\Big\}\nonumber\\
    R &= (\lambda_1+\lambda_4)\left[A_1-\dfrac{A_3B_2}{B_2+D_3}\right]^2+(\lambda_2+\lambda_5)\left[A_2-\dfrac{A_3B_2}{B_2+D_3}\right]^2+\lambda_6(A_1-A_2)^2\\
    U &= \left[(\lambda_1+\lambda_4) (\lambda_2+\lambda_5)+(\lambda_1+\lambda_2+\lambda_4+\lambda_5)\lambda_6\right]^2 \left\{B_1C_2+\dfrac{B_2\left[(B_1-B_2)D_3-B_2C_2\right]}{B_2+D_3}\right\}^2\\
    W &= \left(B_1-\dfrac{B_2^2}{B_2+D_3}\right)^2(\lambda_1+\lambda_4)^2+\left(C_2+\dfrac{B_2D_3}{B_2+D_3}\right)^2(\lambda_2+\lambda_5)^2+2\dfrac{B_2^2D_3^2}{(B_2+D_3)^2}(\lambda_1+\lambda_4)(\lambda_2+\lambda_5)\\
    &+2(B_1-B_2)^2(\lambda_1+\lambda_4)\lambda_6+2C_2^2(\lambda_2+\lambda_5)\lambda_6+(B_1-B_2+C_2)^2\lambda_6^2.\nonumber
\end{align}
\end{subequations}
\section{Thermodynamic Jacobian calculations}\label{appendix:jacobians}
We go through the thermodynamic Jacobian calculations to relate the compressibility of nuclear matter to the susceptibilities, which appear in the bulk viscosity formulas.  For a review of thermodynamic Jacobians, see \cite{swendsen2020introduction}.  Temperature is held constant in all formulas in this Appendix, so we leave it implicit in the Jacobians and derivatives.

Without pions, a few key compressibilities can be written
\begin{subequations}
\begin{align}
    \dfrac{\partial P}{\partial n_B}\bigg\vert_{x_{\mu},\delta\mu_1}&\equiv\dfrac{\partial\left(P,x_{\mu},\delta\mu_1\right)}{\partial\left(n_B,x_{\mu},\delta\mu_1\right)}=\dfrac{\partial\left(P,x_{\mu},\delta\mu_1\right)}{\partial\left(n_B,x_p,x_{\mu}\right)}\dfrac{\partial\left(n_B,x_p,x_{\mu}\right)}{\partial\left(n_B,x_{\mu},\delta\mu_1\right)} = \dfrac{\partial\left(P,\delta\mu_1,x_{\mu}\right)}{\partial\left(n_B,x_{p},x_{\mu}\right)}\dfrac{\partial\left(x_{\mu},x_p\right)}{\partial\left(x_{\mu},\delta\mu_1\right)}\\
    &=\dfrac{\partial\left(P,\delta\mu_1\right)}{\partial\left(n_B,x_{p}\right)}\bigg/ \dfrac{\partial \delta\mu_1}{\partial x_p}\bigg\vert_{n_B,x_{\mu}} = \dfrac{\partial P}{\partial n_B}\bigg\vert_{x_p,x_{\mu}}-\dfrac{\partial P}{\partial x_p}\bigg\vert_{n_B,x_{\mu}}\dfrac{\partial x_p}{\partial\delta\mu_1}\bigg\vert_{n_B,x_{\mu}}\dfrac{\partial\delta\mu_1}{\partial n_B}\bigg\vert_{x_p,x_{\mu}}=\dfrac{\partial P}{\partial n_B}\bigg\vert_{x_p,x_{\mu}}+\dfrac{1}{n_B}\dfrac{A_1^2}{B_1}\nonumber\\
    \dfrac{\partial P}{\partial n_B}\bigg\vert_{x_p-x_{\mu},\delta\mu_2}&\equiv\dfrac{\partial\left(P,x_p-x_{\mu},\delta\mu_2\right)}{\partial\left(n_B,x_p-x_{\mu},\delta\mu_2\right)}=\dfrac{\partial\left(P,x_p-x_{\mu},\delta\mu_2\right)}{\partial\left(n_B,x_p,x_{\mu}\right)}\dfrac{\partial\left(n_B,x_p,x_{\mu}\right)}{\partial\left(n_B,x_p-x_{\mu},\delta\mu_2\right)}\\
    &=\dfrac{\partial\left(P,x_p-x_{\mu},\delta\mu_2\right)}{\partial\left(n_B,x_p,x_{\mu}\right)}\bigg/\dfrac{\partial\left(x_p-x_{\mu},\delta\mu_2\right)}{\partial\left(x_p,x_{\mu}\right)}=\dfrac{\partial P}{\partial n_B}\bigg\vert_{x_p,x_{\mu}}-\left(\dfrac{\partial P}{\partial x_p}\dfrac{\partial \delta\mu_2}{\partial n_B}+\dfrac{\partial P}{\partial x_{\mu}}\dfrac{\partial\delta\mu_2}{\partial n_B}\right)\bigg/\left(\dfrac{\partial\delta\mu_2}{\partial x_{\mu}}+\dfrac{\partial\delta\mu_2}{\partial x_p}\right)\nonumber\\
    &= \dfrac{\partial P}{\partial n_B}\bigg\vert_{x_p,x_{\mu}}+\dfrac{1}{n_B}\dfrac{A_2^2}{B_2+C_2}    \nonumber\\
    \dfrac{\partial P}{\partial n_B}\bigg\vert_{x_p,\delta\mu_1-\delta\mu_2}&\equiv\dfrac{\partial\left(P,x_p,\delta\mu_1-\delta\mu_2\right)}{\partial\left(n_B,x_p,x_{\mu}\right)}\dfrac{\partial\left(n_B,x_p,x_{\mu}\right)}{\partial\left(n_B,x_p,\delta\mu_1-\delta\mu_2\right)} = \dfrac{\partial\left(P,\delta\mu_1-\delta\mu_2\right)}{\partial\left(n_B,x_{\mu}\right)}\bigg/\dfrac{\partial\left(\delta\mu_1-\delta\mu_2\right)}{\partial x_{\mu}}\\
    &= \dfrac{\partial P}{\partial n_B}\bigg\vert_{x_p,x_{\mu}}-\dfrac{\partial P}{\partial x_{\mu}}\dfrac{\partial x_{\mu}}{\partial\left(\delta\mu_1-\delta\mu_2\right)}\dfrac{\partial\left(\delta\mu_1-\delta\mu_2\right)}{\partial n_B}=\dfrac{\partial P}{\partial n_B}\bigg\vert_{x_p,x_{\mu}}-\dfrac{1}{n_B}\dfrac{\left(A_1-A_2\right)^2}{C_1-C_2}\nonumber\\
    &=\dfrac{\partial P}{\partial n_B}\bigg\vert_{x_p,x_{\mu}}-\dfrac{1}{n_B}\dfrac{(A_1-A_2)^2}{B_2-B_1-C_2}.\nonumber
\end{align}
\end{subequations}
For readability, after the first line, we drop the variables that are held constant, but the independent variables are always $\{n_B,x_p,x_{\mu}\}$ (and $T$, which is implicit in this section).  
\section{Partial bulk viscosities without $\lambda_3\rightarrow\infty$ limit}\label{appendix:partialbv}
In this section, we present the formulas for the partial bulk viscosities in dense matter containing thermal pions, without taking the limit of infinitely fast $\lambda_3$.  These expressions are really to be compared with the $\lambda_3\rightarrow\infty$ expressions in the main text.  

The beta equilibration rates are 
\begin{subequations}
\begin{align}
    \gamma_1 &\equiv  -B_1\lambda_1     \\
    \gamma_2 &\equiv  -\left(B_2+C_2\right)\lambda_2     \\
    \gamma_3 &\equiv  -\left(B_2+D_3\right)\lambda_3     \\
    \gamma_4 &\equiv  \left(B_2-B_1-D_3\right)\lambda_4     \\
    \gamma_5 &\equiv  -\left(C_2+D_3\right)\lambda_5     \\
    \gamma_6 &\equiv \left(B_2-B_1-C_2\right)\lambda_6
\end{align}
\end{subequations}
and the partial bulk viscosities are given by
\begin{subequations}
\begin{align}
\zeta_1 &=  -\dfrac{A_1^2}{B_1}\dfrac{\gamma_1}{\gamma_1^2+\omega^2} \label{eq:bv1_l3zero}  \\
\zeta_2 &= -\dfrac{A_2^2}{(B_2+C_2)}\dfrac{\gamma_2}{\gamma_2^2+\omega^2}    \\
\zeta_3 &= -\dfrac{A_3^2}{B_2+D_3}\dfrac{\gamma_3}{\gamma_3^2+\omega^2} \label{eq:zeta3eqn}   \\
\zeta_4 &= \dfrac{(A_1-A_3)^2}{B_2-B_1-D_3}\dfrac{\gamma_4}{\gamma_4^2+\omega^2}    \\
\zeta_5 &= -\dfrac{(A_2-A_3)^2}{C_2+D_3}\dfrac{\gamma_5}{\gamma_5^2+\omega^2}   \\
\zeta_6 &=  \dfrac{(A_1-A_2)^2}{B_2-B_1-C_2}\dfrac{\gamma_6}{\gamma_6^2+\omega^2}. \label{eq:bv6_l3zero}
\end{align}
\end{subequations}
The maximum values of the partial bulk viscosities can be related to the compressibilities via the same procedure as given in Appendix \ref{appendix:jacobians}.  The results are
\begin{subequations}
    \begin{align}
        \zeta_{1,\text{max}} &= \dfrac{n_B}{2\omega}\left( \dfrac{\partial P}{\partial n_B}\bigg\vert_{x_p,x_{\mu},x_{\pi},T}  - \dfrac{\partial P}{\partial n_B}\bigg\vert_{x_{\mu},x_{\pi},\delta\mu_1,T} \right)\label{eq:partial_bv_max_compress_1}\\
        \zeta_{2,\text{max}} &= \dfrac{n_B}{2\omega}\left( \dfrac{\partial P}{\partial n_B}\bigg\vert_{x_p,x_{\mu},x_{\pi},T}  - \dfrac{\partial P}{\partial n_B}\bigg\vert_{x_p-x_{\mu},x_{\pi},\delta\mu_2,T} \right)\\
        \zeta_{3,\text{max}} &= \dfrac{n_B}{2\omega}\left( \dfrac{\partial P}{\partial n_B}\bigg\vert_{x_p,x_{\mu},x_{\pi},T}  - \dfrac{\partial P}{\partial n_B}\bigg\vert_{x_{\mu},x_p-x_{\pi},\delta\mu_3,T} \right)\\
        \zeta_{4,\text{max}} &= \dfrac{n_B}{2\omega}\left( \dfrac{\partial P}{\partial n_B}\bigg\vert_{x_p,x_{\mu},x_{\pi},T}  - \dfrac{\partial P}{\partial n_B}\bigg\vert_{x_p,x_{\mu},\delta\mu_1-\delta\mu_3,T} \right)\\
        \zeta_{5,\text{max}} &= \dfrac{n_B}{2\omega}\left( \dfrac{\partial P}{\partial n_B}\bigg\vert_{x_p,x_{\mu},x_{\pi},T}  - \dfrac{\partial P}{\partial n_B}\bigg\vert_{x_p,x_{\mu}+x_{\pi},\delta\mu_2-\delta\mu_3,T} \right)\\
        \zeta_{6,\text{max}} &= \dfrac{n_B}{2\omega}\left( \dfrac{\partial P}{\partial n_B}\bigg\vert_{x_p,x_{\mu},x_{\pi},T}  - \dfrac{\partial P}{\partial n_B}\bigg\vert_{x_{p},x_{\pi},\delta\mu_1-\delta\mu_2,T} \right).\label{eq:partial_bv_max_compress_6}
    \end{align}
\end{subequations}

\end{widetext}
\bibliography{references}

\end{document}